%% file: modular_Higgs.tex
\RequirePackage{pdf14}

\documentclass{article}

\usepackage{array,setspace,mathrsfs,amsfonts,amsmath,yfonts,dsfont,bbm,colonequals,amscd}
\usepackage{relsize,suffix,mathtools,cancel,bbm,tikz-cd}

\usetikzlibrary{positioning}
\usetikzlibrary{chains}
\usetikzlibrary{arrows,fit,decorations.pathreplacing}
\tikzstyle{every picture}+=[remember picture]
\tikzstyle{na} = [baseline=-.5ex]

\usepackage{./aux/jheppub}

\input{./aux/topmatter}

\input{./sections/titlepage}

\begin{document}

\maketitle

\input{./sections/S1}
\input{./sections/S2}
\input{./sections/S3}
\input{./sections/S4}
\input{./sections/S5}
\input{./sections/S6}
\input{./sections/Acknowledgments}

\appendix

\input{./sections/A1}
\input{./sections/A2}
\input{./sections/A3}
\input{./sections/A4}
\input{./sections/A5}

\bibliographystyle{./aux/JHEP}
\small\baselineskip=.93\baselineskip
\bibliography{./aux/biblio}

\end{document}

%% file: aux/topmatter.tex
\numberwithin{equation}{section}
\newtheorem{conj}{Conjecture}
\newtheorem{thm}{Theorem}

\def\NO#1#2{{\rm NO}(#1,#2)}

\def\bea{\begin{eqnarray}}
\def\eea{\end{eqnarray}}
\def\ceq{\colonequals}

\def\eqq{\equiv}
\def\leqs{\leqslant}
\def\geqs{\geqslant}
\def\vph{\varphi}
\def\vth{\vartheta}
\def\g{\gamma}
\def\G{\Gamma}
\def\th{\theta}
\def\Th{\Theta}
\def\tpdf{\texorpdfstring}

\DeclarePairedDelimiterX\MeijerM[3]{\lparen}{\rparen}%
{\begin{smallmatrix}#1 \\ #2\end{smallmatrix}\delimsize\vert\,#3}

\def\zeromode#1{o\left(#1\right)}

\newcommand\MeijerG[8][]{%
  G^{\,#2,#3}_{#4,#5}\MeijerM[#1]{#6}{#7}{#8}}

\WithSuffix\newcommand\MeijerG*[7]{%
  G^{\,#1,#2}_{#3,#4}\MeijerM*{#5}{#6}{#7}}

\def\VnoH{\VV_+}

\def\SU{\mathrm{SU}}
\def\SO{\mathrm{SO}}
\def\USp{\mathrm{USp}}

\def\Vir{{\rm Vir}}
\def\AD{{\rm AD}}

\def \beg#1{\begin{#1}} 
\def \be{\beg{equation}}
\def \bea{\beg{eqnarray}}
\def \eea{\end{eqnarray}}
\def \ee{\end{equation}}

\def \goodchi{\protect\raisebox{1pt}{$\chi$}}
\def \qq{\mathbbmtt{Q}\,}

\def \af{\mf{a}}
\def \df{\mf{d}}
\def \ef{\mf{e}}
\def \gf{\mf{g}}
\def \uf{\mf{u}}
\def \glf{\mf{gl}}
\def \slf{\mf{sl}}
\def \suf{\mf{su}}
\def \sof{\mf{so}}

\def \ef{\mf{e}}

\def \ceq{\colonequals}

\def \restr#1#2{{\left.\kern-\nulldelimiterspace#1\vphantom{\big|}\right|_{#2}}}

\def \ll{\lambda}

\def \hf{\frac12}

\def \mf{\mathfrak}

\def \nn{\nonumber}

\def \wt{\widetilde}

\def \eg{{\it e.g.}}
\def \ie{{\it i.e.}}

\def \Tr{{\rm Tr}}
\def \STr{{\rm STr}}

\def \zb{{\bar z}}

\def \Cb{\mathbb{C}}
\def \Eb{\mathbb{E}}

\def \Hb{\mathbb{H}}

\def \Ob{\mathbb{O}}
\def \Qb{\mathbb{Q}}

\def \Sb{\mathbb{S}}
\def \Zb{\mathbb{Z}}

\def \BB{\mathcal{B}}
\def \CC{\mathcal{C}}
\def \DD{\mathcal{D}}

\def \FF{\mathcal{F}}
\def \GG{\mathcal{G}}
\def \HH{\mathcal{H}}
\def \II{\mathcal{I}}

\def \MM{\mathcal{M}}
\def \NN{\mathcal{N}}
\def \OO{\mathcal{O}}
\def \PP{\mathcal{P}}
\def \QQ{\mathcal{Q}}
\def \RR{\mathcal{R}}
\def \SS{\mathcal{S}}

\def \VV{\mathcal{V}}
\def \WW{\mathcal{W}}

\def \RRR{\mathfrak{R}}

%% file: sections/titlepage.tex
\title{Vertex operator algebras, Higgs branches, and modular differential equations}

\author[1,2,3]{Christopher Beem}
\author[4]{and Leonardo Rastelli}

\affiliation[1]{Mathematical Institute, University of Oxford, Woodstock Road, Oxford OX2 6GG, United Kingdom}
\affiliation[2]{St. John's College, University of Oxford, St. Giles Road, Oxford OX1 3JP, United Kingdom}
\affiliation[3]{School of Natural Sciences, Institute for Advanced Study, Einstein Drive, Princeton,  NJ 08540, USA}
\affiliation[4]{C.~N.~Yang Institute for Theoretical Physics, Stony Brook University, Stony Brook, NY 11794-3840, USA}

\emailAdd{christopher.beem@maths.ox.ac.uk}
\emailAdd{leonardo.rastelli@stonybrook.edu}

\preprint{YITP-SB-17-27}
\bigskip
\abstract{
Every four-dimensional $\NN=2$ superconformal field theory comes equipped with an intricate algebraic invariant, the associated vertex operator algebra. The relationships between this invariant and more conventional protected quantities in the same theories have yet to be completely understood. In this work, we aim to characterize the connection between the Higgs branch of the moduli space of vacua (as an algebraic geometric entity) and the associated vertex operator algebra. Ultimately our proposal is simple, but its correctness requires the existence of a number of nontrivial null vectors in the vacuum Verma module of the vertex operator algebra. Of particular interest is one such null vector whose presence suggests that the Schur index of any $\NN=2$ SCFT should obey a finite order modular differential equation. By way of the ``high temperature'' limit of the superconformal index, this allows the Weyl anomaly coefficient $a$ to be reinterpreted in terms of the representation theory of the associated vertex operator algebra. We illustrate these ideas in a number of examples including a series of rank-one theories associated with the ``Deligne-Cvitanovi\'c exceptional series'' of simple Lie algebras, several families of Argyres-Douglas theories, an assortment of class $\SS$ theories, and $\NN=4$ super Yang-Mills with $\suf(n)$ gauge group for small-to-moderate values of $n$.
}

\keywords{Supersymmetry, conformal field theory, superconformal field theory, vertex operator algebra, chiral algebra, vector-valued modular forms, superconformal index, Higgs branch, associated variety, Argyres-Douglas theory}

%% file: sections/S1.tex

\section{Introduction and summary}
\label{sec:intro_summary}

In \cite{Beem:2013sza}, a new algebraic invariant of four-dimensional $\NN=2$ superconformal field theories (SCFTs) was introduced --- a vertex operator algebra (VOA) that encodes the spectrum and OPE coefficients of \emph{Schur operators}.\footnote{For succinctness we adopt the terminology ``vertex operator algebra'' throughout this work, though in many instances it would be more accurate to write vertex operator \emph{super}algebra.} The connection between four-dimensional operator algebras and VOAs provides a powerful framework for the analysis of four-dimensional strongly interacting SCFTs. It also leads to surprising predictions for a large new class of VOAs. For example, the rigidity of VOAs generated by affine currents, supercurrents, and stress tensors makes it possible to establish novel unitarity bounds for flavor central charges and Weyl anomaly coefficients \cite{Beem:2013sza,Liendo:2015ofa,Lemos:2015orc,Lemos:2016xke,Cornagliotto:2017dup} and has provided important input into the numerical superconformal bootstrap program in four (and six) dimensions \cite{Beem:2013qxa,Beem:2014zpa,Beem:2015aoa,Lemos:2016xke,Beem:2016wfs}. On the other hand, established results concerning the duality web of $\NN=2$ SCFTs of class $\SS$ imply the existence of a remarkable family of two-dimensional TQFTs valued in vertex operator algebras \cite{Beem:2014rza}.

An important class of questions regarding these VOAs pertains to whether and how they reflect better-understood features of their parent four-dimensional theories. For example, $\NN=2$ SCFTs come equipped with moduli spaces of vacua -- roughly speaking the union of a Higgs branch and a Coulomb branch, though there may be nontrivial intersections between the two -- and in many instances the structure of these finite-dimensional spaces is highly accessible even when other aspects of the theory are not. Furthermore, the low energy dynamics at smooth points in the moduli space of vacua are often tractable. Indeed, interesting steps towards relating moduli space physics to the associated VOA have been taken in \cite{Cordova:2015nma,Cordova:2016uwk,Cordova:2017mhb}, where the spectrum of BPS states on the Coulomb branch has been related to the Schur index of the interacting SCFT.

The focus of the present paper is the relationship between the associated VOA and the Higgs branch of vacua of an $\NN=2$ SCFT. That some relation should exist is \emph{prima facie} plausible since it was shown in \cite{Beem:2013sza} that generators of the Higgs chiral ring must descend to (strong) generators of the associated VOA. Indeed, in several of the examples presented in that work it was found that the generators of the VOA were in one-to-one correspondence with those of the Higgs chiral ring. However, in that same work examples were presented wherein some VOA generators were unrelated to any familiar chiral ring operators. This suggests that the identification of Higgs branch operators must be at least somewhat nuanced. It would be useful to know whether there is exists a sharp characterization of the relationship between the Higgs branch and the VOA that can be formulated intrinsically in terms of the VOA. Abstractly, such a relationship should amount to an operation that takes as its argument a vertex operator algebra and returns a commutative, associative, $\Cb$-algebra. Better yet, since the Higgs branch is holomorphic symplectic, one may hope that the operation will additionally return the holomorphic Poisson bracket.\footnote{One could in principle hope to recover the construction of the full hyperk\"ahler structure on the Higgs branch. We do not discuss this possibility in the present work.}

We propose that such an operation does indeed exist and can be formulated purely in the language of the VOA. That there exists \emph{some} operation to turn VOAs into commutative associative algebras has been well known at least since the work of Zhu \cite{ZhuThesis}. The resulting algebra is sometimes referred to as ``Zhu's commutative algebra'' or ``Zhu's $C_2$ algebra'' or, as we shall refer to it in this paper, ``\emph{the} $C_2$ algebra''. In fact, the $C_2$ algebra automatically comes equipped with a Poisson bracket, so this is an immediate candidate for the Higgs chiral ring as a Poisson algebra. However, it is easy to see that this naive proposal fails. In particular, all strong generators of a VOA necessarily descend to nontrivial generators of the $C_2$ algebra, but we know that in many examples only a strict subset of these operators correspond to operators in the Higgs chiral ring.

A hint towards the correct prescription comes from the empirical fact (provable in Lagrangian theories) that the Higgs chiral ring is not just a commutative, associative $\Cb$-algebra; it is a \emph{reduced} commutative, associative $\Cb$-algebra. This is to say, the Higgs branch is an algebraic variety and the Higgs chiral ring is the coordinate ring of that variety. On the other hand, the $C_2$ algebra has no \emph{a priori} reason to be reduced, and indeed in many examples of interest (including any $C_2$ co-finite vertex operator algebra, about which more later) they are not. Our proposal is that this is the only obstruction to identifying the $C_2$ algebra with the Higgs chiral ring, so it should simply be corrected by reducing the $C_2$ algebra.\footnote{We recall that the operation of reducing a commutative algebra is well defined. It amounts to passing to the quotient with respect to the nilradical, which is the unique ideal comprising all nilpotent elements of the original algebra.} What we are proposing, then, is that the Higgs branch is equivalent to the \emph{associated variety} of the VOA, as introduced by Arakawa in \cite{Arakawa:2010klr}.\footnote{This conjecture was originally announced in \cite{rastelli_string_math}, and has since been featured in a number of papers \cite{Arakawa:2016hkg,Song:2017oew}.} In fact, the reduced $C_2$ algebra is still Poisson, and we propose that it will match the Higgs chiral ring as a Poisson algebra.

A simple argument based on selection rules shows that as a vector space the Higgs chiral ring always injects into the $C_2$ algebra. The more nontrivial part of the conjecture is that every complementary element of the $C_2$ algebra must be nilpotent. In particular, any $C_2$ algebra generator that is not related to a Higgs chiral ring generator should be nilpotent. This nilpotency requirement is nontrivial and requires that the vacuum Verma module built from the strong VOA generators contains certain null vectors. This is particularly interesting because nontrivial null vectors in the vacuum Verma module of a VOA often lead to nice simplifications at the level of observables and representation theory, and this is the second major subject of our study.

We focus our attention on a particular type of null vector that must always be present if our characterization of the Higgs branch is to hold. This is the null vector that is responsible for the stress tensor being nilpotent in the $C_2$ algebra. As we review in Section \ref{sec:modular}, null vectors of this type are known in many (possibly all) instances to give rise to \emph{linear modular differential equations} (LMDEs) that are satisfied by the (super)characters of sufficiently nice VOA modules, the vacuum module in particular. 

The existence of a finite order linear modular differential operator (LMDO) that annihilates the vacuum character of a VOA has important consequences for the modular properties of the character. In the case at hand, we are discovering good modular properties for the (unflavored) Schur limit of the superconformal index of the parent SCFT. The modular properties of this limit of the index were previously investigated in \cite{Razamat:2012uv}, but here we will see a more specific structure. In particular, the unflavored Schur index transforms as a vector-valued (quasi-)modular form of weight zero under ${\rm PSL}(2,\Zb)$ or $\Gamma^0(2)\subset{\rm PSL}(2,\Zb)$, depending on whether the vacuum character is an expansion in integer or half-integer powers of $q$. The modular equation for the unflavored Schur index can also be upgraded to include fugacities for flavor symmetries in the index \cite{flavored_characters}. It is an interesting question whether these properties of the Schur index could be proven independently of considerations of specific null vectors, perhaps by a clever analysis of the $\Sb^3 \times \Sb^1$ partition function that computes the index.

The relationship between null vectors of the type we are considering and modular equations has a somewhat complicated history \cite{Gaberdiel:2007ve,Gaberdiel:2008pr,Gaberdiel:2008ma}. However, recent work of Arakawa and Kawasetsu proves that when the associated variety of a VOA is symplectic,\footnote{More precisely, the associated variety --- which may not be smooth --- should have finitely many symplectic leaves.} then an LMDE for the vacuum character is guaranteed \cite{Arakawa:2016hkg}. Thus the existence of an LMDE for the Schur index is a consequence of our conjecture that the the Higgs branch is the associated variety. In general it is much easier to test for the existence of an LMDE (of order less than some chosen number) that annihilates the Schur index than it is to determine the associated variety of the associated VOA. This then gives us a useful consistency check for our conjecture in many examples.

Having control of the modular behavior of the index also allows for a connection with the so-called ``Cardy behavior'' of the index, \ie, the scaling of the index as $q\to1$ \cite{DiPietro:2014bca}. A generalization of the anomaly-based arguments of  \cite{DiPietro:2014bca} suggests that in this limit the scaling of the index should be determined by the $a_{\rm 4d}$ and $c_{\rm 4d}$ Weyl anomaly coefficients of the SCFT  \cite{Buican:2015ina, DPKR, Ardehali:2015bla} (see also \cite{Cecotti:2015lab}) according to
\begin{equation}
\lim_{q\to1}\II(q)\sim e^{\frac{8\pi^2}{\beta}(c_{4d}-a_{4d})}~,
\end{equation}
where $q\equiv\exp(-\beta)$. The Weyl anomaly coefficient $a_{\rm 4d}$ has previously been largely absent from discussions of the associated VOA of an $\NN=2$ SCFT. We see here that this coefficient makes itself known by way of the non-vacuum modules of the associated VOA whose characters transform among themselves under modular transformations.

These ideas are investigated to varying degrees in an extensive set of examples. An exceptionally well-behaved set of examples are the rank-one theories that are engineered by placing a single D3 brane at an $\NN=2$ $F$-theory singularity. In fact, as was observed in \cite{Beem:2013sza}, from the point of view of the associated VOA it is natural to extend these theories to include two proposed new theories with ${\mathfrak g}_2$ and ${\mathfrak f}_4$ flavor symmetry, respectively. For these theories, our identification of the Higgs branch with the associated variety of the associated VOA has been proven fairly recently in the mathematics literature \cite{Arakawa:2015jya}. Furthermore, the modular differential operator that annihilates the Schur index of these theories can be determined and solved exactly.

We also consider four infinite families of Argyres-Douglas SCFTs. The first series, known as the $(A_1, A_{2n})$ theories, all have no Higgs branch of vacua. By our identification of the Higgs branch, the associated VOAs should be $C_2$ co-finite. Indeed, the VOAs for these SCFTs have previously been identified \cite{rastelli_harvard} as the Virasoro VOA at the values of the central charge appropriate for the $(2, 2n+3)$ non-unitary minimal models. The second series, known as the $(A_1,D_{2n+1})$ theories, all have Higgs branches given by the simplest canonical singularity $\Cb^2/\Zb_2$. The associated VOAs for these theories are $\suf(2)$ affine current algebras at levels $-4n/(2n+1)$. In particular, these are \emph{admissible} levels, and a result of \cite{Arakawa:2010ni} establishes that the associated variety for these VOAs indeed matches the Higgs branch. We further consider the $(A_1, A_{2n-1})$ theories, which have as their Higgs branches the $A$-series of canonical singularities, $\Cb^2/\Zb_n$. We propose that the associated VOAs for these theories are the generalized Bershadsky-Polyakov algebras of \cite{Bershadsky:1990bg,Polyakov:1989dm,Feigin:2004wb,Creutzig:2013pda} at specific values of the central charge. The identification of the Higgs branch with the associated variety for these algebras follows from results in \cite{Arakawa:2010ni}. Finally, we address the $(A_1, D_{2n+2})$ theories. We find a representation of the Higgs branches of these theories as quiver varieties using three-dimensional mirror symmetry. The associated VOAs for these theories were identified in \cite{Creutzig:2017qyf} as certain $\slf(n+2)$ quantum Drinfel'd-Sokolov reductions associated to sub-subregular nilpotent orbits, and by appealing to a theorem of Maffei \cite{Maffei} we verify that the associated varieties of these vertex algebras match the aforementioned quiver varieties.

We further investigate the existence of LMDEs for the unflavored Schur index of a large class of additional SCFTs. These include class $\SS$ theories of $A_1$ type and $\NN=4$ super Yang-Mills theories with low-rank gauge groups, in addition to the rank-three trinion theory $T_4$. For these theories we have for the most part made no effort to explicitly compute the null vectors that gives rise to the differential equations or to determine the associated variety. We take the successful discovery of an LMDE as suggestive evidence for our overall picture. We also comment on some curious patterns in the orders of the LMDEs and the structure of their solutions for the theories we have studied.

%% file: sections/S2.tex

\section{Vertex operator algebras and the Higgs branch}
\label{sec:higgs_from_voa}

We start by reviewing some essential aspects of the correspondence between vertex operator algebras and four-dimensional $\NN=2$ superconformal field theories (see \cite{Beem:2013sza} and \cite{Beem:2014rza} for a more complete treatment). Given an $\NN=2$ SCFT, the associated VOA is recovered by passing to the cohomology of a certain nilpotent fermionic generator of the $\NN=2$ superconformal algebra $\suf(2,2|2)$, which in the conventions of \cite{Beem:2013sza} is given by
\begin{equation}
\qq\ = \QQ^1_{-}+\wt\SS^{2\dot{-}}~. 
\end{equation}
The nontrivial cohomology classes of local operators inserted at the origin $(z=\bar z = w = \bar w = 0)$ have canonical representatives which are the \emph{Schur operators} \cite{Gadde:2011uv}.
These are local operators whose quantum numbers satisfy the relations\footnote{The first condition in \eqref{eq:schur_conditions} implies the second for representations of $\suf(2,2|2)$ that can appear in unitary superconformal theories \cite{Beem:2013sza}.}
\begin{equation}
\begin{split}
\label{eq:schur_conditions}
E - (j_1 + j_2) - 2R &= 0~,\\
r + j_1 - j_2 &= 0~,
\end{split}
\end{equation}
where $E$ is the conformal dimension and $(j_1, j_2, R, r)$ are eigenvalues with respect to appropriate Cartan elements of $\suf(2)_1\times\suf(2)_2\times\suf(2)_R\times\uf(1)_r$, respectively.\footnote{We are adopting conventions where the complex coordinates $z$ and $w$ transform with weights $(j_1,j_2)=(+\hf,+\hf)$ and $(j_1,j_2)=(+\hf,-\hf)$, respectively.} Schur operators are singled out by the fact that they contribute to the Schur limit of the superconformal index. Equivalently, they are operators that are nontrivially annihilated by $\QQ^1_{-}$ and $\wt\QQ_{2\dot{-}}$. Schur operators are always the highest weight states of their respective $\suf(2)_1\times\suf(2)_2\times\suf(2)_R$ modules. The various (unitary) supermultiplets that contain Schur operators and the positioning of Schur operators within those multiplets is summarized in Table \ref{tab:schurTable}. 

\renewcommand{\arraystretch}{1.5}
\begin{table}
\centering
\begin{tabular}{|l|l|l|l|l|}
\hline
\hline
Multiplet					& $\OO_{\rm Schur}$															& $h$			& $r$				& Lagrangian~``letters''				\\
\hline
$\hat\BB_R$					& $\Psi^{11\dots 1}$ 														& $R$			& $0$				& $Q$, $\wt{Q}$							\\
\hline
$\DD_{R(0,j_2)}$			& $\wt{\QQ}^1_{\dot +}\Psi^{11\dots1}_{\dot+\dots\dot+}$ 					& $R+j_2+1$		& $j_2+\frac12$ 	& $Q$, $\wt{Q}$, $\tilde\ll^1_{\dot+}$	\\
\hline
$\bar{\DD}_{R(j_1,0)}$		& ${\QQ}^1_{+}\Psi^{11\dots 1}_{+\dots+}$ 									& $R+j_1+1$		& $-j_1-\frac12$ 	& $Q$, $\wt{Q}$, $\ll^1_{+}$			\\
\hline
$\hat{\CC}_{R(j_1,j_2)}$	& ${\QQ}^1_{+}\wt{\QQ}^1_{\dot+}\Psi^{11\dots1}_{+\dots+\,\dot+\dots\dot+}$	& $R+j_1+j_2+2$	& $j_2-j_1$			& $D_{+\dot+}^{n}Q$, $D_{+\dot+}^{n}\wt{Q}$, $D_{+\dot+}^{n}\ll^1_{+}$, $D_{+\dot+}^{n}\tilde\ll^1_{\dot+}$	\\
\hline
\end{tabular}
\caption{\label{tab:schurTable} Summary of the appearance of Schur operators in short multiplets of the $\NN=2$ superconformal algebra, $\suf(2,2|2)$. The superconformal primary in a supermultiplet is denoted by $\Psi$. There is a single conformal primary Schur operator ${\OO}_{\rm Schur}$ in each listed superconformal multiplet. The holomorphic dimension $h$ and $U(1)_r$ charge $r$ of ${\OO}_{\rm Schur}$ are given in terms of the quantum numbers $(R,j_1,j_2)$ that label the shortened multiplet (left-most column). The schematic form that ${\OO}_{\rm Schur}$ can take in a Lagrangian theory is also indicated in terms of the elementary ``letters'' from which the operator may be built. The complex scalar fields in a hypermultiplet are denoted by $Q$ and $\widetilde Q$, while the left- and right-handed fermions in a vector multiplet are denoted by $\ll_{\alpha}^\II$ and $\tilde \ll_{\dot \alpha}^\II$. Gauge covariant derivatives are denoted by $D_{\alpha \dot \alpha}$.} 
\end{table}

Finite linear combinations of local operators inserted away from the origin cannot define nontrivial $\qq$-cohomology classes unless $w=\bar w=0$. A canonical choice of representatives for local operators inserted on the $w=\bar w=0$ plane, $\Cb_{[z,\zb]}$, is given by \emph{twisted translated} Schur operators,
\begin{equation}
\label{eq:twisted_translated}
\OO(z)\equiv e^{zL_{-1}+\bar{z}(\overline{L}_{-1}+R^-)}\OO_{\rm Sch}(0)e^{-zL_{-1}-\bar{z}(\overline{L}_{-1}+R^-)}~,
\end{equation}
where $L_{-1}$ and $\bar{L}_{-1}$ are the generators of holomorphic and antiholomorphic translations in $\Cb_{[z,\zb]}$, $R^{-}$ is the lowering operator of $\suf(2)_R$, and $\OO_{\rm Sch}(z,\zb)$ is a Schur operator. The OPE of twisted-translated Schur operators, taken at the level of $\qq$-cohomology, is $\slf(2)_z$ covariant and $\slf(2)_{\bar z}$ invariant, with the holomorphic dimension of the twisted-translated operator $\OO(z)$ being determined in terms of the quantum numbers of the corresponding Schur operator according to
\begin{equation}
\label{eq:holomorphic_dimensions}
[L_0,\OO(z)] = h\OO(z)~,\qquad h = \frac{E + j_1 + j_2}{2} = E - R~.
\end{equation}
This holomorphic OPE endows the vector space of Schur operators with the structure of a vertex operator algebra.

It immediately follows from \eqref{eq:schur_conditions} and \eqref{eq:holomorphic_dimensions} that the conformal grading of the VOA is $\hf\Zb_{\geqslant0}$-valued and the only operator with $h=0$ is the identity operator. In any $\NN=2$ SCFT that possesses an energy-momentum tensor, the associated VOA will have its $\slf(2)_z$ symmetry further enhanced to Virasoro symmetry with central charge determined by the Weyl anomaly coefficient $c_{4d}$ of the four-dimensional parent according to $c_{2d}=-12c_{4d}$. Thus for every local $\NN=2$ SCFT the associated VOA has a Virasoro subalgebra that enhances $\slf(2)_z$.

The VOA operator product does not conserve $\suf(2)_R$ charge due to the twisted translation construction, but $\uf(1)_r$ charge is conserved. However, the $\uf(1)_r$ symmetry is a nonlocal symmetry in the VOA, in the sense that there is no local VOA current that generates $\uf(1)_r$ rotations. Finally, we point out that in general, Schur operators may be fermionic, so the VOA may actually be a vertex operator superalgebra. Note, though, that both bosonic and fermionic Schur operators may have conformal weights that are either integers or half integers. In summary:
\begin{quote}\emph{
\noindent \!\!The VOA associated to an $\NN=2$ SCFT is a $\hf\Zb_{\geqslant0}$ graded, conformal vertex operator (super)algebra with a nonlocal $U(1)_r$ symmetry; its underlying vector space is canonically isomorphic as a $\hf\Zb_{\geqslant0}\times\hf\Zb$ graded vector superspace to the space of Schur operators.}
\end{quote}

\subsection{Higgs and Hall-Littlewood chiral rings}
\label{subsec:higgs-chiral-ring}

Some Schur operators also survive more restrictive and algebraically simpler reductions of the SCFT operator algebra. A case of particular importance is that of the $\hat\BB_{R}$ operators detailed in the first row of Table \ref{tab:schurTable}. These obey a more stringent shortening condition $E = 2R$, which implies that they are non-trivially annihilated by the \emph{four} supercharges $\QQ^1_{\alpha}$ and $\wt\QQ_{2\dot\alpha}$ instead of the usual two for Schur operators. Operators obeying this shortening condition form a commutative, associative $\Cb$-algebra known as the \emph{Higgs chiral ring}, which we denote in this work as $\RR_{H}$.%
\footnote{The Higgs chiral ring can be defined by passing to the \emph{simultaneous cohomology} of the four supercharges $\QQ^1_{\alpha}$, $\widetilde \QQ_{2\dot\alpha}$. This means we consider the subset of operators that are annihilated by all four of these supercharges, modulo those operators that are exact with respect to at least one linear combination of the supercharges and are annihilated by all combinations. Multiplication in the chiral ring is defined by taking the usual OPE and working at the level of this simultaneous cohomology, whereupon the insertion point becomes irrelevant. Alternatively, it can be defined by taking the coincident limit of the OPE, which is guaranteed to be non-singular for these operators by unitarity bounds and $\suf(2)_R$ conservation. This is a consistent truncation of the usual $\NN=1$ chiral ring, which is the simultaneous cohomology of two supercharges of the same chirality, say $\QQ^1_{\alpha}$.}

This terminology is related to the general expectation that $\RR_{H}$ will always be the coordinate ring of the Higgs branch of the theory, which we denote by $\MM_{H}$.\footnote{In this work, what we call the Higgs branch is the subspace of the full moduli space of the theory where the UV $\suf(2)_R$ symmetry is broken. This may include so-called ``mixed branches'', where the low energy effective theory includes an unbroken Abelian gauge symmetry, as well as subspaces of the moduli space where the low energy theory is still a nontrivial interacting SCFT.} Recall that the coordinate ring $\RR$ of a complex affine variety $\MM$ is a finitely generated, reduced $\Cb$-algebra from which the variety can be uniquely reconstructed (as a scheme) $\MM=\text{Spec}\,\RR$.%
\footnote{For convenience we recall some basic terminology of commutative algebra. A $\Cb$-algebra is a ring containing the complex numbers as a subring. A ring is said to be reduced if it has no nilpotent elements. The spectrum of a ring is the set of its prime ideals, endowed with the Zariski topology and a structure sheaf making it into a scheme. When dealing with the coordinate ring of an affine variety, the variety can be recovered from the scheme by restricting to closed points.} 
In the interest of keeping track of our assumptions, we will formalize this lore as the following
\begin{conj}[Geometrization of the Higgs chiral ring]
In any $\NN=2$ SCFT, the Higgs chiral ring $\RR_{H} $ is a finitely generated, reduced $\Cb$-algebra. Furthermore, $\RR_{H}$ is the coordinate ring of the Higgs branch $\MM_{H}$ of the moduli space of vacua, \ie,
\begin{equation}
\label{eq:Higgs_conjecture}
\MM_{H} = \text{Spec}\,\RR_{H} ~,\qquad \RR_{H}=\Cb[\MM_H]~.
\end{equation}
\end{conj}
Implicit in this conjecture is the expectation that the Higgs branch is always an affine complex algebraic variety. Both of these conjectures are easily verified in theories with Lagrangian descriptions and both are generally believed to always be true. We will take these conjectures as assumptions without further comment.

The Higgs branch operators are prominent members of the associated VOA. Superconformal selection rules dictate that every generator of the Higgs chiral ring gives rise to a \emph{strong VOA generator}. The strong generators of a VOA are by definition those operators that cannot appear as a non-singular term in any OPE, \ie, they cannot be written as the normally-ordered product of other operators. The strong generators of a VOA and their singular operator product coefficients completely characterize the VOA.\footnote{Strictly speaking, if there are null vectors in the Verma module built from the Fourier coefficients of the strong VOA generators, then one has a choice of whether or not to quotient by them. In the present setting we always remove null states, passing to the unique simple quotient of the Verma module.} Additionally, all Higgs chiral ring operators give rise to VOA operators that are Virasoro primary operators in the case where the four-dimensional theory is local so there is a local VOA stress tensor.

\bigskip

The second and third rows of Table \ref{tab:schurTable} detail more general $\NN=1$ chiral and anti-chiral ring operators that participate in the VOA --- the so-called Hall-Littlewood chiral ring operators \cite{Gadde:2011uv}. As was the case for Higgs branch operators, the Hall-Littlewood chiral ring operators admit a commutative, associative multiplication that can be defined by either taking the coincident limit of the OPE or by working in the simultaneous cohomology of three supercharges. This commutative associative algebra is known as the Hall-Littlewood chiral ring, and we denote it as $\RR_{HL}$. The Hall-Littlewood chiral ring is spanned by the Higgs chiral ring operators in addition to operators lying in $\bar\DD$ type multiplets, while the Hall-Littlewood anti-chiral ring is spanned by the Higgs chiral ring operators along with operators lying in $\DD$ type multiplets. It is believed (and demonstrable in Lagrangian theories) that $\DD$ and $\bar\DD$ multiplets are only present in the spectrum of an SCFT for which there are free Abelian gauge fields present in the low energy effective theory at generic (smooth) points of the Higgs branch. In the case when there are no smooth points on the Higgs branch, there is no clear expectation. As with the Higgs chiral ring, VOA operators associated to generators of the Hall-Littlewood chiral ring are required by superconformal selection rules to be strong VOA generators, and additionally all Hall-Littlewood VOA operators must be Virasoro primary operators.

From Table \ref{tab:schurTable} it is clear that in a Lagrangian theory, the Hall-Littlewood chiral ring operators that are not a part of the Higgs branch chiral ring are operators that include the positive helicity components of the gauginos $\ll^1_{+}$ and $\tilde\ll^1_{\dot{+}}$. This has the interesting consequence that in any Lagrangian theory, the additional elements of the Hall-Littlewood chiral ring that are not Higgs branch operators will all be \emph{nilpotent} as a consequence of being constructed out of elementary fermionic fields. Indeed, in light of our conjecture on the geometrization of the Higgs chiral ring, we expect the additional Hall-Littlewood operators coming from $\bar\DD$ multiplets to be all of the nilpotent elements of $\RR_{HL}$, in which case we have the relation
\begin{equation}
\RR_{H}=(\RR_{HL})_{\rm red.}~.
\end{equation}

\bigskip

Finally, we emphasize that in addition to the operators mentioned above that have interesting lives outside of the VOA, there are many more operators that are not members of any conventionally defined chiral ring. The fourth row of Table \ref{tab:schurTable} details the most general types of Schur operators, which obey less standard shortening conditions. The most important operator in this class is the conserved current for $\suf(2)_R$, which belongs to the stress-tensor multiplet $\widehat{\CC}_{0(0,0)}$ and is universally present in local $\NN=2$ SCFTs. Notably absent from the list of Schur operators are the half-BPS operators in the Coulomb branch chiral ring, which obey the shortening condition $\Delta =|r|$.
 
This quick overview of Schur operators reveals that the VOA has a close connection to the the Higgs branch with little obvious connection to the Coulomb branch, but it also captures a much larger set of operators and observables than the Higgs branch chiral ring. A natural question that arises is whether the VOA operators descending from the Higgs branch chiral ring are in any way special in the VOA --- given the VOA $\VV$, is it possible to reconstruct the Higgs chiral ring $\RR_{H}$, and therefore the Higgs branch $\MM_{H}$? The answer is not immediate. From the two-dimensional viewpoint, there is no obvious distinction between operators that descend from $\widehat\BB$ type operators those that descend from the other kinds of protected multiplets shown in Table \ref{tab:schurTable}. In particular, we have seen $\DD$ and $\bar\DD$ operators can also give rise to strong generators of $\VV$, and in fact (as seen in concrete examples) $\hat\CC$ operators can give strong generators as well. So at the very least, focusing on generators does not appear to be a sufficient strategy.

\subsection{VOA embedding of the Higgs chiral ring}
\label{subsec:higgs_in_voa}

We wish to understand the embedding of the Higgs chiral ring into the associated VOA. We first introduce some general VOA constructions and notation. 

\subsubsection{VOA generalities}
\label{subsubsec:voa_generalities}

Consider a general $\hf\Zb_+$-graded conformal VOA. The underlying vector space $\VV$ is spanned by an infinite set of states $a^i$ each of which defines a vertex operator via the vertex operator map,
\begin{equation}
a^i~~\longmapsto~~ a^i(z)=\sum_{n=-\infty}^{\infty}a^i_{(-h_i-n)}z^n~,
\end{equation}
where the $a^i_{(n)}\in{\rm End}(\VV)$ and $h_i$ is the conformal dimension of $a^i$. There is a unique vacuum vector $\Omega$ of dimension zero such that a state is recovered by acting on the vacuum by the appropriate mode in its Fourier expansion,
\begin{equation}
a^i = \lim_{z\to0}a^i(z)\Omega=a^i_{(-h_i)}\Omega~.
\end{equation}
In addition to the usual OPE algebra on $\VV$, one may consider the non-commutative, non-associative algebra defined by the normally-ordered product,
\begin{equation}
\begin{split}
{\rm NO}&:\VV\times\VV~\rightarrow~\VV~,\\
{\rm NO}&:(a,b)~\mapsto~a_{-h_a}b_{-h_b}\Omega~.
\end{split}
\end{equation}
It is a standard construction to further define a secondary operation denoted as a bracket,
\begin{equation}\label{eq:bracket_filtration}
\{a,b\}\colonequals a_{-h_a+1}b\equiv\oint \left(\frac{dz}{2\pi i}a(z)b(0)\right)\Omega~,
\end{equation}
where $\Omega$ is the vacuum state in the VOA. The operation $\{a,\,\cdot\,\}$ is easily shown by a contour integration argument to be a derivation with respect to the normally-ordered product,
\begin{equation}
\label{eqref:bracket_derivation}
\{a, \NO{b}{c}\}=\NO{b}{\{a,c\}}+\NO{\{a,b\}}{c}~.
\end{equation}

\subsubsection{\texorpdfstring{$\suf(2)_R$}{su(2) R-symmetry} filtration and the associated graded}
\label{subsubsec:filtration_and_algebra}

For the VOAs associated to $\NN=2$ SCFTs, the underlying vector space $\VV$ is the space of Schur operators. In particular, this vector space has a triple grading by $(h, R, r)\in \hf\Zb_+\times\hf\Zb_+\times\hf\Zb$,
\begin{equation}
\VV=\bigoplus_{h,R,r}\VV_{h,R,r}~.
\end{equation}
The normally-ordered product preserves $h$ and $r$ but not $R$, making the $R$ grading unnatural from the point of view of the VOA structure. However, the specifics of the twisted translation construction implies that $R$-charge violation occurs with a definite sign,
\begin{equation}\label{eq:R-violation}
\NO{\VV_{h_1,R_1,r_1}}{\VV_{h_2,R_2,r_2}}\subseteq \bigoplus_{k\geqslant0} \VV_{h_1+h_2,R_1+R_2-k,r_1+r_2}~.
\end{equation}
Consequently, there is a \emph{filtration} by $R$ that is preserved by the normally-ordered product. That is, if we define,
\begin{equation}\label{eq:R-filtration}
\FF_{h,R,r}=\bigoplus_{k\geqslant 0}\VV_{h,R-k,r}~,
\end{equation}
then we have the following filtered property for normally-ordered multiplication,
\begin{equation}\label{eq:R-filtered-product}
{\rm NO}(\FF_{h_1,R_1,r_1},\FF_{h_2,R_2,r_2})\subseteq \FF_{h_1+h_2,R_1+R_2,r_1+r_2}~.
\end{equation}
In addition, the bracket operation obeys
\begin{equation}\label{eq:R-filtered-bracket}
\{\FF_{h,R,r},\FF_{h',R',r'}\}\subseteq \FF_{h+h'-1,R+R'-1,r+r'}~,
\end{equation}
so, the bracket is filtered of tri-degree $(-1,-1,0)$.

Though the normally-ordered product and bracket are in general quite complicated operations, they behave quite will with respect to this filtration. In particular, for the normally-ordered product we have the following properties for the commutator and associator,
\begin{equation}
\begin{split}
\lbrack \FF_{h_1,R_1,r_1}, \FF_{h_2,R_2,r_2}\rbrack_{\rm NO}&\subseteq \FF_{h_1+h_2,R_1+R_2-1,r_1+r_2}~,\\
\lbrack \FF_{h_1,R_1,r_1}, \FF_{h_2,R_2,r_2}, \FF_{h_3,R_3,r_3}\rbrack_{\rm NO}&\subseteq \FF_{h_1+h_2+h_3,R_1+R_2+R_3-1,r_1+r_2+r_3}~.
\end{split}
\end{equation}
Furthermore the symmetrizer and the jacobiator of the bracket obey
\begin{equation}
\begin{split}
\{\FF_{h,R,r},\FF_{h',R',r'}\}_+ &\subseteq \FF_{h+h'-1,R+R'-2,r+r'}~,\\
\{\FF_{h,R,r},\FF_{h',R',r'},\FF_{h'',R'',r''}\} &\subseteq \FF_{h+h'+h''-2,R+R'+R''-3,r+r'+r''}~.
\end{split}
\end{equation}
These properties become very useful upon passing to the associated graded of our filtered VOA,
\begin{equation}
{\rm gr}_\FF\VV=\bigoplus_{h,R,r} \GG_{h,R,r}~,\qquad \GG_{h,R,r}= \FF_{h,R,r}/\FF_{h,R-1,r}~.
\end{equation}
On this space, which is isomorphic as a vector space to $\VV$, the normally-ordered product induces a grade-preserving (with respect to all the gradings) commutative, associative product, and the bracket induces an anti-symmetric bracket of tri-degree $(-1,-1,0)$ that obeys the Jacobi identity. Combined with the derivation property \eqref{eqref:bracket_derivation}, we have a Poisson algebra structure on ${\rm gr}_\FF\VV$.\footnote{In fact, there is a more elaborate structure that arises on this graded space due, for example, to the presence of the spatial derivative $\partial = L_{-1}$, which acts on ${\rm gr}_\FF\VV$ as a derivation of tri-degree $(1,0,0)$. This should make ${\rm gr}_\FF\VV$ into a vertex Poisson algebra \cite{benzvi}. We leave a discussion of this structure to future work.}

In terms of this graded algebra it is easy to describe the Hall-Littlewood and Higgs chiral rings as commutative, associative Poisson algebras.\footnote{To the best of our knowledge, it has not been previously observed that the Hall-Littlewood chiral ring is a Poisson algebra.} The Hall-Littlewood chiral ring is the subalgebra
\begin{equation}
\RR_{HL}=\left(\bigoplus_{R,r}\GG_{R+r,R,r}~,~\NO{\cdot}{\cdot}~,~\{\cdot,\cdot\}\right)~,
\end{equation}
while the Higgs chiral ring is simply the $r=0$ subspace of the Hall-Littlewood chiral ring,\footnote{It is not immediately obvious that the Poisson bracket defined by this construction must match the natural Poisson bracket induced by the holomorphic symplectic structure on the Higgs branch. This follows from an argument analogous to the one given in \cite{Beem:2016cbd}, generalized to the setting of the holomorphic/topological twist of $\NN=2$ theories \cite{Kapustin:2006hi,costello_private}.}
\begin{equation}
\RR_{H}=\left(\bigoplus_{R}\GG_{R,R,0}~,~\NO{\cdot}{\cdot}~,~\{\cdot,\cdot\}\right)~.
\end{equation}

It is possible to reach the Higgs chiral ring without using the intermediate associated graded algebra of the full normally-ordered VOA. This is because the full space of operators that are \emph{not} Higgs chiral ring operators admits a simple expression in terms of the $R$ filtration,
\begin{equation}
\VnoH=\bigoplus_{h>R} \VV_{h,R,r}=\bigcup_{h>R}F_{h,R,r}~.
\end{equation}
There is then a canonical isomorphism between the space of Higgs chiral ring operators and a quotient of $\VV$ by this subspace,
\begin{equation}
\VV_{H}\cong \VV/\VV_+~.
\end{equation}
One quickly verifies from their filtered behaviors that the commutator, associator, symmetrizer, and jacobiator all map into $\VnoH$, and additionally $\VnoH$ is a two-sided ideal with respect to normally-ordered multiplication and also with respect to the secondary bracket, so this quotient reproduces the structure of the Higgs chiral ring as a commutative, associative Poisson algebra. 

Thus the Higgs chiral ring as a Poisson algebra (along with various other nice algebraic structures) can be recovered from the normally-ordered algebra of the VOA by simply taking appropriate vector space quotients. Unfortunately, these quotients require a knowledge of (at least part of) the $R$-filtration on $\VV$, and in general that filtration has not been understood from a purely VOA point of view (though see \cite{Song:2016yfd} for some partial progress in this direction). We therefore turn next to an alternative quotient that produces something very similar to, but not equivalent to, the Higgs chiral ring. This will motivate our proposal for how to recover the Higgs chiral ring without having access to the $R$-filtration.

\subsection{The \tpdf{$C_2$}{C2} algebra}
\label{subsec:higgs_in_c2}

An alternative operation that is intrinsic to a VOA also allows us to extract a Poisson algebra as a quotient of the normally-ordered algebra. To do so, one defines the vector subspace $C_2(\VV)\subset\VV$ as
\begin{equation}
C_2(\VV) \colonequals {\rm Span} \left\{ a^i_{(-h^i-1)}\varphi~,\quad a^i,\varphi \in \VV \right\}~.
\end{equation}
The state $a^i_{(-h_i-n)}\Omega$ is associated to the vertex operator $\partial^n a^i$ by the state/operator map; the vector space $C_2(\VV)$ is, roughly speaking, the space of those normally-ordered composite operators that include any derivatives, though it may include operators that can be written without derivatives if there are appropriate null relations in the VOA.

It turns out that $C_2(\VV)$ is a two-sided ideal with respect to both the normally-ordered product and the secondary bracket, and what's more the associator and commutator in $\VV$ with respect to the normally-ordered product, along with the symmetrizer and Jacobiator with respect to the secondary bracket defined above, all map into $C_2(\VV)$:\footnote{This result actually provides a simpler demonstration that these various operations map into $\VV_+$ as defined above, since we necessarily have $C_2(\VV)\subseteq \VV_+$.}
\begin{equation}\begin{split}
\left[\VV,\VV\right]&\subseteq C_2(\VV)~,\\
\left[\VV,\VV,\VV\right]&\subseteq C_2(\VV)~,\\
\left\{\VV,\VV\right\}_+&\subseteq C_2(\VV)~,\\
\left\{\VV,\VV,\VV\right\}&\subseteq C_2(\VV)~.
\end{split}\end{equation}
Thus the normally-ordered product and secondary bracket induces a commutative, associative Poisson algebra structure on the quotient space
\begin{equation}
\RR_{\VV} \colonequals \VV / C_2(\VV)~.
\end{equation}
This algebra is known as the $C_2$-algebra of $\VV$. If $\VV$ is strongly finitely generated, then $\RR_{\VV}$ has a simple description as the space of polynomials in the generators $\VV$, modulo the ideal induced by null relations.\footnote{There is presently no proof that the associated VOA for every $\NN=2$ SCFT must be strongly finitely generated, but there are no known counterexamples.} If ${\rm dim}\,\RR_{\VV} < \infty$, then $\VV$ is said to be $C_2$-\emph{co-finite}, or to obey the $C_2$ condition. The $C_2$ condition is a necessary condition for rationality.

\subsubsection{Example: Virasoro VOA}
\label{subsubsec:virasoro_example}

To make this construction tangible, let us consider a simple example. The Virasoro VOA has a single strong generator, the stress tensor $T(z)$. The corresponding generator of $\RR_{\VV}$, which we can identify with the equivalence class of $L_{-2}\Omega$ in the quotient $\VV/C_2(\VV)$. At a generic value of the central charge, we have
\begin{equation} 
\label{RVir}
\RR_{\rm Vir} \cong {\rm Span} \left\{ (L_{-2})^k \Omega~,~ k=1, 2, \dots \right\}~.
\end{equation}
If we define $t^k\equiv (L_{-2})^k \Omega$ and denote by $*$ the multiplication on $\RR_{\VV}$ induced by the normally-ordered product on $\VV$, then we have
\begin{equation}
t^{k_1} * t^{k_2} = t^{k_1 + k_2}~,
\end{equation}
so our ring structure is simply given by multiplication of polynomials in one variable. In the generic case this identifies $\RR_{\rm Vir}$ with the freely generated ring on one variable,
\begin{equation}
\RR_{\rm Vir} = \Cb[t]\quad ({\rm generic \;}c)~.
\end{equation}
Additionally, the Poisson bracket on $\RR_{\rm Vir}$ is easily seen to be trivial, regardless of the value of $c$, since we have
\begin{equation}
\{t,t\}~\Longleftrightarrow~ L_{-1}L_{-2}\Omega=L_{-3}\Omega\sim0~.
\end{equation}
As this example illustrates, $R_{\VV}$ need not necessarily be reduced, and the Poisson bracket need not be non-degenerate.

At special values of the central charge there can be null vectors in the Verma module generated by the $L_{-n}$, so in the simple quotient of the Verma module there are relations. Suppose that a null vectors takes the form
\begin{equation}
\label{nullvirasoro}
\NN=(L_{-2})^m \Omega + \sum_{n>2}L_{-n} (\dots) \Omega~,
\end{equation}
for some positive integer $m\geqslant2$ that depends on the value of the central charge. This relation sets $(L_{-2})^m \Omega \sim 0$ in $\VV/C_2(\VV)$, so the $C_2$ algebra becomes
\begin{equation}
\RR_{\rm Vir} = \Cb[t]/\langle t^m \rangle~.
\end{equation}
Thus in such a case $\RR_{\rm Vir}$ is finite-dimensional with ${\rm dim}\,\RR_{\rm Vir} = m$, so the reduced algebra is trivial. An example of this structure that will be relevant in later sections is the Virasoro VOA at $c=-22/5$ (the VOA underlying the $(2,5)$ Virasoro minimal model, also known as the Lee-Yang model). In this algebra there is a null vector
\begin{equation}
\label{eq:LeeYangNull}
\NN_{(2,5)}=(L_{-2})^2 \Omega-\tfrac{5}{3} L_{-4}\Omega~,
\end{equation}
so $m = 2$ and $\RR_{\VV}$ is spanned by $1$ and $t$; we have
\begin{equation}
\RR_{\rm Vir(2, 5)} = \Cb [t]/\langle t^2\rangle~.
\end{equation}

\subsection{Relating ideals and Higgs branch reconstruction}
\label{subsec:relating_ideals}

We have now introduced two quotients of the associated VOA for any $\NN=2$ SCFT that will give rise to two commutative, associative Poisson algebras $\RR_H$ and $\RR_{\VV}$. We would like to understand the relationship between these two constructions.

The first thing that we can demonstrate is that as vector spaces we have a canonical embedding $\RR_{H}\subseteq\RR_{\VV}$. This follows from the fact that $C_2(\VV)\subseteq \VV_+$, which is a simple consequence of the filtered nature of the normally-ordered product and the fact that the derivative operator acts with tri-degree $(1,0,0)$. Nevertheless, we know that in general $C_2(\VV)\neq\VnoH$, since in any theory that is not a free theory we will have $L_{-2}\Omega\in\VnoH$ and $L_{-2}\Omega\notin C_2(\VV)$. The question then becomes whether we can identify the subspace
\begin{equation}
\II_{+}\colonequals \VV_+/C_2(\VV)\subseteq \VV/C_2(\VV)~.
\end{equation}
This subspace is a Poisson ideal in $\RR_{\VV}$, and taking the quotient will reproduce the Higgs chiral ring, $\RR_{\VV}/\II_{+}\cong\RR_H$.

As we have commented earlier, the $C_2$ algebra need not be a reduced algebra, whereas the Higgs chiral ring must be. This means that we must at the very least have the inclusion
\begin{equation}
{\rm Nil}(\RR_{\VV})\subseteq\II_+~,
\end{equation}
where ${\rm Nil}(\RR_{\VV})$ is the \emph{nilradical} of $\RR_{\rm\VV}$ that comprises all nilpotent elements. The most economical guess is then that the nilradical represents the full set of states that must be removed. Though simple to state, proving this property of the VOAs associated to SCFTs appears quite difficult. Nevertheless, we will see that this guess passes many nontrivial consistency checks. We therefore make the following conjecture.
\begin{conj}[Higgs branch reconstruction]
The Higgs branch chiral ring $\RR_{H} $ is equal to the quotient of $\RR_{\VV}$ by its nilradical,
\begin{equation}
\RR_{H} = (\RR_{\VV})_{\rm red}~.
\end{equation}
This is equivalent to the identification of ideals,
\begin{equation}
{\rm Nil}(\RR_{\VV})\cong \VV_+/C_2(\VV)~.
\end{equation}
\end{conj}
For a general VOA $\VV$, Arakawa has defined the \emph{associated variety} $X_{\VV}$ as
\begin{equation}
X_{\VV} \colonequals {\rm Spec}\,(R_\VV)_{\rm red}~.
\end{equation}
Thus our conjecture, combined with the geometrization of the Higgs chiral ring, is equivalent to the statement that the Higgs branch of a four-dimensional $\NN=2$ SCFT can be identified with the associated variety of the associated VOA. In \cite{Arakawa:2016hkg}, strongly finitely generated VOAs whose associated varieties (thought of as Poisson varieties) have finitely many symplectic leaves have been named \emph{quasi-lisse}. The Higgs branch of an $\NN=2$ SCFT will always have finitely many symplectic leaves, so if our conjecture holds then the associated VOA will be quasi-lisse as long as it is strongly finitely generated.

%% file: sections/S3.tex

\section{Modularity and the Schur index}
\label{sec:modular}

The conjecture of the previous section requires that any strong generator $G$ of the associated VOA that is \emph{not} a Higgs chiral ring generator be nilpotent in the $C_2$ algebra. Thus for any such generator there must be a corresponding null vector in the vacuum Verma module $V$ of the chiral algebra of the form\footnote{Here we are distinguishing between $V$, the vacuum Verma module, which may contain null vectors and correspondingly may not be simple as a VOA module, and $\VV$, the vector space underlying the VOA, which is necessarily simple as a VOA module if the VOA itself is simple.}
\begin{equation}
\NN_{G}=(G_{-h_{G}})^k\Omega+\sum_{i}a^i_{-h_i-1}\varphi_i~,\qquad \varphi_i\in\VV~,\quad k\in\Zb_{+}~.
\end{equation}
The existence of null vectors in $V$ will generally depend delicately on the structure constants of the VOA, so we expect this to be a strong constraint on the VOA associated to any given SCFT.

This idea manifests in an especially interesting way when we turn our attention to the Virasoro subalgebra of the associated VOA. The stress tensor of the associated VOA arises from the $\widehat{\CC}_{0(0,0)}$ supermultiplet in four dimensions, so the subspace of $\VV$ generated by acting with Virasoro generators on the vacuum lies entirely within $\VV_+$. However, with the exception of free theories, states of the form $L_{-2}^k\Omega$ do not lie in $C_2(V)$, where we note that we are referring to the $C_2$ subspace of the Verma module $V$, not its simple quotient. Therefore if the reconstruction conjecture is true, then there must always exist some null state in $V$ of the form
\begin{equation}\label{eq:stress_tensor_null}
\NN_T=(L_{-2})^k\Omega+\sum_i a^i_{-h_i-1}\varphi_i~.
\end{equation}
In other words, we require that $(L_{-2})^k\Omega\in C_{2}(\VV)$ for some positive integer $k$.

This property of a vertex operator algebra has been historically linked with the existence of LMDEs for the characters of sufficiently nice $\VV$-modules \cite{Gaberdiel:2007ve,Gaberdiel:2008pr,Gaberdiel:2008ma}. The literature on this topic is somewhat complicated, though recently it has clarified significantly in the special case of quasi-lisse VOAs \cite{Arakawa:2016hkg}. We will take a moment to describe the situation as it pertains to our investigation.

\subsection{Null vectors and differential equations}
\label{subsec:null_vectors_diff_eq_review}

Suppose that \eqref{eq:stress_tensor_null} holds in a given vertex operator algebra $\VV$. It follows that there is also a null vector in the Verma module of the VOA of the form\footnote{See Appendix \ref{app:modular_recursion} for a discussion of square-bracket modes versus ordinary modes. The difference between square-brackets ordinary brackets will not be important for the reader who only wants to get a general sense of the arguments presented in this section.}
\begin{equation}\label{eq:null_form}
\NN_{[T]}=(L_{[-2]})^k\Omega-\varphi~,\qquad \varphi\in C_{[2]}(V)~.
\end{equation}
Because this is a null state, correlation functions that include insertions of $\NN_{[T]}$ must vanish. In particular, the torus one-point function of $\NN$ must vanish, which implies the following trace formula on $\VV$,
\begin{equation}
\STr_{\VV}\left(o(\NN_{[T]})q^{L_0-\frac{c}{24}}\right)=\STr_{\VV}\left(o(L_{[-2]}^k\Omega)q^{L_0-\frac{c}{24}}\right)-\STr_{\VV}\left(o(\varphi)q^{L_0-\frac{c}{24}}\right)=0~.
\end{equation}
The trace of $o((L_{[-2]})^k\Omega)$ can be evaluated using the trace formulae of Appendix \ref{app:modular_recursion} in terms of LMDOs acting on the vacuum character $\goodchi_{\VV}(q)=\STr_{\VV}(q^{L_0-\frac{c}{24}})$. On the other hand, since $\varphi$ is in $C_{[2]}(V)$, the trace of $o(\varphi)$ can also be rewritten using the recursion relations supplied in Appendix \ref{app:modular_recursion} in terms of traces of zero modes of operators of lower conformal dimension with coefficients given by (twisted) Eisenstein series.

In principle, for every term that appears in the trace after implementing this recursion, one of three things will happen:
\begin{enumerate}
\item[$(i)$] The term is the zero mode of an element of $C_{[2]}(\VV)$, in which case the recursion relations can be applied again.
\item[$(ii)$] The term is of the form $o((L_{[-2]}^{r})\Omega)$ with $r<k$, in which case the trace can be evaluated in terms of a modular differential operator of lower order than the first acting on the vacuum character.
\item[$(iii)$] The resulting zero mode is of the form $o(G^1_{[-h_1]}\cdots G^n_{[-h_n]}\Omega)$ where the $G^i$ are strong generators and at least one $G^i\neq T$. In this case, without additional knowledge, we will be unable to go further, though it may turn out that additional arguments allow the trace of such a zero mode to be evaluated.
\end{enumerate}
Because $\varphi$ has finite conformal dimension and each step in the recursion produces traces of zero modes of operators of strictly lower conformal dimension than the preceding one, this recursion algorithm ultimately terminates. We note that the requirement that case $(iii)$ not arise while performing the algorithm is equivalent to the technical condition described in, \eg, \cite{Gaberdiel:2008pr} that there be a vector of the form 
\begin{equation}
(L_{[-2]})^k\Omega+\sum_{i=1}^{k} g_i(q)(L_{[-2]})^{r-i}\Omega \quad\in\quad O_q(\VV)~,
\end{equation}
where $O_q(\VV)$ is the vector subspace of $\VV\otimes\Cb[E_4(q),E_6(q)]$ spanned by elements of the form
\begin{equation}
a_{[-h_a-1]}b+\sum_{k\geqslant2}G_{2k}(q)a_{[2k-h_a-1]}b~,\quad a,b\in\VV~.
\end{equation}

As a standard example, let us turn again to the Virasoro algebra with central charge $c=-22/5$. We now consider the null vector
\begin{equation}
\NN_{[T]}=(L_{[-2]})^2\Omega-\tfrac53 L_{[-4]}\Omega~.
\end{equation}
Using the recursion relations in Appendix \ref{app:modular_recursion}, we have
\begin{equation}
\begin{split}
\Tr_{\VV}\left(o(L_{[-2]}L_{[-2]}\Omega)q^{L_0-c/24}\right)&=\left(D_q^{(2)}+\frac{c}{2}\Eb_4(\tau)\right)\left(\Tr_{\VV}(q^{L_0-c/24})\right)~,\\
\Tr_{\VV}\left(o(L_{[-4]}\Omega)q^{L_0-c/24}\right)&=0~,
\end{split}
\end{equation}
From which we deduce the modular equation
\begin{equation}
\left(D_q^{(2)}-\tfrac{11}{5}\Eb_4(\tau)\right)\goodchi_{{\rm Vir}(2,5)}(q)=0~.
\end{equation}
In this case, it is impossible to encounter the obstruction of case $(iii)$ because the only strong generator is the stress tensor itself.

It has recently been shown by Arakawa and Kawasetsu \cite{Arakawa:2016hkg} that for the special case of a quasi-Lisse VOA, in which case a null vector of the relevant type necessarily exists, the vacuum character is always the solution of a LMDE. This is a slightly more nuanced statement than anything we have been describing here. In particular, it does not guarantee that the order of the relevant LMDO is the same as the power of $L_{[-2]}$ appearing in the minimal null vector of the relevant type. In principle it is possible that one will encounter the obstruction of possibility $(iii)$ when applying the recursion algorithm to the case of the minimal null vector, but the proof then implies that there will nevertheless be some other null vector at a higher level for which the recursion will proceed cleanly. This means that means that the veracity of our Higgs branch conjecture for the associated variety would imply the existence of such a LMDE.

\subsection{Modular equations for Schur indices and Cardy behavior}
\label{subsec:modular_equations_for_indices}

Though the proof of \cite{Arakawa:2016hkg} means that our Higgs branch conjecture implies the existence of an LMDE, we will find substantial evidence for the existence of LMDEs even in theories where the Higgs branch conjecture is not easy to verify. Therefore we wish to make the following \emph{a priori} independent conjecture, that may be true even if the Higgs branch conjecture eventually fails or requires modification:
\begin{conj}[Modularity]
The (appropriately normalized) Schur index of any four-dimensional $\NN=2$ SCFT solves a finite-order, monic, holomorphic, (twisted) modular differential equation.
\end{conj}
The normalization in question is the one corresponding to the standard normalization of a torus partition function in two-dimensional conformal field theory, namely we have\footnote{The importance of this prefactor for good modular behavior was previously pointed out in \cite{Razamat:2012uv}.}
\begin{equation}
\II_{\rm Schur}(q)\colonequals q^{c_{4d}/2}\STr_{\HH}(q^{E-R}) = \STr_{\VV} (q^{L_0-c_{2d}/24})~.
\end{equation}
The question of whether the differential operator in question will be modular or twisted modular depends on whether there are operators with half-integer conformal weight in the VOA. We remark, however, that even if there are operators of half-integer weight, it may be the case that no twisted Eisenstein series appear during the implementation of the recursion relations, so there may still be an untwisted modular differential operator. This can be seen, for example, in the case of $\NN=4$ supersymmetric Yang-Mills theory with gauge algebra $\suf(3)$, as discussed below in Section \ref{subsec:examples_n4}. The precise class of twisted modular equations in question are detailed in Appendix \ref{app:eisenstein_and_modular}.

An immediate corollary of our modularity conjecture is that the vacuum character of the associated chiral algebra of any SCFT should transform in a finite-dimensional representation of the modular group, \ie, it should transform as a part of a vector-valued (quasi)-modular form of weight zero. As we will see, there may be logarithms appearing in other entries in this vector-valued form, which is the reason for the ``quasi'' qualification. We can therefore understand the behavior of the Schur index as an analytic function of $q$ in in terms of the other solutions that make up the vector-valued modular form.

\subsubsection{Cardy behavior}
\label{subsubsec:cardy_untwisted}

Of particular interest is the behavior of the Schur index in the limit $q\to1$. A generalization \cite{Buican:2015ina, DPKR, Ardehali:2015bla} of the arguments of \cite{DiPietro:2014bca} implies that this behavior is determined by the four-dimensional Weyl anomalies $a_{4d}$ and $c_{4d}$. In particular, if we write $q = e^{2\pi i\tau}$, then we have
\begin{equation}
\label{eq:cardyeq}
\lim_{\tau\to0}\log \II_{\rm Schur}(q) \sim \frac{4 \pi i(c_{4d}-a_{4d})}{\tau}~.
\end{equation}
On the other hand, if the Schur index transforms as part of a vector valued modular form comprising the solutions of the appropriate LMDE, when we can define the $S$-transformed nome
\begin{equation}
\tilde q \colonequals \exp\left(\frac{-2\pi i}{\tau}\right)~,
\end{equation}
and we will have
\begin{equation}
\II_{\rm Schur}(q) = \sum_{i}\SS_{0i}\goodchi_i(\tilde q)~,
\end{equation}
where the matrix elements $\SS_{0i}$ are rational numbers and the $\goodchi_i(\tilde q)$ are either the solutions of the LMDE (in the $PSL(2,\Zb)$-modular case) or solutions of the conjugate LMDE in the $\Gamma^0(2)$-modular case. These solutions will, in good cases, be defined by power series in $\tilde q$,
\begin{equation}
\goodchi_i(\tilde q)=\tilde q^{-c_{2d}/24+h_i}(1+\ldots)~,
\end{equation}
where the $\ldots$ are subleading as $\tilde q\to0$. More generally, there may solutions that are logarithmic at leading order,
\begin{equation}
\goodchi_i(\tilde q)=\tilde q^{-c_{2d}/24+h_i}((\log\tilde q)^k+\ldots)~.
\end{equation}
Regardless of the situation with logarithms, in the ``high temperature'' limit $\tau\to0$ the Schur index/vacuum character will behave as
\begin{equation}
\lim_{\tau\to0}\log \II_{\rm Schur}(q) \sim \frac{\pi i c_{\rm eff}}{12\tau}+\ldots~,
\end{equation}
where the effective central charge is defined by\footnote{Here we are assuming that all of the $\SS_{0i}$ in the modular transformation are generically nonzero. In principle, there can be instances where this is not the case. We will return to this in some of the examples below. We note that in the context of unitary two-dimensional conformal field theory, the vacuum module itself is always the representation of lowest dimension because the $h_i>0$, thus $h_{\min}=0$. This is not generally the case for non-unitarity chiral algebras of the type we are studying.}
\begin{equation}
c_{\rm eff} = c_{2d} - 24 \min\!{}_i(h_i)~.
\end{equation}
Comparing with the predicted high temperature behavior, we have the relation
\begin{equation}
\label{eq:ceff_equation}
c_{\rm eff} = 48 (c_{4d} - a_{4d})~,
\end{equation}
from which we find an expression for the Weyl anomaly $a_{4d}$ in terms of $h_{\min}$,
\begin{equation}
\label{eq:a_anomaly_equation}
a_{\rm 4d}=\frac{h_{\min}}{2}-\frac{5c_{2d}}{48}~.
\end{equation}
This relation is particularly interesting in light of the Hofman-Maldacena bound \cite{Hofman:2008ar,Hartman:2016dxc,Hofman:2016awc},
\begin{equation}
\frac54\geqslant \frac{a_{4d}}{c_{4d}}\geqslant \frac12~.
\end{equation}
This then implies \cite{Cecotti:2015lab}
the following inequalities relating $h_{\min}$ and the Virasoro central charge of the VOA associated to a unitary $\NN=2$ SCFT,
\begin{equation}
\frac{c_{2d}}{8}\leqslant h_{\min} \leqslant 0~.
\end{equation}
The left inequality is saturated for the free hypermultiplet SCFT, while the right inequality is saturated for the free vector multiplet. It is interesting to note that this is already an effective constraint on, for example, the collection of non-unitary Virasoro minimal model VOAs that can be related to unitary $\NN=2$ SCFTs. The $(5,8)$, $(7,11)$, $(8,13)$, and $(9,14)$ minimal VOAs cannot arise, just to name a few.

\subsubsection{Sample analysis: second order equations}
\label{subsubsec:order_two}

As an example of how this structure plays out, let us consider the hypothetical case where the vacuum character satisfies of a $\Gamma$-modular equation of degree two. This is the simplest possibility, because the unique LMDE of degree one admits only the constant solution. The form of such a second-order equation is completely determined by a single numerical coefficient and takes the form\footnote{Here, as elsewhere, we assume that our modular equation is \emph{holomorphic} and \emph{monic}, as it will be if it arises from the recursion argument outlined above. Nevertheless, it can be interesting to relax this assumption \cite{Mathur:1988na}.}
\begin{equation}
\label{eq:secondordereq}
\left(D^{(2)}_q+\lambda \Eb_4(\tau)\right)\goodchi(q)=0~.
\end{equation}
The vacuum and module characters will take the form
\begin{equation}
\begin{split}
\goodchi_0(q)&=q^{-c_{2d}/24}(1+a_1q+a_2q^2+\ldots)~,\\
\goodchi_1(q)&=q^{-c_{2d}/24+h}(1+b_1q+b_2q^2+\ldots)~,
\end{split}
\end{equation}
and upon acting with the LMDE the coefficient $\lambda$ gets related to the Virasoro central charge $c_{2d}$ and the conformal weight of the primary state in the non-vacuum module $h_1$ according to
\begin{equation} 
\label{eq:c2dlambda}
\lambda = -\frac{5}{4}\left(\frac{c_{2d}^2}{4} + c_{2d}\right) ~,\qquad h_1 = \frac{c_{2d}+2}{12}~,
\end{equation}
which means we have two cases:
\begin{equation}
h_{\min}=
\begin{cases}
\frac{c_{2d}+2}{12}~,\qquad c_{2d}<-2~,\\
~~~0~~~\,,\qquad c_{2d}>-2~.
\end{cases}
\end{equation}
The Weyl anomaly $a_{4d}$ is then determined by \eqref{eq:a_anomaly_equation} to be
\begin{equation}
a_{4d} = 
\begin{cases}
\frac{1}{12} - \frac{c_{2d}}{16}	&=~\frac{1}{12}+\frac{3c_{4d}}{4}~,\quad c_{2d}<-2~,\\
-\frac{5c_{2d}}{48}					&=~\frac{5c_{4d}}{4}~,~\quad\qquad c_{2d}>-2~,
\end{cases}~,
\end{equation}
for any SCFT whose Schur index satisfies a second order modular equation. In the second case, the upper Hofman-Maldacena bound is saturated so one should actually require $c_{2d}\leqslant-2$, with saturation occurring for free vector multiplets.

If we further assume that the theory in question has a one-dimensional Coulomb branch with the Coulomb branch chiral ring generator having dimension (and $\uf(1)_r$ charge) given by $r$, and also that the Shapere-Tachikawa formula holds, which in the rank-one case takes the form \cite{Shapere:2008zf}
\begin{equation}
2 a_{4d} - c_{4d} = \frac{2r-1}{4}~,
\end{equation}
then we deduce am expression for $r$ in terms of the Weyl anomaly coefficient $c_{4d}$,
\begin{equation}
\label{eq:r2ndorder}
r = c_{4d} +5/6~.
\end{equation}

\subsubsection{Sample analysis: half-integer case}
\label{subsubsec:cardy_twisted}

Now we return to the more general case when the vertex algebra is $\frac12\Zb$ graded but not $\Zb$ graded. The $n$-dimensional space of solutions of the corresponding twisted LMDE of order $n$ is now spanned by solutions of the form
\begin{equation}
\goodchi_i(q)\sim q^{-\frac{c}{24}+h_i}(1+O(q^{\frac12}))~.
\end{equation}
However, in this case this set of characters do not transform amongst themselves under arbitrary modular transformations, but only under elements of the congruence subgroup $\Gamma^0(2)$, as described in Appendix \ref{app:eisenstein_and_modular}. In particular, the element $S:\tau\mapsto-1/\tau$ does not belong to $\Gamma^0(2)$, so the small $\beta$ behavior of the Schur index will not be controlled by other solutions of the vacuum modular equation.

Nevertheless, the vacuum character will still transform in a definite way under the full modular group $\Gamma$ because the modular equation itself transforms in a definite way under arbitrary modular transformations. The twisted LMDOs arising in the half-integer graded case will be of the form
\begin{equation}
\DD^{(n)}=\left(D_q^{(n)}+\sum_{k=1}^{n-1}\left(
\sum_{\substack{r+s=k\\r\leqslant s}}c_{r,s}\Theta_{r,s}(\tau)
\right)D_q^{(n-k)}\right)~,
\end{equation}
where the $\Gamma^0(2)$-modular forms $\Theta_{r,s}(\tau)$ are defined in Appendix \ref{app:eisenstein_and_modular}. Now the reasoning of the integer-weighted case goes through in the case of half-integer weights, but now the differential operator transforms under the $S$-transformation into the \emph{conjugate differential operator},
\begin{equation}
\wt{\DD}^{(n)}=\left(D_q^{(n)}+\sum_{k=1}^{n-1}\left(
\sum_{\substack{r+s=k\\r\leqslant s}}c_{r,s}\wt{\Theta}_{r,s}(\tau)
\right)D_q^{(n-k)}\right)~,
\end{equation}
where the $\Gamma_0(2)$-modular forms $\wt{\Theta}_{r,s}(\tau)$ are also defined in Appendix \ref{app:eisenstein_and_modular}. So instead of the characters $\goodchi_i(q)$, the effective central charge will be determined in terms of the \emph{conjugate characters} $\tilde\goodchi_i(q)$ that are annihilated by this conjugate operator.

For illustrative purposes, let us consider the half-integral analogue of the calculation above for second-order twisted LMDEs. A generic second order equation of the type outlined above will have now have three free coefficients. Let us fix two by demanding that we consider theories with no free fields and no flavor symmetries (which implies that the $O(q^{1/2})$ and $O(q)$ terms in the vacuum character vanish). With these assumptions, the most general possible second order equation is parameterized by the central charge $c$ and is given by
\begin{equation}
\label{eq:half_int_second_order}
\DD^{(2)}=D_q^{(2)}+\tfrac{22+5c}{102}\Theta_{0,1}(\tau)D_q^{(1)}-\tfrac{c(24+7c)}{4896}\Theta_{0,2}(\tau)+\tfrac{c(45+11c)}{4896}\Theta_{1,1}(\tau)~.
\end{equation}
Passing to the conjugate equation and solving the corresponding indicial equation, we find that the conformal weights of the conjugate characters are related to the Virasoro central charge by
\begin{equation}
2\tilde{h}^2-\tfrac{1}{34}\tilde{h}(9c_{2d}+26)+\tfrac{1}{544}c_{2d}(5c_{2d}+22)==0~.
\end{equation}
It turns out that this gives values of $a_{4d}$ that are always compatible with the Hofman-Maldacena bounds so long as $\tilde h$ is real. Reality of $\tilde h$ requires
\begin{equation}
c_{2d}\geqslant\frac{47-17 \sqrt{17}}{4}~.
\end{equation}

We know of no SCFTs that should fit into this category. Nevertheless, it is interesting that upon searching for solutions of the LMDE given by \eqref{eq:half_int_second_order} with moderate values (absolute value less than $100$) for the degeneracy at level $3/2$, we find only two solutions for which the coefficients are integers. One has no states at level $3/2$ -- in fact it has nonzero degeneracy only at integer levels -- and is the $(2,5)$ Virasoro vacuum character. We will return to that example below. The other solution is genuinely half-integer graded and we do not recognize it. The character takes the form
\begin{equation}
\begin{split}
\goodchi_0(q)=q^{7/32}
\big(1&-q^{3/2}+q^2-q^{5/2}+q^3-q^{7/2}+2 q^4-2 q^{9/2}+2 q^5-3 q^{11/2}+4 q^6\\
&-4 q^{13/2}+4 q^7-6 q^{15/2}+7 q^8-7 q^{17/2}+8 q^9-10 q^{19/2}+12 q^{10}+\ldots\big)~.
\end{split}
\end{equation}
The implied values for the various interesting quantities are then
\begin{equation}
c_{2d}=-\frac{21}{4}~, \qquad h_1=-\frac{1}{4}~,\qquad \tilde h_{1} = -\frac{7}{32}~,\qquad \tilde{h}_2 = -\frac{3}{32}~,\qquad a_{4d}=\frac{53}{64}~.
\end{equation}
In fact, this central charge and character can be seen to coincide with the vacuum character of the $\NN=1$ super Virasoro VOA at the central charge relevant for the $(1,4)$ model \cite{Kac:1978ge,Deka:2004bf}. There can be no unitary four-dimensional theory associated to this VOA, because fermionic Schur operators must come in pairs as a consequence of $\suf(2,2|2)$ representation theory and CPT, and in this VOA there is a single fermionic generator of dimension $3/2$. There could nevertheless be a non-unitary four-dimensional theory associated to this VOA. With no flavor symmetry and a second-order modular equation, it would be something of a minimal theory.

\subsection{Comments on additional solutions}
\label{subsec:interpretation}

It is natural to ask whether the other solutions of the modular differential equations can be interpreted in terms of four-dimensional physics. The additional solutions of the LMDE for a VOA vacuum character --- or more precisely, the other entries of the vector-valued modular form containing the vacuum character --- are generally expected to be characters of nontrivial modules over the VOA in question. In the case of rational VOAs, this is necessarily the case. But even in the irrational case (which is the generic case for VOAs coming from four-dimensional SCFTs) the derivation of a modular differential equation for the vacuum character starting with a null vector of the form \eqref{eq:null_form} goes through identically when the original trace is evaluated not in the vacuum representation but in a sufficiently nice non-vacuum module.

More precisely, suppose there exists a module $M$ over $\VV$ that is a conformal highest weight module with finite-dimensional weight spaces under a diagonalizable action of $L_0$. If it is true that case $(iii)$ never occurs during the recursive derivation of the modular equation, then the (super-)character of the module,
\begin{equation}
\goodchi_{M}(q)=\STr_{M}\left(q^{L_0-\frac{c}{24}}\right)~,
\end{equation}
will solve the same modular differential equation as the vacuum character. In principle one can further relax these conditions and consider the case where the action of $L_0$ is not diagonalizable, but acts with a nontrivial Jordan block structure on finite-dimensional generalized eigenspaces. In this case, characters can be generalized to \emph{pseudo-traces} \cite{Miyamoto:2003xp}, which involve logarithms.

More generally, there may be reasonably nice modules for the VOAs that we are studying that do not have finite-dimensional (generalized) $L_0$ eigenspaces, but do have finite-dimensional weight spaces upon further refining by additional flavor fugacities. This will be the case, for example, for the Kac-Wakimoto admissible-level $\suf(2)$ affine current algebras discussed in Section \ref{subsec:admissible_higgs}, as well as for certain more complicated cases such as the $\sof(8)$ current algebra at level $k_{2d}=-2$. In such cases, at least two possible phenomena are known to arise \cite{Mukhi:1989bp,creutzig_private}:
\begin{itemize}
\item Taking sums and differences of simple characters --- treated as analytic functions rather than formal power series --- yields a quantity that is finite when flavor fugacity is set to zero.
\item The module characters as analytic functions cannot be arranged into a combination that is finite in the limit of zero flavor fugacities, but there is nevertheless a regularization of the singular behavior that yields a ``fake'' character that may contain logarithms even if the original characters did not.
\end{itemize}
It turns out that both of these phenomena arise in our examples, though we leave a careful discussion of the modular behavior of flavored characters to future work \cite{flavored_characters}.

From the point of view of four-dimensional physics, there is a guaranteed source of modules over $\VV$: $\NN=(2,2)$ superconformal surface operators. When such surface operators are oriented so as to fill the two directions orthogonal to the VOA plane, \ie, when the span the $\Cb_{[w,\bar{w}]}$ plane and intersect that VOA plane at the origin $z=\zb=0$. In the presence of such a surface operator, the cohomological construction of VOA operators and OPEs still goes through in any neighborhood away from the origin of the VOA plane. On the other hand, one may develop a bulk-defect OPE which produces, upon passing to the usual cohomology, a module structure for the bulk VOA. The relationship between surface defect Schur indices and vertex algebra module characters has been explored recently in \cite{Cordova:2017mhb} using the remarkable Coulomb branch technology for the calculation of the defect indices. It turns out that in general, surface operators in four-dimensions can give rise to more elaborate things than pure VOA modules -- for example, canonical surface defects in class $\SS$ theories generally give rise to rather interesting twisted modules \cite{twisted_modules}.

The above discussion leaves out the other principal defect of interest in four-dimensional theories: line operators. Line operators are initially an unattractive candidate to furnish VOA modules because they do not preserve the right supersymmetries for the cohomological construction of the VOA to go through. Nevertheless, there has been suggestive work showing that superconformal indices of four-dimensional $\NN=2$ theories in the presence of line defects can be rewritten in terms of characters of modules for the associated VOA \cite{Cordova:2015nma}. An indirect explanation for this phenomenon has been put forward in \cite{Cordova:2017mhb}, but it remains to be seen whether a more direct module structure incorporating supersymmetric line defects is possible. In particular, it seems likely that line defects can more naturally be thought of as modules over the vertex Poisson algebra described in Section \ref{subsec:higgs_in_voa} rather than over the full VOA \cite{line_defects}.

%% file: sections/S4.tex

\section{The Deligne-Cvitanovi\'c exceptional series}
\label{sec:deligne}

Our first set of examples is a collection of nine vertex algebras that exhibit a number of remarkable properties. They are the vertex algebras associated to the \emph{Deligne-Cvitanovi\'c (DC) exceptional series} of simple Lie algebras\footnote{These Lie algebras are singled out by their peculiar representation-theoretic properties \cite{Deligne}. For example, one is able to write closed-form expressions (as rational functions of $h^\vee$) for the dimensions of certain finite-dimensional representations that appear in multiple tensor products of the adjoint representation. See the appendix of Cvitanovi\'c's book \cite{Cvitanovic:2008zz} for an account of the curious history of this list of Lie algebras.}
\begin{equation}
\af_0 \subset \af_1 \subset \af_2 \subset \gf_2 \subset \df_4 \subset \mf{f}_4 \subset \ef_6 \subset \ef_7 \subset \ef_8~,
\end{equation}
at the negative levels
\begin{equation}
k_{2d} =-\frac{h^\vee}{6}-1~, 
\end{equation}
where $h^\vee$ is the dual Coxeter number. The theory attached to $\af_0$ (the trivial Lie algebra) is the Virasoro VOA with central charge $c_{2d}=-22/5$, which is the value corresponding to the Lee-Yang, or $(2,5)$, minimal model. Eight of these vertex algebras, namely the cases $\{\af_0\,,\af_1\,,\af_2\,,\df_4\,,\ef_6\,,\ef_7\,,\ef_8\}$, are known to be associated to physical four-dimensional theories. They are the rank-one SCFTs that arise on the worldvolume of a single $D3$ brane at an F-theory singularity -- indeed, one recognizes in this part of the list the classification of singular fibers of an elliptic K3 surface. On the other hand, the four-dimensional interpretation of the $\gf_2$ and $\mf{f}_4$ cases remains unclear.

There are several ways to understand the significance of this list of affine current algebras and their corresponding four-dimensional SCFTs. Perhaps the most physically interesting is via unitarity bounds for the central charges of interacting SCFTs \cite{Beem:2013sza, Liendo:2015ofa, Lemos:2015orc}. The DC affine algebras are singled out as the only simple affine current VOAs associated to (putative) four-dimensional theories that saturate \emph{simultaneously} the unitarity bounds for $c_{4d}$ and $k_{4d}$. From this fact one can deduce a variety of interesting properties, as we will explain.

\subsection{Exceptional series from central charge unitarity bounds}
\label{subsec:unitarity_bounds}

We first review the logic behind the derivation of the unitarity bounds on central charges from the associated VOA. The starting point is the fact that the VOA for any local $\NN=2$ SCFT with simple flavor algebra $\gf_F$ contains a ``universal'' vertex operator subalgebra generated by a stress tensor $T$ of central charge $c_{2d} = -12c_{4d}$ and $\hat{\gf}_F$ affine currents $J^A$ at level $k_{2d} = -k_{4d}/2$. The vertex algebra correlators $\langle TTTT \rangle$, $\langle J^{A}J^{B}TT \rangle$ and $\langle J^{A}J^{B}J^{C}J^{D}\rangle$ are then meromorphic functions that are completely fixed by the choice of Lie algebra $\gf_F$ and parameters $c_{2d}$ and $k_{2d}$. These meromorphic correlators can be decomposed into $\slf(2)$ global conformal blocks weighted by three-point couplings. The meromorphic correlators capture the protected part of four-point functions of $\suf(2)_R$ currents and moment map operators in the four-dimensional SCFT.

In general, a given $\slf(2)$ primary operator of the vertex algebra may descend from a linear combination of various protected operators of the four-dimensional theory that transform in different $\suf(2,2|2))$ representations. Fortunately, for this simple set of correlators one is able to completely resolve the ambiguity under the assumptions that the theory possesses a unique stress tensor and has \emph{no} conserved currents of spin greater than two. The latter condition holds in any \emph{interacting} CFT \cite{Maldacena:2011jn}. With these assumptions in place, one uncovers a precise relationship between two-dimensional and four-dimensional three-point couplings. 

One next demands that the four-dimensional theory be unitary, in which case the appropriately defined three-point couplings must be real, and it is this reality condition that gives rise to several inequalities for the central charges $k_{4d}$ and $c_{4d}$. Each such inequality is saturated when a certain three-point coupling is zero, which means that a certain protected four-dimensional operator that would be allowed by selection rules in the relevant OPE is absent. In the VOA the avatar of this absence is the presence of a certain null vector. We now review the most stringent unitarity bounds that arise upon carrying out this analysis.

\begin{enumerate}
\item[(i)] For fixed $k_{4d} < 2h^\vee$, one finds an \emph{upper} bound on $c_{4d}$ \cite{Beem:2013sza}, 
\begin{equation}
\label{eq:upperc}
c_{4d} \leqslant \frac{k_{4d}\,{\rm dim}\;{\gf}_F}{12(2 h^\vee - k_{4d})}~,\quad{\rm when}\quad k_{4d} < 2h^\vee~.
\end{equation}
This bound arises from consideration of the leading non-singular term in the OPE of two affine currents in the flavor-singlet channel. Recall that the affine currents descend from moment map operators in four dimensions, which lie in $\hat\BB_1$ multiplets. 

There are two multiplets that contribute as $\slf(2)$ primaries to the leading non-singular OPE of two $J$s: namely the stress tensor multiplet $\hat\CC_{0(0,0)}$ and the Higgs chiral ring multiplet $\hat\BB_2$. The three-point coupling $\langle\hat\BB_1 \hat {\BB}_1 \hat {\CC}_{0(0, 0)} \rangle$ of the stress tensor multiplet is fixed by a conformal Ward identity in terms of the Weyl anomaly coefficient $c_{4d}$. Thus the squared OPE coefficient of the $\hat\BB_2$ multiplet is fixed in terms of $c_{4d}$ and $k_{4d}$. Imposing its positivity leads to the bound \eqref{eq:upperc}. 

Saturation of this bound implies the \emph{absence} of a flavor singlet $\hat\BB_2$ multiplet in the $\hat\BB_1 \times \hat\BB_1$ OPE. In the conventions relevant for the VOA, saturation of the bound amounts to the central charge obeying the Sugawara relation
\begin{equation}
c_{2d} = \frac{k_{2d}\;\dim\;\gf_F}{h^\vee + k_{2d}}~,
\end{equation}
so the stress tensor $T$ is identified with the Sugawara stress tensor of the affine current algebra.\footnote{In this case, in the affine current subalgebra there is no null state associated to the saturation of the unitarity bound. However, if one naively takes the appropriate vertex operator subalgebra to be generated by the affine currents in addition to an independent stress tensor, then there is a null state enforcing the Sugawara construction for the stress tensor,
\begin{equation}
\NN_{\rm Sug}=\left(L_{-2}-\frac{1}{k+h^\vee}J_{-1}^AJ_{-1}^A\right)\Omega~.
\end{equation}
}

\item[(ii)] There is additionally a \emph{lower bound} for $k_{4d}$, which depends only on the choice of flavor algebra $\gf_F$ \cite{Beem:2013sza},
\begin{equation}
k_{4d} \geqslant k_{4d}^{\rm min} (\gf_F)~, \quad \gf_F \neq \af_1~.
\end{equation}
The bound arises from consideration of the leading non-singular term in the OPE of two affine currents in flavor \emph{non-singlet} channels. By assumption, the four-dimensional theory contains a unique $\hat\CC_{0(0,0)}$ multiplet which is a flavor singlet, so there are no possible contributions from $\hat\CC_{0(0,0)}$ multiplets in the non-singlet channels. One can therefore compute the three-point coupling $\langle\hat\BB_1\hat\BB_1\hat\BB_2\rangle$ unambiguously as a function of $k_{4d}$ for each choice of non-singlet flavor representation $\RRR$ in which the $\hat\BB_2$ can transform. The allowed representations $\RRR$ are the ones that appear in the symmetrized tensor product of two copies of the adjoint representations. Each choice of $\RRR$ gives rise to a different bound. In Table \ref{tab:bounds}, the most stringent bound for each simple flavor algebra $\gf_F$ is displayed, along with the representation $\RRR$ responsible for the bound. Saturation of one of these bound implies the \emph{absence} of a $\hat\BB_2$ multiplet in the representation $\RRR$ in the OPE $\hat\BB_1\times\hat \BB_1$. In the four-dimensional theory this is a Higgs chiral ring relation. In the vertex algebra, this translates into the statement that at the relevant value of $k_{2d}$ there is a level-two null state in the representation $\RRR$ of the form
\begin{equation}
\label{eq:NJJ}
\NN^\RRR_{[JJ]} = c_{AB}^{\phantom{AB}\RRR} J_{-1}^A J_{-1}^B\,\Omega~,
\end{equation}
where the coefficients $c_{AB}^{\phantom{AB}\RRR}$ accomplish the projection to the representation $\RRR$. Note that the case $\gf_F = \af_1$ is exceptional: the bound given by the analysis above is gives $k_{4d} \geqslant - 2$, which is automatically obeyed (and can never be saturated) in any unitary theory.

\begin{table}[t] 
\centering
\begin{tabular}{lllc}
\hline
\hline
$\gf_F$ 		&					& Bound 							& Representation $\RRR$ 					\\[.5ex] 
\hline 
$\SU(N)$ 	& $N \geqslant 3$ 	& $k_{4d}\geqslant N$				& $\mathbf{N^2-1}_{\mathrm{sym}}$			\\[.3ex]
$\SO(N)$ 	& $N = 4,\ldots,8$	& $k_{4d}\geqslant 4$ 		 		& $\mathbf{\frac{1}{24} N(N-1)(N-2)(N-3)}$	\\[.3ex]
$\SO(N)$ 	& $N \geqslant 8$ 	& $k_{4d}\geqslant N-4$ 			& $\mathbf{\frac12 (N+2)(N-1)}$				\\[.3ex]
$\USp(2N)$	& $N \geqslant 3$ 	& $k_{4d}\geqslant N+2$				& $\mathbf{\frac12 (2N+1)(2N-2)}$			\\[.3ex]
$G_2$ 	 	&					& $k_{4d}\geqslant \frac{10}{3}$	& $\mathbf{27}$								\\[.3ex]
$F_4$ 	 	&					& $k_{4d}\geqslant 5$				& $\mathbf{324}$							\\[.3ex]
$E_6$ 	 	&					& $k_{4d}\geqslant 6$			 	& $\mathbf{650}$							\\[.3ex]
$E_7$ 	 	&					& $k_{4d}\geqslant 8$				& $\mathbf{1539}$							\\[.3ex]
$E_8$ 	 	&					& $k_{4d}\geqslant 12$				& $\mathbf{3875}$							\\[.3ex]
\hline
\end{tabular}
\caption{Unitarity bounds for the anomaly coefficient $k_{4d}$ arising from positivity of the $\hat\BB_2$ three-point function in non-singlet channels \cite{Beem:2013sza}.\label{tab:bounds}}
\end{table}

\item[(iii)] Finally, there is a \emph{lower bound} on $c_{4d}$ for fixed $k_{4d}$ \cite{Lemos:2015orc},\footnote{We mention here that there is a generalization of the bound \eqref{eq:clower} to the case where the flavor algebra is reductive, $\gf_F = \prod_i \gf_F^{(i)}$,
\begin{equation}
c_{4d} \geqslant \frac{11}{60} \left(1 + \sqrt{1 + \frac{180}{121} \sum_i \, \frac{k^{(i)}_{4d} \, {\rm dim} \,{G}^{(i)}_F} {3 k^{(i)}_{4d} - h^{\vee (i)}}} \right)~.
\end{equation}
This result is valid also when $U(1)$ factors are present, with the formal assignment $h^\vee \equiv 0$ for $U(1)$. The proof is analogous to the case with simple flavor algebra.} 
\begin{equation}
\label{eq:clower}
c_{4d} \geqslant \frac{11}{60} \left(1 + \sqrt{1 + \frac{180}{121}\frac{k_{4d} \, {\rm dim} \,{G}_F}{3 k_{4d} - h^\vee}} \right)~.
\end{equation}
This bound arises upon consideration of flavor-singlet $\slf(2)$ primaries of holomorphic dimension $h = 4$ that appear in the OPE decomposition of the mixed correlator system $\langle TTTT\rangle$, $\langle TTJ^AJ^B\rangle$ and $\langle J^AJ^BJ^CJ^D\rangle$. There are \emph{a priori} two distinct primaries that can appear. One can choose a basis such that the first primary is defined to be the one that appears in the $T \times T$ OPE, and the second is orthogonal to the first. In general the $J \times J$ OPE will contain a linear combination of both primaries. From the four-dimensional viewpoint, these states correspond to two distinct flavor-singlet $\hat\CC_{1(\frac12,\frac12)}$ multiplets, appearing in the OPEs of two moment maps and of two $\suf(2)_R$ currents. Imposing positivity of the squared OPE coefficients of both multiplets leads to the bound \eqref{eq:clower}. When the bound is saturated, the ``second'' $\hat\CC_{1(\frac12,\frac12)}$ multiplet is absent, \ie, precisely the same multiplet is appearing in both the OPE of the $\suf(2)_R$ currents and that of the moment maps. In the vertex algebra, saturation of this bound implies the existence of a null state of the form
\begin{equation}
\label{eq:NT}
\NN_{[T]} = \left((L_{-2})^2 + \alpha L_{-4} + \beta J_{-3}^A J_{-1}^A + \gamma J_{-2}^AJ_{-2}^A \right)\Omega~,
\end{equation}
where $\alpha$, $\beta$, and $\gamma$ are coefficients that can in principle be determined by direct computation. In particular, we see that when the bound \eqref{eq:clower} is saturated, then we have
\begin{equation}
(L_{-2})^2\Omega \in C_2(\VV)~.
\end{equation}
\end{enumerate}

In general it is impossible for all three of the above bounds to be saturated for a single VOA. The choice of a simple Lie algebra $\gf_F$ determines $h^\vee$ and ${\rm dim}\,\gf_F$, but to saturate all three bounds requires that the two variables $k_{2d}$ and $c_{2d}$ satisfy three equations. The equations are generically independent and so no solution exists. The exceptions are precisely $\gf_F = \af_1\,,\af_2\,,\gf_2\,,\df_4\,,\mf{f}_4\,,\ef_6\,,\ef_7\,,\ef_8$. (The case $\gf_F = \af_1$ is somewhat less special because there is no meaningful lower bound on $k_{4d}$ so we are only imposing two equations). In every case, the specified values of $k_{4d}$ and $c_{4d}$ are expressed as the same functions of $h^\vee$,
\begin{equation}
k_{4d} = \frac{h^\vee}{3}+2~,\qquad c_{4d} = \frac{h^\vee}{6} + \frac{1}{6}~.
\end{equation}
For the trivial Lie algebra $\gf_F = \af_0$, \ie, in the case of no continuous flavor symmetry, bounds (i) and (ii) are of course meaningless but bound (iii) still applies \cite{Liendo:2015ofa},
\begin{equation}
c_{4d} \geqslant \frac{11}{30}~.
\end{equation}
The bound is saturated by the Virasoro algebra at central charge $c_{2d} = - 22/5$, which is the value relevant for the $(2,5)$ minimal model. We will consider this vertex algebra as belonging to the DC series, with the formal assignment $h^\vee = 6/5$.

\subsection{Properties of the Deligne-Cvitanovi\'c algebras}
\label{subsec:properties}

Saturation of bound (iii) implies the existence of a null state of the form \eqref{eq:NT}. Following the methods outlined in section \ref{sec:modular}, one derives a second-order LMDE for the full modular group $\Gamma$ for the vacuum character of the VOA, which necessarily takes the form given in \eqref{eq:secondordereq}. (In this case, the recursion relation \eqref{eq:recursion_general} truncates after a single step). In principle, the coefficient $\lambda$ in front of the Eisenstein series can be computed from the precise form of the null state, but it is also easily fixed using \eqref{eq:c2dlambda} in terms of the central charge,
\begin{equation}
\lambda = -5\left(\frac{c_{2d}^2}{4} + c_{2d}\right) =-5(h^\vee +1)(h^\vee -1)~.
\end{equation}
We conclude that the vacuum characters of the DC series of vertex algebras obey a uniform set of second order modular differential equations,
\begin{equation}
\label{eq:delinge_series_diffeq}
\left(D_q^{(2)}-5(h^{\vee}+1)(h^\vee-1)\Eb_4(q)\right)\goodchi(q)=0~.
\end{equation}
The explicit solutions of this equation have been discussed recently in \cite{Arakawa:2016hkg}; we also derive them (in a slightly different form) in Appendix \ref{app:deligne_solutions}. We note that by solving the indicial equation, or by using \eqref{eq:c2dlambda}, the values of $h_1$ (the conformal weight of the second solution to the indicial equation) are immediately found to take the very simple form
\begin{equation}
h_1=h_{\min}=-\frac{h^\vee}{6}~.
\end{equation}
It is interesting to note that for $\gf=\df_4,\ef_6,\ef,\ef_8$, this is an integer, and the second solution to \eqref{eq:delinge_series_diffeq} is logarithmic.

In Table \ref{tab:deligne} we summarize this data and more for the the vertex algebras associated to the full Deligne-Cvitanovi\'c series. The values of the conformal anomaly coefficient $a_{4d}$ are derived assuming that the Cardy behavior of the Schur index is controlled by \eqref{eq:cardyeq}. Finally, assuming that the parent four-dimensional theory has a dimension-one Coulomb branch and that the Shapere-Tachikawa formula holds, one derives the $\uf(1)_r$ charge $r$ of the unique Coulomb chiral ring generator, see \eqref{eq:r2ndorder}. For the cases $\{\af_0\,,\af_1\,,\af_2\,,\df_4\,,\ef_6\,,\ef_7\,,\ef_8\}$, the values of $a_{4d}$ and $r$ calculated this way are found in agreement with the known values for the rank-one F-theory SCFTs. For the cases of $\gf_2$ and $\mf{f}_4$, the values of $r$ are problematic, as they do not fit the Kodaira classification of possible defect angles for a dimension-one scale-invariant Coulomb branch geometry (see, \eg, \cite{Argyres:2015ffa, Argyres:2015gha, Argyres:2016xmc} for a recent discussion). This might be an indication that these vertex algebras do not descend from actual $4d$ SCFTs. Alternatively, one of the assumptions in the calculation of $r$ might be wrong, \eg, the $4d$ theories might be of higher rank, violate the Tachikawa-Shapere formula, or perhaps have a Coulomb branch of exotic type \cite{Argyres:2017tmj}.\footnote{It has been argued that the hypothetical theory with ${\mathfrak f}_4$ flavor symmetry cannot exist due to a mismatch of global anomalies \cite{Shimizu:2017kzs}.}

\renewcommand{\arraystretch}{1.5}
\begin{table}
\centering
\begin{tabular}{|c|c|c|c|c|c|c|}
\hline \hline
~$\gf$~ 	& ~$h^\vee$~& ~$k_{2d}= \frac{-h^\vee-6}{6}$~ & ~$c_{2d}=-2-2h^\vee$~ & ~$h_1=-\frac{h^\vee}{6}$~ & ~$a_{4d} = \frac{5+3h^{\vee}}{24}$~ & ~$r =\frac{h^\vee+6}{6}$ \\ 
\hline 
$\af_0$		& $\frac65$ & $-\frac65$ & $-\frac{22}{5}$ & $-\frac{1}{5}$ & $\frac{43}{120}$ & $\frac{6}{5}$\\
\hline 
$\af_1$		& $2$		& $-\frac43$ & $-6$ 	& $-\frac{1}{3}$ & $\frac{11}{24}$ 	& $\frac{4}{3}$\\ 
\hline
$\af_2$		& $3$		& $-\frac32$ & $-8$		& $-\frac{1}{2}$ & $\frac{7}{12}$ 	& $\frac{3}{2}$ \\
\hline
$\gf_2$		& $4$ 		& $-\frac53$ & $-10$	& $-\frac{2}{3}$ & $\frac{17}{24}$ 	& $\frac{5}{3}^{\star}$\\
\hline
$\df_4$		& $6$ 		& $-2$		 & $-14$	& $-1$ 			 & $\frac{23}{24}$ 	& 2\\
\hline
$\mf{f}_4$	& $9$ 		& $-\frac52$ & $-20$	& $-\frac{3}{2}$ & $\frac{4}{3}$ 	& $\frac{5}{2}^{\star}$\\
\hline
$\ef_6$		& $12$ 		& $-3$		 & $-26$	& $-2$ 			 & $\frac{13}{8}$	& $3$\\
\hline
$\ef_7$		& $18$ 		& $-4$		 & $-38$	& $-3$ 			 & $\frac{59}{24}$ 	& $4$\\
\hline
$\ef_8$		& $30$ 		& $-6$		 & $-62$	& $-5$ 			 & $\frac{95}{24}$ 	& $6$\\
\hline
\end{tabular}
\caption{\label{tab:deligne} 
The Deligne-Cvitanovi\'c series of simple Lie algebras, the data of the associated vertex algebras and the data of their (putative) parent $4d$ SCFTs. The $\af_0$ entry is formally a member of the list corresponding to the trivial Lie algebra. It corresponds to the VOA of the $(2,5)$ Virasoro minimal model. As the $4d$ interpretation of the $\gf_2$ and $\mf{f}_4$ cases is still unclear, the values of $a_{4d}$ and $r$ for these entries are formal/conjectural. In particular even if these theories exist, the values of $r$ may be different if the various assumptions we have made about their Coulomb branches do not hold.}
\end{table}

Saturation of bounds (i)-(iii) also has immediate consequences for the associated variety of these VOAs. For any affine current VOA $V_k(\gf)$, the generators of the $C_2$ algebra are the equivalence classes of the adjoint valued currents,
\begin{equation}
J^A_{-1}\Omega \in \VV ~~\Longrightarrow~~ j^A \in \RR_\VV~,\qquad A=1,\ldots,\dim\gf~,
\end{equation}
with the Poisson bracket determined by the structure constants of $\gf$,
\begin{equation}
\{j^A,j^B\} = f^{AB}_{\phantom{AB}C}j^C~.
\end{equation}
Thus the $C_2$ algebra is necessarily a Poisson subalgebra of the symmetric algebra $S(\gf_{\Cb})$. In the DC examples, the null state implied by the saturation of bound (ii) implies the relation
\begin{equation}
\restr{(j\otimes j)}{\RRR}=0~.
\end{equation}
Furthermore, saturation of bounds (i) and (iii) imply the relations,
\begin{equation}
\frac{1}{k_{2d}+h^\vee}\restr{(j\otimes j)}{\bf 1}=t~,\qquad t^2 = 0~.
\end{equation}
Thus, in the reduced algebra that defines the associated variety, we have that we must quotient $\gf_\Cb$ by the ideal generated by the singlet factor and the $\RRR$-valued factor in the square of $j^A$. Now the list of DC simple Lie algebras can also be characterized as those simple Lie algebras for which the symmetrized product of two adjoint representations contains at most three representations,
\begin{equation}
\label{eq:symDC}
{\rm Sym}^2 ({\bf adj}) = {\bf 1} \oplus {\RRR} \oplus (2 {\bf adj} ) \quad {\rm for \; DC \; Lie \; algebras}~.
\end{equation}
Here ${\bf 1}$ denotes the singlet, $(2 {\bf adj})$ is the representation whose Dynkin indices are twice those of the adjoint representation, and ${\RRR}$ can be read in Table \ref{tab:bounds} for all DC algebras except $\af_1$, for which it is absent.\footnote{The $\af_1$ case is special: the decomposition of the symmetrized product of two adjoint representations contains only two terms, ${\rm Sym}^2 ({\bf 3})= {\bf 1} \oplus {\bf 5}$, where ${\bf 5} \equiv (2 {\bf adj} )$. Recall that bound (ii) is trivial for $\af_1$. Saturation of bound (i) is sufficient to conclude that the product of two moment maps contains only $\hat {\BB}_2$ operators in the $(2 {\bf adj} )$ representation.} 

Now for any simple Lie algebra, one may define the ``Joseph ideal" $\II_2$ in the symmetric algebra $S(\gf)$ using the following decomposition of the symmetrized product of two adjoint representations,
\begin{equation}
{\rm Sym}^2 ({\bf adj}) = (2 {\bf adj} ) \oplus \II_2~.
\end{equation}
Thus we see that for the DC examples, the ideal we have shown must be removed in passing to the associated variety is precisely the Joseph ideal. What's more, there can be no further quotient in the associated variety because for negative levels, states of the form $(J^{h.w.}_{-1})^n\Omega$ cannot appear in any null vectors, and any further quotient of the associated variety would require such nulls.\footnote{Here by $J^{h.w.}$ we denote the $\gf_\Cb$ highest weight.} Consequently, we have the elegant result,
\begin{equation}
(\RR_\VV)_{\rm red} = S(\gf_\Cb)/\II_2\cong \overline{\Ob_{\min}(\gf_\Cb)}~,
\end{equation}
where we have recognized that the variety defined by removing the Joseph ideal is the \emph{minimal nilpotent orbit} of $\gf_\Cb$. This result was also derived as a part of a much more general computation of associated varieties for affine current VOAs in \cite{Arakawa:2015jya}. For the special $\af_0$ case (the Virasoro vertex algebra at $c_{2d}= -22/5$) we have discussed the calculation of the associated variety in \ref{subsubsec:virasoro_example}: the stress tensor is the only generator, and its nilpotency in $C_2(\VV)$ implies that the associated variety is just a point, which in a formal sense can be taken to be the minimal nilpotent orbit of the trivial Lie algebra.

This result is in precise agreement with physical expectations. As we have already mentioned, the four-dimensional theories associated to the DC vertex algebras of type $\{\af_0\,,\af_1\,,\af_2\,,\df_4\,,\ef_6\,,\ef_7\,,\ef_8\}$, are the rank-one SCFTs that arise on the worldvolume of a single $D3$ brane at an F-theory singularity. The F-theory picture makes it clear that the Higgs branch of these theories is the centered one-instanton moduli space for $\gf_\Cb$, which is well-known to be isomorphic to the $\gf_\Cb$ minimal nilpotent orbit.\footnote{The $\af_0$ case is of course special; there is no moment map operator to begin with and the Higgs branch is trivial.} While there is no known four-dimensional theory associated the the DC vertex algebras of type $\gf_2$ and $\mf{f}_4$, if such theories exist their Higgs branch is predicted to be the minimal nilpotent orbit of the corresponding flavor algebras. In fact, the full list of DC vertex algebras was obtained in \cite{Beem:2013sza} by a slightly different route from the one we have followed here, namely by demanding that the Higgs branch of the putative four-dimensional theory be given by the one-instanton moduli space for $\gf_\Cb$. This restricts the list of possible flavor algebras to the DC series, as for any other algebra there would be a forth representation appearing on the right hand side of \eqref{eq:symDC}. The values of $c_{2d}$ and of $k_{2d}$ must then saturate the bounds (i) and (ii) in order have the Joseph relations imposed.\footnote{For $\gf_F = \af_1$, saturation of the central charge bound (i) is sufficient to obtain a Higgs branch isomorphic to the one-instanton moduli space, for any value of $k_{4d}$. The value of $k_{4d}$ can be fixed by further imposing bound (iii), as we have done in the discussion above. However, as we will see in Section \ref{subsec:admissible_higgs}, there can be theories with different values of $k_{4d}$ that still have the $\af_1$ one-instanton moduli space for their Higgs branch/associated variety.}

%% file: sections/S5.tex

\section{Argyres-Douglas theories}
\label{sec:argyres_douglas}

An interesting class of test cases for the ideas outlined in the previous sections are Argyres-Douglas (AD) theories \cite{Argyres:1995jj,Xie:2012hs}. Importantly, recent developments have rendered the Schur indices of these theories accessible due to a relation with the wall-crossing properties of the spectrum of BPS particles on the Coulomb branch \cite{Cordova:2015nma}. The same indices have also been predicted using the connection between the superconformal index of class $\SS$ theories and two-dimensional $q$-deformed Yang-Mills  correlators \cite{Buican:2015ina, Buican:2015hsa, Buican:2015tda, Song:2015wta, Buican:2017uka}. These developments have led to a number of proposals for vertex operator algebras associated to certain Argyres-Douglas theories \cite{Cordova:2015nma,Creutzig:2017qyf, Song:2017oew}.\footnote{See also \cite{Buican:2014hfa, Buican:2016arp, wenbin, Buican:2017fiq} for other recent developments concerning  Argyres-Douglas theories and their associated VOAs, and \cite{Maruyoshi:2016aim, Agarwal:2016pjo, Agarwal:2017roi, Benvenuti:2017bpg} for an interesting new  Lagrangian perspective on Argyres-Douglas theories.}
Here we explore the proposed Higgs branch reconstruction and modular finiteness conditions for several infinite families of Argyres-Douglas vertex operator algebras. The general framework outlined in this paper should be equally applicable to the much larger landscape of Argyres-Douglas theories explored in the aforementioned works.

\subsection{\tpdf{$(A_1,A_{2n})$}{(A1,A2n)} theories: trivial Higgs branch}
\label{subsec:AD_trivial_higgs}

Our first examples are the $(A_1, A_{2n})$ Argyres-Douglas theories (these are also referred to as the $I_{2,2n+1}$ theories in \cite{Xie:2013jc}). As four-dimensional theories, these have rank $n$ (\ie, $n$-dimensional Coulomb branches) and trivial Higgs branches. Their Weyl anomaly coefficients are given by
\begin{equation}\label{eq:A1_Aeven_central_charges}
c_{4d}=\frac{n(6n+5)}{6(2n+3)}~,\qquad a_{4d}=\frac{n(24n+19)}{24(2n+3)}~.
\end{equation}
The simplest case of $n=1$ coincides with the $\af_0$ entry in the Deligne-Cvitanovi\'{c} series discussed above.

We have proposed that VOAs associated to these Argyres-Douglas theories are the non-unitary $(2,2n+3)$ Virasoro VOAs \cite{rastelli_harvard},\footnote{See also the earlier \cite{Cecotti:2010fi} for a relationship between VOAs and Argyres-Douglas theories.}
\begin{equation} \label{Viridentification}
\goodchi[\AD_{(A_1,A_{2n})}]=\Vir_{2,2n+3}~.
\end{equation}
For these Virasoro VOAs, the two-dimensional central charge takes the values
\begin{equation}
\label{eq:virasoro_minimal_central_charge}
c_{2d}=-\frac{2n(6n+5)}{2n+3}~.
\end{equation}
The match between \eqref{eq:A1_Aeven_central_charges} and \eqref{eq:virasoro_minimal_central_charge}, along with the aforementioned membership of the $n=1$ theory in the Deligne-Cvitanovi\'c series, was the original motivation for the proposed identification. The match has been extended to the level of the Schur index/vacuum character in \cite{Cordova:2015nma,Buican:2015ina}.

These examples constitute a simple example of the proposed Higgs branch/associated variety identification. It is well known that for the $(p,q)$ Virasoro VOA, there is a null state at level $(p-1)(q-1)$ and that the coefficient of $(L_{-2})^{\frac12(p-1)(q-1)}\Omega$ is nonzero in that null state. Following the discussion in \ref{subsubsec:virasoro_example}, the $C_2$ algebras for these VOAs are therefore given by
\begin{equation}
\RR_{\Vir_{p,q}}=\Cb[t]/\langle t^{\frac12(p-1)(q-1)}\rangle~.
\end{equation}
Every element other than the identity is nilpotent in these rings, so the reduced rings are trivial,
\begin{equation}
\left(\RR_{\Vir_{p,q}}\right)_{\rm red}\cong\Cb~,
\end{equation}
and the associated variety is just a point,
\begin{equation}
X_{\Vir_{p,q}}=\,{\rm Spec}\,\Cb~\cong~\Cb^0~,
\end{equation}
which matches with the moduli space physics of these theories. The triviality of the associated variety is equivalent to the statement that $\Vir_{p,q}$ is a $C_2$-co-finite VOA, and indeed the minimal-model Virasoro VOAs are the first textbook examples of $C_2$-co-finite VOAs. We observe that the $C_2$-co-finite case, which for some time was the case of primary interest in the mathematical literature, only accounts for SCFTs with trivial Higgs branches.

In terms of modular differential equations, the null state that truncates $\RR_{\Vir_{2,2n+3}}$ gives rise to an LMDE of order $n+1$ for the vacuum character, with the $n$ additional solutions being the characters of the non-vacuum modules $\phi_{r,1}$ for $r=2,\ldots,n+1$. Of these, the module with the smallest conformal dimension is $\phi_{n+1,1}$, which has dimension
\begin{equation}
h_{\min} = h_{n+1,1} = -\frac{n(n+1)}{2(2n+3)}~,
\end{equation}
which, upon substituting into \eqref{eq:a_anomaly_equation} along with the expression for the Virasoro central charge \eqref{eq:virasoro_minimal_central_charge}, does indeed reproduce the predicted value for $a_{4d}$ given in \eqref{eq:A1_Aeven_central_charges}.

The proposal (\ref{Viridentification})  has been vastly generalized in \cite{Cordova:2015nma}. The AD theories $(A_{k-1},A_{N-1})$ with $k$ and $N$ relatively prime are identified
with the $(k, k+n)$ ${\cal W}_k$ algebra,
\be
\goodchi[\AD_{(A_{k-1},A_{N-1})}]={\cal W}_{k} (k, k+N)~ , \quad  (k, N) = 1\,.
\ee
These vertex algebras are well-known to be $C_2$-co-finite, matching again physical expectations about  the parent $AD$ theories, which have trivial Higgs branches.

\subsection{\tpdf{$(A_1,D_{2n+1})$}{(A1,D(2n+1)} theories: \tpdf{$\Cb^2/\Zb_2$}{C2/Z2} Higgs branches}
\label{subsec:admissible_higgs}

Our next family of examples are the Argyres-Douglas theories of type $(A_1,D_{2n+1})$, also denoted $(I_{2,2n-1},F)$ in \cite{Xie:2013jc}. For this infinite family of SCFTs, the Higgs branch of vacua always coincides with the (closure of the principal) nilpotent orbit of $\suf(2)$, \ie, with the orbifold $\Cb^2/\Zb_2$. As a complex algebraic variety this Higgs branch is realized as
\begin{equation}
\MM_H=\Cb^2/\Zb_2 \equiv \Cb[j^1,j^2,j^3]/\langle j^1j^1+j^2j^2+j^3j^3\rangle~.
\end{equation}
The $j^A$ are the moment maps for the holomorphic $\suf(2)$ action on $\MM_H$, and in particular the Poisson bracket is determined by its behavior for these generators, which is determined by the $\suf(2)$ commutation relations,
\begin{equation}
\label{eq:su2_poisson}
\{j^A,j^B\}=i\epsilon^{ABC}j^C~.
\end{equation}
Note that for for the associated variety/Higgs branch correspondence to hold in these examples the VOA cannot be $C_2$-co-finite, and so in particular these cannot be rational VOAs. As four-dimensional SCFTs, these are rank-$n$ theories with Weyl anomaly coefficients given by \cite{Xie:2012hs,Xie:2013jc}
\begin{equation}
\label{eq:A1DOdd_central_charges}
c_{4d}=\frac{n}{2}~,\qquad a_{4d}=\frac{n(8n+3)}{8(2n+1)}~.
\end{equation}

It was proposed on the basis of matching central charges and 
Higgs branch generators)that the VOAs for these theories
are (the simple quotients of) the $\suf(2)$ affine current VOA at certain levels \cite{rastelli_delzotto_unpublished}, 
\begin{equation}
\goodchi[\AD_{(A_1,D_{2n+1}}]= V_{\frac{-4n}{2n+1}}(\suf(2))~.
\end{equation}
Note that the first member of this family coincides with the $\mathbf{a}_1$ DC vertex algebra. The superconformal indices of the  $(A_1,D_{2n+1})$ theories have been matched with the vacuum characters of these VOAs \cite{Cordova:2015nma}.

The levels of these affine current algebras are all \emph{admissible} in the sense of Kac and Wakimoto \cite{Kac:1988qc}, but not integrable.\footnote{Kac-Wakimoto admissible levels for $\suf(2)$ are levels of the form $k=-2+\frac{p}{q}$ where $(p,q)=1$ and $p\geqs 2$.} The admissible levels for affine current algebras are interesting precisely because even though they do not yield rational VOAs (their representation category are known to not be semi-simple) their vacuum characters nevertheless transform in finite-dimensional representations of ${\rm PSL}(2,\Zb)$.

The associated varieties of affine current VOAs at admissible levels have been determined in \cite{Arakawa:2010ni}, and for the levels in question it was found that the associated variety is given by the closure of the principal nilpotent orbit of $\suf(2)$. For illustrative purposes, let us see how this works explicitly for the simplest cases $n=1,2$.

The VOAs are strongly generated by the currents $J^A(z)$, $A=1,2,3$. The $C_2$ algebra is then necessarily a quotient by some ideal of the free commutative algebra generated by the representatives of these currents in the quotient $\VV/C_2(\VV)$, which we shall denote by $j^A$ as we expect these to be identified with the moment maps on the Higgs branch. In other words, we have
\begin{equation}
\RR_{\VV}=\Cb[j^1,j^2,j^3]/\II~.
\end{equation}
In addition, the structure of the affine current VOA guarantees that the secondary bracket behaves as dictated by \eqref{eq:su2_poisson} for the generators of $\RR_{\VV}$. Identifying the ideal $\II$ is not entirely trivial for general the general case. However for small values of $n$ it is not difficult to perform the exercise by hand. In particular, one finds null states in the vacuum Verma modules whose images in the $C_2$ algebra are given by
\begin{equation}
\begin{split}
\NN^A&=j^A(j^1j^1+j^2j^2+j^3j^3)^{\phantom{2}}~,\qquad n=1~,\\
\NN^A&=j^A(j^1j^1+j^2j^2+j^3j^3)^2~,\qquad n=2~.
\end{split}
\end{equation}
We note here that the relation $\sum_{A}j^Aj^A=0$ does \emph{not} hold in $\RR_\VV$ in either case. On the other hand, we do have the relations
\begin{equation}
\begin{split}
\left(\sum_{A}j^Aj^A\right)^2&=0~~~{\rm in}~~~\RR_\VV~,\qquad n=1~,\\
\left(\sum_{A}j^Aj^A\right)^3&=0~~~{\rm in}~~~\RR_\VV~,\qquad n=2~.
\end{split}
\end{equation}
Thus the quadratic Casimir is in the nilradical of $\RR_{\VV}$, and by passing to the associated variety (\ie, forgetting nilpotents), we recover the correct Higgs chiral ring as a Poisson algebra. It is natural to conjecture that in general the quadratic Casimir will be nilpotent in $\RR_\VV$ of degree $n+1$. Since the quadratic Casimir is the image of the Sugawara stress tensor in $\RR_{\VV}$, our previous considerations would therefore lead us to expect a modular equation of order $n+1$. We will see that this is in fact the case.

In general, there are in fact $2n$ admissible, non-vacuum modules whose characters appear in the modular transformations of the vacuum character for these algebras. The primaries in these modules have $\suf(2)$ spin and conformal dimensions given by \cite{Kac:1988qc}
\begin{equation}
j_k = -\frac{k+1}{2n+1}~,\qquad h_k = \frac{j_k(j_k+1)(2n+1)}{2}~,\qquad k=0,\ldots,2n-1.
\end{equation}
Of these, the modules with minimal conformal dimension are the $k=n-1$ and $k=n$ cases, for which we have
\begin{equation}
\label{eq:hmin_kacmoody}
h_{\min}=-\frac{n(n+1)}{2(2n+1)}~.
\end{equation}
The characters for these modules are given by
\begin{equation}
\goodchi_k(q,a)=\Tr_{M_k} \left(q^{L_0-c/24}a^{J^3_0}\right)=\frac{\Theta_{2n-2k-1,4n+2}(\tau,\alpha)-\Theta_{-2n-2k-3,4n+2}(\tau,\alpha)}{\Theta_{1,2}(\tau,\alpha)-\Theta_{-1,2}(\tau,\alpha)}~,
\end{equation}
where $a=\exp(2\pi i \alpha)$ and 
\begin{equation}
\Theta_{n,m}(\tau,\alpha)=\sum_{r\in \Zb+(n/2m)}q^{mr^2}a^{mr}~.
\end{equation}
These characters are meant to be thought of in terms of their series expansions in $q$ and $a^{-1}$.

These characters are not well defined in the limit $a\to1$ because the horizontal $\suf(2)$ representations with fixed conformal dimension within these modules are infinite-dimensional. Nevertheless, it turns out that differences of characters, when considered as analytic functions of the fugacities as opposed to formal power series, can have finite limits when the flavor fugacities are set to zero. In particular, the following combinations of characters are well-defined in the unflavored limit \cite{Mukhi:1989bp},
\begin{equation}
\goodchi^+_k(q,q)= \goodchi_{k}(q,q)-\goodchi_{2n-1-k}(q,q)~,\qquad k=0,\ldots,n-1~.
\end{equation}
In each case, the two simple characters appearing in the linear combination have the same conformal dimension, so the full set of conformal dimensions are realized by the characters that make sense when $a=0$. It is these combinations of characters, along with the vacuum character, that transform into one another under modular transformations and that solve an order $n+1$ LMDE. Indeed, using the value for $h_{\min}$ from \eqref{eq:hmin_kacmoody} in \eqref{eq:a_anomaly_equation}, along with the value of the Virasoro central charge, we recover the expression for the Weyl anomaly $a_{4d}$ given in \eqref{eq:A1DOdd_central_charges}.


\subsection{\tpdf{$(A_1, A_{2n-1})$}{(A1,A(2n-1))} theories: Kleinian singularities for Higgs branches}
\label{subsec:canonical_singularities}

The $(A_1,A_{2n-1})$ Argyres-Douglas theories (\ie, the $I_{2,2n}$ theories) have rank $(n-1)$ and their Higgs branches are the $A$-type Kleinian singularities (also called canonical singularities),
\begin{equation}
\MM_{H}(A_1,A_{2n-1})\cong\Cb^2/\Zb_{n}~.
\end{equation}
As algebraic varieties, these singularities can all be realized as hypersurfaces in $\Cb^3_{[x,y,z]}$,
\begin{equation}
\Cb^2/\Zb_n\cong{\rm Spec}\big(\Cb[x,y,z]/\langle xy+z^{n}\rangle\big)~,
\end{equation}
with the Poisson bracket acting on the generators according to
\begin{equation}
\{z,x\}=x~,\qquad \{z,y\}=-y~,\qquad \{x,y\} = n z^{n-1}~.
\end{equation}
The Weyl anomaly coefficients are given by
\begin{equation}
a_{4d}=\frac{12n^2-5n-5}{24(n+1)}~,\qquad c_{4d} = \frac{3n^2-n-1}{6(n+1)}~.
\end{equation}
For $n=2$ this family coincides with our previous examples (indeed the $(A_1,A_3)$ theory is the same thing as the $(A_1, D_3)$ theory), but for $n\geqs3$ these are distinct theories from those considered above.

A strong hint of what VOAs should be associated to these SCFTs comes from (part of) the Grothendieck-Brieskorn-Slodowy theorem. The relevant part of the statement of the theorem, which we paraphrase from \cite{Namikawa:slodowy}, is as follows:
\begin{thm}[Grothendieck-Brieskorn-Slodowy \cite{Slodowy}]
Let $\gf$ be a simple complex Lie algebra of type $ADE$ and $f\in\gf$ a point in the subregular nilpotent orbit. Denote by $\SS_{f}$ the Slodowy slice transverse to $f$. Further denote the nilpotent cone in $\gf$ by $\NN$. Then the intersection $\SS_{f}\cap\NN$ is a Kleinian surface singularity associated to the same Dynkin diagram as $\gf$.
\end{thm}
Thus the Higgs branches in question can be realized as the intersections of nilpotent cones with subregular Slodowy slices for $\gf=\slf(n)$, $n\geqs2$.

The relevance of this theorem arises from the results of Arakawa on the relationship between quantum Drinfel'd-Sokolov reduction of affine current VOAs and intersections of nilcones with Slodowy slices. In particular, Arakawa has proven the following general result, which we state in a more limited form relevant for our purposes:
\begin{thm}[Arakawa \cite{Arakawa:2010ni}]\label{thm:arakawa}
Let $\VV$ be an affine current VOA $V_k(\gf)$ of level $\tilde k$ with $\gf$ a simple Lie algebra. Let $X_{\VV}$ denote the associated variety of $\VV$. For $f$ a nilpotent element of $\gf$, let $H_f^{\frac{\infty}{2}+0}(\VV)$ denote the VOA obtained by generalized quantum Drinfel'd-Sokolov reduction of $\VV$ associated to the $\slf(2)\subset\gf$ defined by $f$. Then the associated variety of the reduced VOA is obtained from that of $\VV$ by intersecting with the Slodowy slice $\SS_f$,
\begin{equation}
X_{H_f^{\frac{\infty}{2}+0}(\VV)}=X_{\VV}\cap \SS_{f}~.
\end{equation}
\end{thm}
Consequently, for any $\tilde k$ such that $X_{V_{\tilde k}(\gf)}=\NN_{\gf}$ --- \ie, for which the associated variety is the closure of the principal nilpotent orbit in $\gf$ --- the associated variety of the subregular quantum Drinfel'd-Sokolov reduction for that algebra will have as its associated variety the corresponding Kleinian singularity.

The associated varieties of affine current VOAs have been studied in detail in \cite{Arakawa:2010ni} for the case where $k$ is an admissible level (in the sense of Kac and Wakimoto).\footnote{For $\gf=\slf(n)$, a level $\tilde k$ is Kac-Wakimoto admissible if it is of the form $\tilde k=-n+\frac{p}{q}$ with $(p,q)=1$ and $p\geqs n$.} There it was found that the associated variety of the $V_{\tilde k}(\slf(n))$ VOA for $\tilde k$ admissible is the closure of the principal nilpotent orbit if and only if
\begin{equation}\label{eq:arakawa_criterion}
\tilde k=-n+\frac{p}{q}~, \qquad (p,q)=1~,\qquad p,q\geqs n~.
\end{equation}
Hence for such an admissible level, subregular quantum Drinfel'd-Sokolov reduction will yield a VOA with the Kleinian singularity as its associated variety. 

This construction yields candidates for the VOAs associated to all of the $(A_1, A_{2n-1})$ Argyres-Douglas fixed points. The Virasoro central charge of the VOA obtained by subregular quantum Drinfel'd-Sokolov reduction is fixed in terms of $n$ and $k$ according to
\begin{equation}
c_{n,\tilde k}=-\frac{(\tilde k (n-1)+(n-2) n) \left(\tilde k (n-2) n+(n-3) n^2+1\right)}{\tilde k+n}~.
\end{equation}
Equating this with the expected central charges above, we have
\begin{equation}
\label{eq:qds_k_vals}
\tilde k=-\frac{n^2}{n+1}~,\quad {\rm or}\quad \tilde k=-\frac{n^3-3n^2-2n-1}{n^2-2n}~.
\end{equation}
It is easy to see that the former solution is both admissible and satisfies Arakawa's criterion \eqref{eq:arakawa_criterion}. It is less immediately obvious, but nevertheless true, that the latter solution is also admissible and satisfies \eqref{eq:arakawa_criterion}. We immediately observe that for the case of $n=2$, where there is no subregular orbit, the former solution reproduces the level $\tilde k=-4/3$ that is appropriate for the $(A_1, A_3)$ theory, whereas the second solution becomes singular. We will see below that for other small values of $n$ it is also the former solution that is compatible with what is known about these theories. All in all, we conjecture the identification\footnote{The same conjecture has been made independently in \cite{Creutzig:2017qyf}.}
\be
\goodchi[\AD_{(A_1,A_{2n-1})}]=H_{f_{\rm subreg}}^{\frac{\infty}{2}+0}\left(V_{\frac{-n^2}{n+1}}(\suf(n))\right)\,.
\ee
The subregular Drinfel'd-Sokolov reductions at the particular levels $k=-n^2/(n+1)$ have been the subject of significant investigation previously \cite{Feigin:2004wb,Creutzig:2013pda}. Among the reasons that they have attracted such interest is that they admit several alternative formulations aside from their definition via hamiltonian reduction.

\subsubsection{\tpdf{$(A_1,A_5)$}{(A1,A5)}: Bershadsky Polyakov}
\label{subsubsec:bershadsky_polyakov}

The first example is the subregular Drinfel'd-Sokolov reduction of $\suf(3)$. This is the famous Bershadsky-Polyakov algebra \cite{Polyakov:1989dm,Bershadsky:1990bg}. For generic $\tilde k$ this is a strongly finitely generated $\WW$ algebra with generators $Z$, $X$, $Y$, and $T$ of dimensions $1$, $\frac32$, $\frac32$ and $2$, respectively. As a $\WW$-algebra, this VOA is defined by the singular OPEs
\small
\begin{equation}
\begin{split}
T(z)T(w)&\sim \frac{\frac{k(14-9k)}{4(2+k)}}{(z-w)^4}+\frac{T(w)}{(z-w)^2}+\frac{T'(w)}{z-w}~,\\
T(z)X(w)&\sim \frac{\frac32 X(w)}{(z-w)^2}+\frac{X'(w)}{z-w}~,\\
T(z)Y(w)&\sim \frac{\frac32 Y(w)}{(z-w)^2}+\frac{Y'(w)}{z-w}~,\\
Z(z)Z(w)&\sim \frac{k/2}{(z-w)^2}~,\\
Z(z)X(w)&\sim \frac{X(w)}{(z-w)}~,\\
Z(z)Y(w)&\sim \frac{-Y(w)}{(z-w)}~,\\
X(z)Y(w)&\sim \frac{\frac38(3k^2-2k)}{(z-w)^3}+\frac{\frac34(3k-2)Z(w)}{(z-w)^2}+\frac{3(ZZ)(w)+\frac38(3k-2)Z'(w)-\frac34 (k+2)T(w)}{z-w}~.
\end{split}
\end{equation}
\normalsize
where the $\uf(1)$ level $k$ is related to the parent $\suf(3)$ level according to
\begin{equation}
k = \tfrac23(2\tilde k+3)~.
\end{equation}
The two possible values of $\tilde k$ in \eqref{eq:qds_k_vals} are $\tilde k=-9/4$ and $\tilde k=7/3$, which translates to $\uf(1)$ levels $k=-1$ and $k=46/9$, respectively. Thus the second solution is ruled out by four-dimensional unitarity, so we have $k=-1$. At this level, there is a null state at level three that encodes the Higgs branch relation,
\begin{equation}
\NN_{H}=\left(X_{-\frac32}Y_{-\frac32}+Z_{-1}Z_{-1}Z_{-1}-\tfrac32 L_{-2}Z_{-1}-3Z_{-2}Z_{-1}+\tfrac38 L_{-3}+\tfrac{11}{8}Z_{-4}\right)\Omega~.
\end{equation}
In the $C_2$ algebra, this implies that we have the relation
\begin{equation}
xy+z^3=\tfrac{3}{2}tz~.
\end{equation}
The secondary bracket on $\RR_{\VV}$ is also easily read off from the singular OPEs,
\begin{equation}
\{z,x\}=x~,\qquad \{z,y\}=-y~,\qquad \{x,y\}=3z^2-\tfrac34 t~,
\end{equation}
and $t$ is central with respect to the bracket. Consequently we will recover the Higgs branch relation as a Poisson variety as long as the stress tensor is nilpotent in the $C_2$ algebra. To observe this nilpotence, we must go to level six, where there is a null state such that
\begin{equation}
(L_{-2})^3\Omega\in C_2(\VV)~\Longrightarrow~t^3=0~.
\end{equation}
We have not been able to put the null state in a particularly beautiful form, so we do not display it here.

Curiously, this null state does \emph{not} lead directly to a third-order $\Gamma^0(2)$-modular differential equation -- it suffers from the obstruction outlined in Section \ref{subsec:modular_equations_for_indices}. Instead, the vacuum character is annihilated by a \emph{fourth-order} $\Gamma^0(2)$-modular differential operator which we display here:
\small
\begin{equation}\label{eq:A1A5_diffop}
\begin{split}
\DD_{(A_1,A_5)}=D_q^{(4)}&+
5\Eb_2{\textstyle \left[\genfrac{}{}{0pt}{}{-1}{+1}\right]}(\tau)
D_q^{(3)}-
25\Eb_4{\textstyle \left[\genfrac{}{}{0pt}{}{+1}{+1}\right]}(\tau)D_q^{(2)}-
\left(
\tfrac{1550}{3}\Eb_6{\textstyle \left[\genfrac{}{}{0pt}{}{+1}{+1}\right]}(\tau)+
\tfrac{2675}{6}\Eb_6{\textstyle \left[\genfrac{}{}{0pt}{}{-1}{+1}\right]}(\tau)
\right)D_q^{(1)}\\
&-
\left(
\tfrac{10115}{3}\Eb_8{\textstyle \left[\genfrac{}{}{0pt}{}{+1}{+1}\right]}(\tau)
+\tfrac{21455}{6}\Eb_8{\textstyle \left[\genfrac{}{}{0pt}{}{-1}{+1}\right]}(\tau)
-\tfrac{455}{2}\Eb_4{\textstyle \left[\genfrac{}{}{0pt}{}{+1}{+1}\right]}(\tau)\Eb_4{\textstyle \left[\genfrac{}{}{0pt}{}{-1}{+1}\right]}(\tau)
\right)~.
\end{split}
\end{equation}
\normalsize
The conformal weights of the additional solutions to this twisted modular equation can be determined by solving its indicial equation, while the conformal weights relevant for the high temperature limit are determined by solving the indicial equation for the conjugate differential operator. The result is the following weights,
\begin{equation}
h_i=\left\{-\frac12, -\frac38, -\frac14, 0\right\}~,\qquad \tilde h_i=\left\{-\frac{9}{16}, -\frac{5}{16}, -\frac{1}{16}, \frac{7}{16}\right\}~,
\end{equation}
from which we deduce
\begin{equation}
\tilde h_{\rm min}=-\frac{9}{16}~,\qquad a_{4d}=\frac{11}{12}~,
\end{equation}
which verifies the results from \cite{Xie:2013jc}.

An additional observation is that the vacuum character is actually annihilated by a third-order differential operator, but it is only modular with respect to the smaller conjugacy subgroup $\Gamma(2)$. This differential operator involves modular forms that cannot be produced by the recursion relation described in Appendix \ref{subsec:app_recursion_relations}, and is given by
\small
\begin{equation}\label{eq:A1A5_diffop_extra}
\begin{split}
\widehat{\DD}_{(A_1,A_5)}&=D_q^{(3)}+
\left(
\tfrac32\Eb_2{\textstyle \left[\genfrac{}{}{0pt}{}{-1}{+1}\right]}+
3\Eb_2{\textstyle \left[\genfrac{}{}{0pt}{}{+1}{-1}\right]}\right)
D_q^{(2)}\\
&-\left(
\tfrac{37}{4}\Eb_4{\textstyle \left[\genfrac{}{}{0pt}{}{+1}{+1}\right]}+
\tfrac{21}{2}\Eb_4{\textstyle \left[\genfrac{}{}{0pt}{}{-1}{+1}\right]}-
6\Eb_4{\textstyle \left[\genfrac{}{}{0pt}{}{+1}{-1}\right]}
\right)
D_q^{(1)}\\
&-
\left(
\tfrac{385}{32}\Eb_6{\textstyle \left[\genfrac{}{}{0pt}{}{+1}{+1}\right]}+
\tfrac{141}{4}\Eb_6{\textstyle \left[\genfrac{}{}{0pt}{}{-1}{+1}\right]}+
82\Eb_6{\textstyle \left[\genfrac{}{}{0pt}{}{+1}{-1}\right]}+
\tfrac{67}{2}\Eb_2{\textstyle \left[\genfrac{}{}{0pt}{}{-1}{+1}\right]}\Eb_4{\textstyle \left[\genfrac{}{}{0pt}{}{+1}{-1}\right]}+
23\Eb_2{\textstyle \left[\genfrac{}{}{0pt}{}{+1}{-1}\right]}\Eb_4{\textstyle \left[\genfrac{}{}{0pt}{}{-1}{+1}\right]}
\right)~.
\end{split}
\end{equation}
\normalsize
The solutions of the fourth-order $\Gamma^0(2)$-modular equation (and its conjugate) that do not solve the third-order $\Gamma(2)$-modular equation and its conjugate are those with $h=-\frac38$ and $\tilde h=\frac{7}{16}$. It is interesting to observe that the conjugate solution with positive conformal dimension, which is something we will not see in any other examples, does not survive the more stringent test of solving the third-order equation. That said, at present we have no specific understanding of the role played by the third-order equation.

\input{sections/S5_A1A7}
\input{sections/S5_A1A9}
\input{sections/S5_A1Deven}


%% file: sections/S5_A1A7.tex

\subsubsection{\tpdf{$(A_1,A_7)$}{(A1,A7))}: First generalized Bershadsky-Polyakov}
\label{subsubsec:A1A7_Argyres_Douglas}

For general level, the subregular Drinfel'd-Sokolov reduction of $V_{\tilde{k}}(\slf(4))$ yields a $\WW$-algebra that is strongly generated by currents $\{Z,X,Y,T,W\}$ where $Z$ is a $\uf(1)$ current, $X$ and $Y$ are have weight two and charge $\pm1$ under $Z$, $T$ is the stress tensor and $W$ is an additional generator of conformal weight three. However when $\tilde{k}=-16/5$, $W$ becomes null, so the only generators are the stress tensor and the currents associated to Higgs chiral ring generators. It is at this value that the subregular reduction is equivalent to the generalized Bershadsky-Polyakov algebra described in \cite{Creutzig:2013pda}.

It is intriguing that there is actually a one-parameter family of $\WW$-algebras with this smaller list of generators. The general form of the singular OPEs for this family of $\WW$ algebras are as follows,
\small
\begin{eqnarray}
T(z)T(w)&\sim& \frac{-\frac{3k^2-8k+2}{2+k}}{(z-w)^4}+\frac{T(w)}{(z-w)^2}+\frac{T'(w)}{z-w}~,\nonumber\\
T(z)X(w)&\sim& \frac{2X(w)}{(z-w)^2}+\frac{X'(w)}{z-w}~,\nonumber\\
T(z)Y(w)&\sim& \frac{2 Y(w)}{(z-w)^2}+\frac{Y'(w)}{z-w}~,\nonumber\\
T(z)Z(w)&\sim& \frac{Z(w)}{(z-w)^2}+\frac{Z'(w)}{z-w}~,\nonumber\\
Z(z)Z(w)&\sim& \frac{\frac{k}{2}}{(z-w)^2}~,\\
Z(z)X(w)&\sim& \frac{X(w)}{(z-w)}~,\nonumber\\
Z(z)Y(w)&\sim& \frac{-Y(w)}{(z-w)}~,\nonumber\\
X(z)Y(w)&\sim& \frac{\frac32 k^2(k-2)}{(z-w)^4}+\frac{3k(k-2)Z}{(z-w)^3}
+\frac{-k(k+2)T(w)+4(k-1)(ZZ)(w)+\frac32 k(k-2)Z'(w)}{(z-w)^2}\nonumber\\
&&\qquad+\frac{4(ZZZ)(w)-2(k+2)(TZ)(w)+4(k-1)(Z'Z)(w)-\frac{k}{2}(k-2)T'(w)+\frac12(k^2+4)Z''(w)}{z-w}~.\nonumber
\end{eqnarray}
\normalsize
The $C_2$ algebra for this VOA is generated by $x$, $y$, $z$, and $t$ with Poisson brackets given by
\begin{equation}
\{z,x\}=x~,\qquad \{z,y\}=-y~,\qquad \{x,y\}=4z^3-2(k+2)tz~,
\end{equation}
and as usual $t$ Poisson commutes with everything.

For $k=-1/3$ and $k=-4/5$, there is a null state at level four that could be a shadow of the Higgs branch relation. Because it is complicated, we only write the resulting relation in the $C_2$ algebra,
\begin{equation}
\begin{split}
xy+z^4-\tfrac{10}{3}tz^2+\frac{10}{9}t^2&=0~,\qquad k=-\tfrac{1}{3}~,\\
xy+z^4-\tfrac{30}{11}tz^2-\frac{3}{11}t^2&=0~,\qquad k=-\tfrac{4}{5}~.
\end{split}
\end{equation}
The relevant values of the Virasoro central charge for these solutions are given $c_{2d}=-6$ for $k=-1/3$ and $c_{2d}=-86/5$ for $k=-4/5$. Thus it is the latter solution that we identify with the $(A_1,A_7)$ Argyres-Douglas theory. Indeed, it is this solution which arises from the subregular Drinfel'd-Sokolov reduction at $\tilde k=-16/5$.

The Schur index for this theory has been given in \cite{Cordova:2015nma}. We have found a sixth-order LMDE that is solved by this index,
\small
\begin{equation}
\label{eq:A1A7_diffop}
\begin{split}
\DD_{(A_1,A_7)}=D_q^{(6)}
&-77\,
\Eb_4(\tau)\,
D_q^{(4)}
-2156\,
\Eb_6(\tau)\,
D_q^{(3)}
-\tfrac{384461}{25}\,
\Eb_4(\tau)^2\,
D_q^{(2)}\\
&-\tfrac{4296908}{25}\,
\Eb_4(\tau)\Eb_6(\tau)\,
D_q^{(1)}
-\left(
\tfrac{1145859}{5}\,\Eb_4(\tau)^3+
\tfrac{1970584}{5}\,\Eb_6(\tau)^2
\right)~.
\end{split}
\end{equation}
\normalsize
This suggests that there is a relation $t^6=0$ in the $C_2$ algebra, so after passing to the reduced algebra we should recover the Higgs branch chiral ring, as is guaranteed by the abstract arguments above.

The additional solutions of the indicial equation for this LMDO give module weights
\begin{equation}
h_i=\left\{-\frac45,\,-\frac35,\,-\frac25,\,-\frac15,\,0,\,\frac15\right\}~,
\end{equation}
from which we deduce
\begin{equation}
h_{\min} = -\frac45~,\qquad a_{4d}=\frac{167}{120}~,
\end{equation}
which does indeed matches with the correct physical value.

%% file: sections/S5_A1A9.tex

\subsubsection{\tpdf{$(A_1,A_{2n-1})$}{(A1,A(2n-1))} for \tpdf{$n\geqs5$}{n>=5}}
\label{subsubsec:A1A9_Argyres_Douglas}

For the $(A_1,A_9)$ theory, and presumably for higher $n$ as well, it is no longer the case that the $\WW$-algebra generated by $X$, $Y$, $Z$, and $T$ sits in a continuous family of algebras. On the contrary, for this case the algebra with these strong generators can only be closed at certain discrete values of $c_{2d}$. Here we briefly summarize the results for the $(A_1,A_9)$ case. 

Here there is only one solution to the $\WW$-algebra bootstrap equations that is compatible with four-dimensional unitarity:
\small
\allowdisplaybreaks
\begin{align}
T(z)T(w)&\sim \frac{-\frac{23}{2}}{(z-w)^4}+\frac{T(w)}{(z-w)^2}+\frac{T'(w)}{z-w}~,\nonumber\\
T(z)X(w)&\sim \frac{\frac52X(w)}{(z-w)^2}+\frac{X'(w)}{z-w}~,\nonumber\\
T(z)Y(w)&\sim \frac{\frac52Y(w)}{(z-w)^2}+\frac{Y'(w)}{z-w}~,\nonumber\\
T(z)Z(w)&\sim \frac{Z(w)}{(z-w)^2}+\frac{Z'(w)}{z-w}~,\nonumber\\
Z(z)Z(w)&\sim \frac{-\frac13}{(z-w)^2}~,\nonumber\\
Z(z)X(w)&\sim \frac{X(w)}{(z-w)}~,\\
Z(z)Y(w)&\sim \frac{-Y(w)}{(z-w)}~,\nonumber\\
X(z)Y(w)&\sim \frac{\frac{35}{9}}{(z-w)^5}+\frac{-\frac{35}{3}Z}{(z-w)^4}
+\frac{-\frac{35}{27}T(w)+\frac{140}{9}(ZZ)(w)-\frac{35}{6}Z'(w)}{(z-w)^3}\nonumber\\
&\qquad+\frac{-\frac{35}{3}(ZZZ)(w)+\frac{35}{9}(TZ)(w)+\frac{140}{9}(Z'Z)(w)-\frac{35}{54}T'(w)-\frac{35}{9}Z''(w)}{(z-w)^2}\nonumber\\
&\qquad+\frac{5(ZZZZ)(w)-5(TZZ)(w)+\frac{5}{18}(TT)(w)+\frac{35}{18}(TZ)'(w)-\frac{35}{2}(Z'ZZ)(w)}{z-w}\nonumber\\
&\qquad+\frac{\frac{85}{24}(Z'Z')(w)+\frac{125}{12}(Z''Z)(w)-\frac{5}{18}T''(w)-\frac{35}{24}Z'''(w)}{z-w}~.\nonumber
\end{align}
\normalsize
We see that for this solution the Virasoro central charge is given by $c_{2d}=-23$, matching four-dimensional expectations. In this case we have bootstrapped the singular OPEs and found null states whose images in the $C_2$ algebra are given by
\begin{equation}
ty=0~,\qquad xy+z^5+\tfrac56 t^2z -\tfrac{10}{3}tz^3=0~.
\end{equation}
The second of these obviously corresponds to the Higgs chiral ring relation for this theory, while the first is somewhat mysterious from a four-dimensional point of view. It is the first null that is responsible for the closure of the operator algebra.

We have also derived modular differential equations for the Schur index of this theory. As in the $(A_1,A_5)$ case, we find a lower-order modular equation for the smaller modular group $\Gamma(2)$. In particular, we find a sixth-order $\Gamma(2)$-modular differential equation. We have also found a two-dimensional family of ninth-order $\Gamma^0(2)$-modular differential equations. The minimal solution weight for the conjugate equations is given by 
\begin{equation}
\tilde h_{\min}=-\frac{25}{24}~~\Longrightarrow~~a_{4d}=\frac{15}{8}~.
\end{equation}

%% file: sections/S5_A1Deven.tex

\subsection{\tpdf{$(A_1, D_{2n+2})$}{(A1,D(2n+2))} theories: Slodowy slices to nilpotent orbits}
\label{subsec:slodowy_slices}

To complete our discussion of Argyres-Douglas theories, we consider the final infinite family of type $(A_1, \Gamma)$, namely the theories of type $(A_1, D_{2n+2})$ for $n\geqslant1$ (\ie, the $(I_{2,2n},F)$ theories). These are rank-$n$ theories that are predicted to have four-dimensional central charges given by
\begin{equation}
    a_{4d} = \frac{n}{2}+\frac{1}{12}~,\qquad c_{4d} = \frac{n}{2}+\frac{1}{6}~.
\end{equation}
We are not aware of a reference where the Higgs branches of these theories are written down explicitly, but they can be readily determined by passing to three dimensions. Three-dimensional mirrors to the circle compactification of these theories were proposed in \cite{Xie:2012hs}; they are abelian quiver gauge theories associated to the following simple quivers:
\begin{equation}\label{diag:coulomb_quiver}
    \begin{tikzpicture}[font=\footnotesize]
        \begin{scope}[auto, every node/.style={draw, minimum size=1cm}, node distance=0.6cm];
            \node[rectangle] (flav1) at (0,0) {$n+2$};
            \node[circle, right=of flav1] (gauge1) {$1$};
            \node[circle, right=of gauge1] (gauge2) {$1$};
            \node[rectangle, right=of gauge2] (flav2) {$1$};
        \end{scope}
        \draw (flav1)  -- (gauge1);
        \draw (gauge1) -- (gauge2);
        \draw (gauge2) -- (flav2);
    \end{tikzpicture}
\end{equation}
Consequently the Higgs branches of our Argyres-Douglas theories are the Coulomb branches of the IR fixed points of these quiver gauge theories. But these three-dimensional gauge theories further admit three-dimensional abelian Lagrangian mirrors that are easily determined using the technology of \cite{Cremonesi:2014uva}\footnote{We are grateful to Sergio Benvenuti for pointing this out to us.}. Thus the Higgs branches of the $(A_1,D_{2n+2})$ Argyres-Douglas theories are identified with the Higgs branches of the three-dimensional $U(1)^{n}$ gauge theories associated to the following quivers:
\begin{equation}\label{diag:higgs_quiver}
    \begin{tikzpicture}[font=\footnotesize]
        \begin{scope}[auto, every node/.style={draw, minimum size=.8cm}, node distance=0.5cm];
            \node[rectangle] (flav1) at (0,0) {$2$};
            \node[circle, right=of flav1] (gauge1) {$1$};
            \node[circle, right=of gauge1] (gauge2) {$1$};
            \node[draw=none, right=of gauge2] (dots) {$\cdots$};
            \node[circle, right=of dots] (gaugenm2) {$1$};
            \node[circle, right=of gaugenm2] (gaugenm1) {$1$};
            \node[rectangle, right=of gaugenm1] (flavnm1) {$1$};
        \end{scope}
        \draw (flav1)    -- (gauge1);
        \draw (gauge1)   -- (gauge2);
        \draw (gauge2)   -- (dots);
        \draw (dots)     -- (gaugenm2);
        \draw (gaugenm2) -- (gaugenm1);
        \draw (gaugenm1) -- (flavnm1);
        \draw[decoration={brace,raise=18pt,amplitude=10pt},decorate]
        (gauge1) -- node[above=28pt] {$n$ times} (gaugenm1);
    \end{tikzpicture}
\end{equation}
As we will see, this description of the Argyres-Douglas Higgs branches as quiver varieties is precisely reproduced by an investigation of the associated variety of the associated vertex algebras.

In \cite{Creutzig:2017qyf} the vertex algebras associated to this family were proposed to be the to-called $\WW_{n+1}$ algebras, which are the sub-subregular%
\footnote{The sub-subregular embedding of $\slf(2)\hookrightarrow\slf(n)$ is well-defined for $n\geqslant3$ and for $n\geqslant4$ can be understood as the regular embedding of $\slf(2)\hookrightarrow\slf(n-2)\hookrightarrow\slf(n)$, where the latter embedding is the obvious one where the fundamental decomposes as ${\bf n}\mapsto{\bf (n-2)}\oplus{\Cb}\oplus{\Cb}$. For $n=3$ the sub-subregular embedding is the trivial embedding.} %
quantum Drinfeld-Sokolov reductions of the affine current algebras $V_{k}(\suf(n+2))$ with $k=-\frac{(n^2+2n)}{n+1}$. These are strongly finitely generated vertex operator algebras that, for generic level, are generated by operators of dimension $h=2,3,\ldots,n$ along with affine $\glf(2)$ currents and operators of dimension $h=(n+1)/2$ transforming in the fundamental and anti-fundamental representations of $\glf(2)$.

The associated variety of the $\WW_{n+1}$ algebra can be determined using Theorem \ref{thm:arakawa} and results on intersections of Slodowy slices with nilpotent orbits from \cite{Maffei} as follows. The associated variety for the affine current algebra $V_{k}(\suf(n+2))$ for the relevant level is the subregular nilpotent orbit of $\slf(n+2)$, which is specified by the partition $[n+1,1]$ \cite{Arakawa:2010ni}. By the aforementioned theorem, the associated variety for the $\WW_{n+1}$ algebra can therefore be described as the intersection
\begin{equation}
    X_{\WW_p} = \overline{\Ob_{[n+1,1]}}\cap \SS_{[n,1,1]}~.
\end{equation}
A theorem of Maffei (proving a conjecture of Nakajima \cite{Nakajima:1994nid}) shows that this intersection is isomorphic to a quiver variety. We state a (slightly weaker version of the) full theorem of Maffei here for convenience (see also \cite{Henderson} for a useful summary):
\begin{thm}[Maffei \cite{Maffei}]
    Let $v=(v_1,\ldots,v_{\ell-1})$ and $d=(d_1,\ldots,d_{\ell-1})$ be two $(\ell-1)$-tuples of non-negative integers, and further define the tuple $r(d,v)=(r_1,\ldots,r_{\ell})$ according to
    \begin{equation}
        \begin{split}
            r_1 &= \left(\sum_{i=1}^{\ell-1}d_i\right) -v_1~,\\
            r_j &= \left(\sum_{i=j}^{\ell-1}d_i\right) -v_j+v_{j-1}~,\\
            r_\ell &= v_{\ell-1}~.
        \end{split}
    \end{equation}
    so $\sum_{i=1}^{\ell} r_i = N = \sum_{i=1}^{\ell-1}i\times d_i$. To $r(d,v)$ we assign a partition of $N$ denoted $\lambda_r$ as follows: let $\rho_1\geqslant \rho_2 \geqslant \ldots \geqslant \rho_{\ell}$ be a permutation of $r(d,v)$. Then $\lambda_r=[1^{\rho_1-\rho_2}2^{\rho_2-\rho_3}\ldots (\ell-1)^{\rho_{\ell-1}-\rho_{\ell}}(\ell)^{\rho_\ell}]$. The framed $A$-type quiver variety with gauge nodes having ranks given by $v$ and flavor nodes having ranks given by $d$ is isomorphic as an algebraic variety to the intersection of the (closure of the) minimal nilpotent orbit of $\slf(N)$ associated to the partition $\lambda_r$ with the Slodowy slice to a nilpotent element $x$ specified by the partition $[1^{d_1}\ldots (\ell-1)^{d_{\ell-1}}]$.
\end{thm}

Applying this to our family of intersections, we take $x$ to be specified by $[n,1,1]$ and $\lambda_r=[n+1,1]$, from which we determine $r=(2,1,\ldots,1)$, and $d=(2,0,\ldots,0,1)$ with $\ell=n+1$, and finally $v=(1,\ldots,1)$. This describes precisely the quiver variety associated to the quiver given in \eqref{diag:higgs_quiver} when $n\geqslant2$. The special case $n=1$ is degenerate, as here we are discussing the closure of the minimal nilpotent orbit of $\slf(3)$ with no Slodowy intersection. In this case we have $\lambda_r=[2,1]$, $r=(2,1)$, $d=(3)$, and $v=(1)$. The quiver is therefore the well-known quiver describing the minimal nilpotent orbit of $\slf(3)$,
\begin{equation}\label{diag:degenerate_quiver}
    \begin{tikzpicture}[font=\footnotesize]
        \begin{scope}[auto, every node/.style={draw, minimum size=1cm}, node distance=0.6cm];
            \node[rectangle] (flav1) at (0,0) {$3$};
            \node[circle, right=of flav1] (gauge1) {$1$};
        \end{scope}
        \draw (flav1) -- (gauge1);
    \end{tikzpicture}
\end{equation}
which indeed matches the $n=1$ case of the quiver in \eqref{diag:higgs_quiver}. We therefore find perfect agreement between the Higgs branches of these Argyres-Douglas theories and the associated varieties of the proposed associated vertex algebras.

%% file: sections/S6.tex

\section{Further results}
\label{subsec:further_data}

In general, it is easier to test whether the Schur index of some $\NN=2$ SCFT is the solution to a monic modular differential equation of fixed order than it is to solve for the associated variety of the corresponding vertex operator algebra. This has enabled us to collect a substantial amount of data in support of the conjecture that Schur indices should generally satisfy such differential equations. Below we summarize our results in this area. We will see that the order of the monic differential equation behaves somewhat erratically within several nice families of SCFTs, but there may be some order underlying the chaos.

\subsection{\texorpdfstring{$A_1$}{A1} class \texorpdfstring{$\SS$}{S}}
\label{subsec:class_S_examples}

The unflavored Schur indices of class $\SS$ theories of type $A_1$ admit a general, simple expression from which it is easy to generate the $q$-series expansion to very high order \cite{Gaiotto:2012xa},
\begin{equation}
\label{eq:a1_index}
\II^{\af_1}_{g,s}=(q;q)_{\infty}^{2g-2-2s}\sum_{k=0}^{\infty}\left(
\frac{(k + 1)^s q^{\frac{k}{2}(2 g - 2 + s)}}
{(1 - q^{k + 1})^{2 g - 2 + s}}\right)~.
\end{equation}
The $a$ and $c$ Weyl anomalies are given by
\begin{equation}
a_{g,s}=\frac{53}{24}(g-1)+\frac{19}{24}s~,\qquad c_{g,s}=\frac{13}{6}(g-1)+\frac{5}{6}s~.
\end{equation}
The associated VOAs for these theories admit a variety of cohomological presentations \cite{Beem:2013sza,Beem:2014rza}, and there is a proposal for the list of generators in a $\WW$-algebraic presentation at genus zero \cite{Beem:2014rza,Lemos:2014lua}. Nevertheless, for all cases other than $g=0$, $s=3,4$ and $g=1, s=1$, the explicit $\WW$-algebra has yet to be constructed.

We note in advance that for $2g-2+s$ even, the Schur index has only integer powers of $q$. It is expected (though not proven) that the vertex operator algebras for these theories are genuinely $\Zb$-graded. Furthermore, for genus zero theories, it is expected that the vertex operator algebra is purely bosonic.

\subsubsection*{Genus zero, three punctures}

The simplest case is the three-punctured sphere, for which the vertex operator algebra is four copies of the symplectic boson VOA,
\begin{equation}
q^{abc}(z)q^{a'b'c'}(w)=\frac{\varepsilon^{aa'}\varepsilon^{bb'}\varepsilon^{cc'}}{z-w}+O((z-w)^0))
\end{equation}
This is a bosonic, $\frac12\Zb$-graded VOA. The $C_2$ algebra and associate variety for this VOA are simple: the $C_2$ algebra has no nilpotent elements and is freely generated by equivalence classes of the states $q_{-\frac12}^{abc}\Omega$, with the Poisson bracket defined on the generators according to
\begin{equation}
\{q^{abc},q^{a'b'c'}\}=\varepsilon^{aa'}\varepsilon^{bb'}\varepsilon^{cc'}~.
\end{equation}
The associated variety is therefore just the affine space $\Cb^8$ with the canonical holomorphic Poisson bracket. Thus in this instance, the identification of the Higgs branch with the associated variety is trivial.

The stress tensor is a composite and lies in $C_2(\VV)$ by construction,
\begin{equation}
L_{-2}\Omega=\frac12\varepsilon_{aa'}\varepsilon_{bb'}\varepsilon_{cc'}(q^{abc})_{-\frac32}(q^{a'b'c'})_{-\frac12}\Omega\quad \in \quad C_2(\VV)~.
\end{equation}
Taking the trace of the (square-bracket) zero-modes of both sides and applying the recursion relations given in Appendix \ref{subsec:app_recursion_relations}, we find
\begin{equation}
\begin{split}
\Tr_{\VV}\Big(\zeromode{L_{[-2]}\Omega}q^{L_0-c/24}\Big)&=\phantom{-}\tfrac12\varepsilon_{aa'}\varepsilon_{bb'}\varepsilon_{cc'}\Tr_\VV\Big(o\left((q^{abc})_{[-\frac32]}(q^{a'b'c'})_{[-\frac12]}\Omega\right)q^{L_0-c/24}\Big)~,\\
\PP_2\left(\Tr_{\VV}q^{L_0-c/24}\right)&=-\tfrac12\varepsilon_{aa'}\varepsilon_{bb'}\varepsilon_{cc'}\varepsilon^{aa'}\varepsilon^{bb'}\varepsilon^{cc'}\Eb_2{\textstyle \left[\genfrac{}{}{0pt}{}{-1}{+1}\right]}(\tau) \Tr_{\VV}\left(q^{L_0-c/24}\right)
\end{split}
\end{equation}
which is a first-order modular differential equation for the vacuum character,
\begin{equation}
\left(D_q^{(1)}-4\,\Eb_2{\textstyle \left[\genfrac{}{}{0pt}{}{-1}{+1}\right]}(\tau)\right)\chi_{0,3}^{\af_1}(q)=0~.
\end{equation}
The conjugate LMDO is
\begin{equation}
\wt{\DD}^{\af_1}_{\CC_{0,3}}=\left(D_q^{(1)}-4\,\Eb_2{\textstyle \left[\genfrac{}{}{0pt}{}{+1}{-1}\right]}(\tau)\right)~,
\end{equation}
from which we find that $\tilde h_{\min}=-\frac12$. Inserting this and the value $c_{2d}=-4$ into \eqref{eq:a_anomaly_equation}, we recover the expected Weyl anomaly
\begin{equation}
a_{4d} = \frac16~.
\end{equation}

\subsubsection*{Genus zero, four punctures}

The case of the four punctured sphere was discussed above in Section \ref{sec:deligne} -- it is the affine current algebra $\widehat{\sof(8)}_{-2}$. The vacuum character obeys a second order modular differential equations,
\begin{equation}
\left(D_q^{(2)}- 175\Eb_{4}(\tau)\right)\II_{0,4}^{\af_1}(q)=0~.
\end{equation}
The non-vacuum solution of this equation gives $h_{\min}=-1$, from which we rederive the Weyl anomaly for this theory
\begin{equation}
a_{4d}=\frac{23}{24}~.
\end{equation}

\subsubsection*{Genus one, one puncture}

The genus one theory with one puncture is $\NN=4$ super Yang-Mills theory with $\suf(2)$ gauge algebra and an extra free hypermultiplet. The pure $\suf(2)$ $\NN=4$ theory will be discussed in more detail below in Section \ref{subsec:examples_n4}. We note here that the unflavored Schur index of the class $\SS$ theory is annihilated by the twisted LMDE
\begin{equation}
\DD^{\af_1}_{\CC_{1,1}}=D_q^{(2)}-4\,\Eb_2{\textstyle \left[\genfrac{}{}{0pt}{}{-1}{+1}\right]}(\tau) D_q^{(1)} -11\,\Eb_{4}(\tau)+16\,\Eb_4{\textstyle \left[\genfrac{}{}{0pt}{}{-1}{+1}\right]}(\tau)~.
\end{equation}
From the dual LMDO, we find $h_{\min}=-\frac12$, which along with $c_{2d}=-10$ gives the correct value for the Weyl anomaly
\begin{equation}
a_{4d}=\frac{19}{24}~.
\end{equation}

\renewcommand{\arraystretch}{1.5}
\begin{table}
\centering
\begin{tabular}{|c|c|c|c|c|c|}
\hline \hline
~$\CC_{g,s}$~  & ~${\rm ord}(\DD)$~ & ~Modular Group~  & ~Indicial roots $h_i$~ & ~Conjugate roots $\tilde h_i$~ & ~$\dim V_{\DD}$~  \\ 
\hline 
$\CC_{0,3}$ & $1$ & $\Gamma^0(2)$ & $0$ & $(-\frac12)$ & $0$\\
\hline 
$\CC_{0,4}$ & $2$ & $\Gamma$ & $-1, 0$ & --- & $0$\\
\hline
$\CC_{0,5}$ & $4$ & $\Gamma^0(2)$  & $(-1)_3, 0$ & $(-\frac32), (-\frac12)_3$& $0$\\
\hline
$\CC_{0,6}$ & $6$ & $\Gamma$ & $-2, (-1)_4, 0$ & --- & $0$\\
\hline
$\CC_{0,7}$ & $13$ & $\Gamma^0(2)$  & $(-2)_5, (-1)_3, 0, (\star)_4$ & $-\frac52, (-\frac32)_5, -(\frac12)_3, (\star)_4$& $2$\\
\hline
$\CC_{0,8}$ & $16$ & $\Gamma$ & $-3, (-2)_6, (-1)_4, 0, (\star)_4$ & --- & $0$\\
\hline
$\CC_{1,1}$ & $2$ & $\Gamma^0(2)$  & $-\frac12, 0$ & $(-\frac12)_2$ & $0$\\
\hline
$\CC_{1,2}$ & $4$ & $\Gamma$ & $(-1)_2, -\frac13, 0$ & --- & $0$\\
\hline
$\CC_{1,3}$ & $6$ & $\Gamma^0(2)$  & $-\frac32, (-1)_3, -\frac12, 0$ & $(-\frac32)_2, (-\frac12)_4$ & $0$\\
\hline
$\CC_{1,4}$ & $9$ & $\Gamma$ & $(-2)_2, (-1)_5, (0)_2$ & --- & $0$\\
\hline
$\CC_{2,0}$ & $6$ & $\Gamma$ & $(-1)_4, (0)_2$ & --- & $0$\\
\hline
$\CC_{2,1}$ & $11$ & $\Gamma^0(2)$  & $(-\frac32)_4, (-1)_3, 0, (\star)_3$ & $(-\frac32)_4, (-\frac12)_4, (\star)_3$ & $1$\\
\hline
\end{tabular}
\caption{\label{tab:A1ClassS} Summary of modular differential operators that annihilate the vacuum characters of class $\SS$ vertex operator algebras of type $\af_1$. Provided are the order of the minimal modular differential operator that annihilates the vacuum character, the modular subgroup under which the corresponding differential equation is invariant, the list of solutions of the indicial equation, and the dimension of $V_{\DD}$, the vector space of modular differential operators at the given order that mutually annihilate the vacuum character.}
\end{table}

\subsubsection*{Comments on the general case}

The above results and those for the more many other cases in which the associated VOA has not been constructed are summarized in Table \ref{tab:A1ClassS}. The explicit modular differential operators that annihilate the various indices can be found in Appendix \ref{app:class_S_modular}. The module weights reported are the $h_i$ such that $-c/24+h_i$ is a solution to the indicial equation for the LMDE. A couple of observations may be worth recording. 

\begin{enumerate}
\item[$\bullet$] The first is that in \emph{all} cases we have studied, the additional solutions of the modular equations are mostly logarithmic. Our expectation is that these logarithms are relics of regularizing ill-defined characters in the limit where flavor fugacities are set to zero. Consequently, in light of our previous discussions, this suggests that for these class $\SS$ theories there are no superconformal surface defects with finite-dimensional spaces of operators at fixed level in the Schur index.

\item[$\bullet$] For genus zero with seven punctures and genus two with one puncture, we have found that at the minimal order for which there exists an LMDO annihilating the vacuum character, there is actually a continuous family of such LMDOs (we have denoted the vector space of such LMDOs as $V_\DD$ in Table \ref{tab:A1ClassS}). It would be of some interest to understand whether there is a multiplicity of null vectors leading to this positive-dimensional space of LMDOs, and if so whether there is any four-dimensional interpretation of the phenomenon.

\item[$\bullet$] Relatedly, for a number of high-rank examples (genus zero with seven and eight punctures and genus two with one puncture), we have found complex, irrational roots to the indicial and, when relevant, conjugate indicial equations. In the cases mentioned in the previous bullet, these extra solutions depend on the choice of LMDO in $V_\DD$, so the actual vector-valued modular form that includes the vacuum character is of lower dimension than the LMDE would suggest and does not include the solutions with complex weights. For genus zero with eight punctures, we also take the complex irrational roots as an indication that the space of solutions to the differential equations in question are too large. In this case, algebraically independent LMDOs of higher order that annihilate the same character could exclude the additional solutions. It would be interesting to find these additional operators and investigate their four-dimensional meaning.

\item[$\bullet$] Taking for granted the behavior postulated in the previous bullet, we find that the dimensions of the vector-valued modular forms for the genus zero theories with $s=3$, $4$, $5$, $6$, $7$, and $8$ punctures are $1$, $2$, $4$, $6$, $9$, and $12$. From this, we optimistically conjecture that the dimension for a general number of punctures is given by
\begin{equation}
\dim\left({\rm v.v.m.f.}\right)(s)=\lfloor (s-1)^2/4\rfloor~.
\end{equation}
We have excluded the existence of an LMDE of order less than or equal to $16$ for genus zero with nine punctures, so for our conjecture to hold the phenomenon of additional LMDOs must apply in this case.

\item[$\bullet$] We finally observe that with the exception of the irrational/additional indicial roots described in the previous two bullets, all values of $h_i$ and $\tilde h_i$ are negative. We suspect this is a general phenomenon. It would be interesting to explain this phenomenon in terms of the physics of surface operators, perhaps as a consequence of unitarity.

\end{enumerate}

\subsection{\texorpdfstring{$\NN=4$}{N=4} super Yang-Mills}
\label{subsec:examples_n4}

The $\NN=4$ gauge theory with $\suf(N)$ gauge group has an associated $\frac12\Zb$-graded VOA. In all cases this VOA has the small $\NN=4$ superconformal algebra as a subalgebra. For general $N$, a list of generators has been proposed for a $\WW$-algebraic description of this VOA in \cite{Beem:2013sza}, but the structure constants have not been determined. Only the $\suf(2)$ case has been constructed as a $\WW$-algebra in the literature.

\subsubsection{\texorpdfstring{$\suf(2)$}{su(2)} gauge algebra}
\label{subsubsec:n4_su2_summary}

For the special case of the $\suf(2)$ gauge theory, a fairly complete analysis is possible. Here the Higgs branch is the $\slf(2)$ nilpotent orbit $\Cb^2/\Zb_2$. The Higgs branch chiral ring is generated by the $\suf(2)$ moment map $\mu^A$, $A=1,2,3$, which is subject to the Joseph relation
\begin{equation}
\label{eq:n4_Higgs_relations}
\restr{(\mu\otimes\mu)}{\bf 1}=0
\end{equation}
The Hall-Littlewood chiral ring and anti-chiral ring have additional (fermionic) generators of dimension $5/2$, $\omega^\alpha$, and $\tilde\omega^\alpha$ arising from the extra supercurrents, where $\alpha$ is an $\suf(2)_F$ doublet index. There are additional chiral ring relations involving these generators,
\begin{equation}
\label{eq:n4_HL_relations}
\restr{(\mu\otimes \omega)}{\bf 2}=\restr{(\mu\otimes\tilde{\omega})}{\bf 2}=0~,\quad (\omega\otimes \omega)=(\tilde\omega\otimes\tilde \omega)=(\omega\otimes\tilde \omega) =0~.
\end{equation}

The associated VOA is precisely the small $\NN=4$ algebra with $c=-9$, with no extra generators \cite{Beem:2013sza}. This VOA is generated by $\suf(2)$ affine currents $J^A(z)$ and supercurrents $G^{\alpha}(z)$ and $\tilde G^{\alpha}(z)$, their singular OPEs take the form
\begin{equation}
\begin{split}
T(z)T(w)&\sim\frac{\frac{c}{2}}{(z-w)^4}+\frac{2T(w)}{(z-w)^2}+\frac{T'(w)}{z-w}~,\\
J^A(z)J^B(w)&\sim \frac{\frac{c}{12}\kappa^{AB}}{(z-w)^2}+\frac{f^{AB}_{\phantom{AB}C}J^C(w)}{z-w}~,\\
J^A(z)G^\alpha(z)&\sim \frac{(\sigma^A)_{\beta}^{\phantom{\beta}\alpha} G^\beta(w)}{z-w}~,\\
J^A(z)\tilde{G}^\alpha(z)&\sim \frac{(\sigma^A)_{\beta}^{\phantom{\beta}\alpha} \tilde G^\beta(w)}{z-w}~,\\
G^\alpha(z)\tilde{G}^\beta(w)&\sim\frac{\frac{c}{3}\varepsilon^{\alpha\beta}}{(z-w)^3}+\frac{-4(\sigma_A)^{\alpha\beta}J^A(w)}{(z-w)^2}+\frac{\varepsilon^{\alpha\beta}T(w)-2(\sigma_A)^{\alpha\beta}J^A(w)}{z-w}~.
\end{split}
\end{equation}
For this value of the central charge, the stress tensor is not an independent strong generator. Rather it is a composite obtained from the affine currents by the Sugawara construction,
\begin{equation}
\label{eq:n4_su2_sug}
T(z)=2\kappa_{AB}(J^A J^B)(z)~.
\end{equation}

The current and supercurrent generators of the VOA correspond to Higgs branch chiral ring and Hall-Littlewood (anti-)chiral ring generators in the four-dimensional theory. The relation \eqref{eq:n4_su2_sug} is the VOA avatar of the Higgs branch relation \eqref{eq:n4_Higgs_relations}. Additionally, there are null states at dimensions $h=5/2$ and $h=3$,
\begin{equation}
\begin{split}
(\NN_{JG})^\alpha&=\left((\sigma_A)_{\beta}^{\phantom{\beta}\alpha}J^A_{-1}G^\beta_{-3/2}-\tfrac12G^\alpha_{-5/2}\right)\Omega~.\\
(\NN_{J\tilde{G}})^\alpha&=\left((\sigma_A)_{\beta}^{\phantom{\beta}\alpha}J^A_{-1}\tilde{G}^\beta_{-3/2}-\tfrac12\tilde G^\alpha_{-5/2}\right)\Omega~.\\
(\NN_{G\tilde G})^A&=\left((\sigma^A)_{\alpha\beta}G^\alpha_{-3/2}\tilde G^\beta_{-3/2}+2f^A_{\phantom{A}BC}J^B_{-2}J^C_{-1}+2J^A_{-3}-2L_{-2}J^A_{-1}\right)\Omega~,\\
\NN_{G\tilde G}&=\left(\varepsilon_{\alpha\beta}G^\alpha_{-3/2}\tilde G^\beta_{-3/2}+L_{-3}\right)\Omega~,\\
\NN_{GG}&=\varepsilon_{\alpha\beta}\left(G^\alpha_{-3/2}G^\beta_{-3/2}\right)\Omega~,\\
\NN_{\tilde G\tilde G}&=\varepsilon_{\alpha\beta}\left(\tilde G^\alpha_{-3/2}\tilde G^\beta_{-3/2}\right)\Omega~,
\end{split}
\end{equation}
that are related to the additional chiral ring relations of \eqref{eq:n4_HL_relations}.

The $C_2$ algebra is generated by the equivalence classes of the strong generators, 
\begin{equation}
j^A\sim J^A_{-1}\Omega~, \quad
\omega^\alpha\sim G^\alpha_{-3/2}\Omega~,\quad
\tilde\omega^\alpha\sim\tilde{G}^\alpha_{-3/2}\Omega~.
\end{equation}
The fermionic generators $\omega$ and $\tilde\omega$ are automatically nilpotent, so the only issue in determining the associated variety is what happens to the moment maps. Their fate is sealed by a null vector at level four that relates the the square of the stress tensor to an element of $C_2(\VV)$,
\begin{equation}\label{eq:n4_su2_null}
\NN_{T}=\Big(
\left(L_{-2}\right)^2
+\varepsilon_{\alpha\beta}\left(\tilde G^\alpha_{-5/2}G^\beta_{-3/2}-G^\alpha_{-5/2}\tilde G^\beta_{-3/2}\right)
-\kappa_{AB}\left(J^A_{-2}J^B_{-2}\right)-\tfrac12 L_{-4}\Big)\Omega~.
\end{equation}
This null vector gives rise to the relation
\begin{equation}
\left(\sum_A j^Aj^A\right)^2=0\quad {\rm in}\quad\RR_\VV~.
\end{equation}
Thus we see that the element of $R_{\VV}$ corresponding to the Higgs branch relation in four dimensions is indeed nilpotent, and so vanishes in the reduced algebra. 

As in the DC series discussed in Section \ref{sec:deligne}, there can be no further relations that would remove the currents themselves from the reduced algebra. Such a removal would require a relation of the form $(j^+)^n=0$ in $\RR_\VV$, which would require a null state in $\suf(2)$ affine current subalgebra of the form $J_{-1}^{+})^n\Omega=0$, and such null states cannot exist at negative level. Thus we do recover the identification of the associated variety with the Higgs branch in this example.

\smallskip

The null state \eqref{eq:n4_su2_null} gives rise to a second-order, $\Gamma^0(2)$-modular differential equation for the vacuum character of this theory,
\begin{equation}
\DD^{\NN=4}_{\suf(2)}=D_q^{(2)}-2\Eb_2{\textstyle \left[\genfrac{}{}{0pt}{}{-1}{+1}\right]}(\tau) D_q^{(1)} -18\Eb_{4}(\tau)+18\Eb_4{\textstyle \left[\genfrac{}{}{0pt}{}{-1}{+1}\right]}(\tau))~.
\end{equation}
This equation has a single non-vacuum solution with $h=-\frac12$ -- this is a logarithmic solution. The conjugate differential operator is given by
\begin{equation}
\widetilde{\DD}^{\NN=4}_{\suf(2)}=D_q^{(2)}-2\Eb_2{\textstyle \left[\genfrac{}{}{0pt}{}{+1}{-1}\right]}(\tau) D_q^{(1)} -18\Eb_{4}(\tau)+18\Eb_4{\textstyle \left[\genfrac{}{}{0pt}{}{+1}{-1}\right]}(\tau))~,
\end{equation}
which has a two dimensional kernel, for which $\tilde h_{\min}=-\frac38$. Upon substitution into \eqref{eq:a_anomaly_equation}, this reproduces the correct value of $a_{4d}=3/4$ for the Weyl anomaly of the four-dimensional theory.

\renewcommand{\arraystretch}{1.5}
\begin{table}
\centering
\begin{tabular}{|c|c|c|c|c|}
\hline \hline
~$N$~  & ~${\rm ord}(\DD)$~ & ~Modular Group~  & ~Dimensions $h_i$~ & Conjugate dimensions $\tilde h_i$~~  \\ 
\hline 
$2$ & $2$ & $\Gamma^0(2)$ & $-\frac12, 0$ & $(-\frac38)_2$\\
\hline 
$3$ & $4$ & $\Gamma$ & $(-1)_3, 0$ & -----\\
\hline
$4$ & $6$ & $\Gamma^0(2)$  & $(-2)_2, (-\frac{3}{2})_3, 0$ & $(-\frac{15}{8})_4, (-\frac{7}{8})_2$\\
\hline
$5$ & $9$ & $\Gamma$ & $(-3)_5, (-2)_3, 0$ & -----\\
\hline
$6$ & $12$ & $\Gamma^0(2)$  & $(-\frac{9}{2})_3, (-4)_5, (-\frac{5}{2})_3, 0$ & $(-\frac{35}{8})_6, (-\frac{27}{8})_4, (-\frac{11}{8})_2$\\
\hline
$7$ & $16$ & $\Gamma$ & $(-6)_7, (-5)_5, (-3)_3, 0$ & -----\\
\hline
\end{tabular}
\caption{\label{tab:N4} Summary of modular differential operators that annihilate the vacuum characters of vertex operator algebras for $\NN=4$ super Yang-Mills theories. Provided are the order of the minimal modular differential operator that annihilates the vacuum character, the modular subgroup under which the corresponding differential equation is invariant, and the lists of solutions of the indicial equation and, when relevant, the conjugate indicial equation. The notation $(h_i)_{d_i}$ represents that the dimension $h_i$ occurs with multiplicity $d_i$.}
\end{table}

\subsubsection{Comments on the general case}
\label{subsubsec:n4_general_comments}

For higher rank cases, the associated VOAs have not been constructed as $\WW$-algebras in the literature, and we have not endeavored to analyze their null states. From the known expressions for the superconformal index, though, we have been able to find modular differential operators that annihilate the Schur index for $\gf=\suf(n)$ with $n=2,\ldots,7$. Relevant information about these differential operators is collected in Table \ref{tab:N4}, while the full expressions for the differential operators can be found in Appendix \ref{app:N4_modular}. Here we make some elementary, but potentially meaningful, observations regarding these operators.

\begin{enumerate}

\item[$\bullet$] First, we note that the values of $h_{\min}$ or, when relevant, $\tilde h_{\min}$ in all examples correctly reproduce the expected relation $a_{4d}=c_{4d}$ when inserted into Equation \ref{eq:a_anomaly_equation}.

\item[$\bullet$] We also note that, as in the class $\SS$ case, all of the additional solutions in addition to the vacuum character of the modular equations have $h_i<0$, and they are all potentially logarithmic (we have not attempted to construct all solutions, so it may be that in some cases there are non-logarithmic solutions despite their integer/half-integer separation from the vacuum dimension.

\item[$\bullet$] From the explicit forms of the differential operators given in Appendix \ref{app:N4_modular}, we see that for  $n=3$, $5$, and $7$, there is no constant term in the differential operators, so a constant function is always a solution. This is perhaps suggestive of the existence of additional differential operators of higher degree that would eliminate the constant solution from the modular orbit of the vacuum character.

\end{enumerate}

\subsection{\texorpdfstring{$T_4$}{T4} theory}
\label{subsec:T4_example}

As a final example, we consider the $T_4$ trinion theory. The Schur index for this theory is written most simply in the TQFT form of \cite{Gadde:2009kb,Gaiotto:2012xa}, but for our purposes it is best to have an expansion to high orders in $q$,
\begin{equation}
\begin{split}
\II_{T_4}(q)&=
   1+
   45 q+   
   128 q^{\frac32}+
   1295 q^2+
   5632 q^{\frac52}+
   33117 q^3+
   148352 q^{\frac72}+
   707340 q^4+
   2993664 q^{\frac92}\\
   &+
   12613923 q^5+
   49769216 q^{\frac{11}{2}}+
   191923893 q^6+
   708246016 q^{\frac{13}{2}}+
   2545387192 q^7\\
   &+
   8845957248 q^{\frac{15}{2}}+
   29966750747 q^8+
   98752864256 q^{\frac{17}{2}}+
   317881941694 q^9\\
   &+
   999327596160 q^{\frac{19}{2}}+
   3075532233083 q^{10}+
   9270593078784 q^{\frac{21}{2}}+
   27412890263961 q^{11}\\
   &+
   79570344948352 q^{\frac{23}{2}}+
   226982031641227 q^{12}+
   636756053977088 q^{\frac{25}{2}}+
   1758243618100910 q^{13}\\
   &+
   4781763625305472 q^{\frac{27}{2}}+
   12817731868201647 q^{14}+
   33884429064923648 q^{\frac{29}{2}}\\
   &+
   88392539111437047 q^{15}+
   227657529787627648 q^{\frac{31}{2}}+
   579191787392656267 q^{16}\\
   &+
   1456250341802891776 q^{\frac{33}{2}}+
   3620067533783343295 q^{17}+
   8901049365742734336 q^{\frac{35}{2}}\\
   &+
   21656077076478143385 q^{18}+
   52154251986491389568 q^{\frac{37}{2}}+
   124371739163793345678 q^{19}\\
   &+
   293775740780474832896 q^{\frac{39}{2}}+
   687553975761374631611 q^{20}+
   O(q^{41/2})
\end{split}
\end{equation}
A $\WW$-algebraic presentation has been proposed for the associated VOA in \cite{Lemos:2014lua}; the proposal is a $\frac12\Zb-$graded bosonic vertex operator algebra. We can find a sixth-order $\Gamma$-modular differential operator that annihilates this character,
\begin{equation}
\begin{split}
\DD_{T_4}&=D_{q}^{(6)}
-\tfrac12\Theta_{0,1}D_{q}^{(5)}
-\left(
\tfrac{631}{144}\Theta_{0,2}
-\tfrac{721}{288}\Theta_{1,1}
\right)D_{q}^{(4)}-
\left(
\tfrac{77}{48}\Theta_{0,3}
-\tfrac{29}{48}\Theta_{1,2}
\right)D_{q}^{(3)}\\
&+
\left(
\tfrac{2647}{20736}\Theta_{0,4}
+\tfrac{113507}{10368}\Theta_{1,3}
-\tfrac{137717}{13824}\Theta_{2,2}
\right)D_{q}^{(2)}
-\left(
\tfrac{6599}{124416}\Theta_{0,5}
-\tfrac{1774301}{124416}\Theta_{1,4}
+\tfrac{1597979}{124416}\Theta_{2,3}
\right)D_{q}^{(1)}\\
&+\left(
\tfrac{39}{4096}\Theta_{0,6}
-\tfrac{195}{4096}\Theta_{1,5}
+\tfrac{14379}{2048}\Theta_{2,4}
-\tfrac{54579}{8192}\Theta_{3,3}
\right)~,
\end{split}
\end{equation}
and the conjugate modular equation is obtained by replacing $\Theta_{p,q}\leftrightarrow \widetilde{\Theta}_{p,q}$. The solutions to the indicial and conjugate indicial equations are given by
\begin{equation}
\begin{split}
h&=\left\{-4,-\tfrac72,-3,-3,-3,0\right\}~,\\
\tilde{h}&=\left\{-5,-4,-3,-2,-2,-2\right\}~,
\end{split}
\end{equation}
so the additional solutions will generally be logarithmic. By examining the high temperature behavior, we find
\begin{equation}
\tilde{h}_{\rm min}=-5~,\quad c_{\rm eff}=42~, \quad a_{4d}=\frac{45}{8}~,
\end{equation}
which matches the expected value for $a_{4d}$ \cite{Benini:2009gi}.

%% file: sections/Acknowledgments.tex

\section*{Acknowledgments}

It is a great pleasure to thank Tomoyuki Arakawa, Mathew Bullimore, Clay C\'ordova, Kevin Costello, Thomas Creutzig, Maxime Gabella, Matthias Gaberdiel, Abhijit Gadde, Davide Gaiotto, Madalena Lemos, Pedro Liendo, Andrew Linshaw, Wolfger Peelaers, Shu-Heng Shao, David Simmons-Duffin, Jaewon Song, and Edward Witten for helpful discussions. C.B. gratefully acknowledges support from the Frank and Peggy Taplin Fellowship at the IAS. The work of C.B. work was additionally supported by a grant \#494786 from the Simons Foundation. L.R. is supported in part by the National Science Foundation under Grant No. NSF PHY-1620628. During the conception and completion of this work, the authors benefited from multiple stays at the Aspen Center for Physics, which is supported by National Science Foundation grant PHY-1607611.

%% file: sections/A1.tex

\section{Modular forms and linear modular differential operators}
\label{app:eisenstein_and_modular}

In this appendix we collect useful facts, definitions, and conventions regarding modular forms and modular differential operators. See any standard reference, \eg, \cite{ZagierBook}, for further details.
\vspace{12pt}

\noindent Let the modular parameter $\tau\in\Hb$ take values in the upper half plane. The modular group $\G\eqq{\rm PSL}(2,\Zb)$ acts on $\Hb$ according to
\begin{equation}
\tau\mapsto\frac{a\tau+b}{c\tau+d}~,\qquad a,~b,~c,~d~\in~\Zb~,\qquad ad-bc=1~.
\end{equation}
This group is generated by the elements
\begin{equation}
S:\tau\mapsto-\frac{1}{\tau}~,\qquad T:\tau\mapsto \tau+1~,
\end{equation}
subject to the relations $S^2 = (ST)^3 = 1$.

We define the \emph{nome} $q\ceq e^{2\pi i \tau}$ and let $\g\in\G$ act on $q$ in the natural way,
\begin{equation}
\g\circ q=e^{2\pi i \frac{a\tau+b}{c\tau+d}}~.
\end{equation}
The principal congruence subgroups of $\G$ are defined as follows
\begin{equation}
\G(N)\ceq\left\{\begin{pmatrix}~a~&~b~\\~c~&~d~\end{pmatrix}\in\G~,\qquad a\eqq d\eqq\pm1~,~~b\eqq c\eqq 0\quad {\rm mod}~N\right\}~.
\end{equation}
Congruence subgroups are subgroups of $\G$ that themselves contain $\G(N)$ as a subgroup for some $N$. Some standard congruence subgroups that will be useful for our purposes are
\begin{eqnarray}
\G^1(N)&\ceq&\left\{\begin{pmatrix}~a~&~b~\\~c~&~d~\end{pmatrix}\in\G~,\qquad a\eqq d\eqq 1~,~~ b \eqq 0\quad {\rm mod}~N\right\}~,\\
\G^0(N)&\ceq&\left\{\begin{pmatrix}~a~&~b~\\~c~&~d~\end{pmatrix}\in\G~,\qquad b\eqq 0\quad {\rm mod}~N\right\}~.
\end{eqnarray}
The congruence subgroups groups $\G_0(N)$ and $\G_1(N)$ are somewhat more conventional and are related to the groups with upper indices by an overall conjugation by the element $S\in\G$. The case of relevance for this work is $N=2$, and for this value these groups are not distinct: $\G^0(2)\eqq\G^1(2)$~.

A modular form of weight $k$ for $\G$ is a \emph{holomorphic} function $f:\Hb\to\Cb$ that transforms according to
\begin{equation}
f\left(\frac{a\tau+b}{c\tau+d}\right)=(c\tau+d)^k f(\tau)~,\qquad 
\begin{pmatrix}~a~&~b~\\~c~&~d~\end{pmatrix}~\in~\G~,
\end{equation}
that is additionally finite as $\Im(\tau)\to+\infty$. There can be no non-zero modular forms for $\G$ with odd degree due to the requirement that the central element of ${\rm SL}(2,\Zb)$ act trivially. Any modular form has a convergent Fourier expansion in $q$ and is finite in the limit $q\to0$, \ie,
\begin{equation}
f(\tau)=\sum_{n=0}^{\infty}a_n q^n~.
\end{equation}
The vector space of modular forms of weight $k$ is denoted $M_k(\G,\Cb)$. A fundamental result in the theory of modular forms is that $M_k(\G,\Cb)$ is finite dimensional for any $k$.

Similarly, a modular form of weight $k$ for a congruence subgroup of $\wt\G\subset\G$ is a holomorphic function on the upper half plane such that
\begin{equation}
f\left(\frac{a\tau+b}{c\tau+d}\right)=(c\tau+d)^k f(\tau)~,\qquad \begin{pmatrix}~a~&~b~\\~c~&~d~\end{pmatrix}~\in~\wt\G~,
\end{equation}
and in addition is finite as $\Im(\tau)\to+\infty$ and for $\tau\in\Qb$. Any modular form for $\wt\G\in\{\G(N),\G^1(N),\G^0(N)\}$ has a Fourier series expansion in $q^\frac{1}{N}=e^{\frac{2\pi i\tau}{N}}$,
\begin{equation}
f(\tau)=\sum_{n=0}^{\infty}a_n q^{n/N}~.
\end{equation}
The modular forms of weight $k$ for a subgroup $\wt\G$ of the modular group are denoted $M_k(\wt\G,\Cb)$.

\subsection{Theta functions}

The classical Jacobi theta constants are defined as
\begin{equation}
\begin{split}
\vth_{00}(\tau)&\equiv\th_3(\tau)\ceq\sum_{n=-\infty}^{\infty}q^\frac{n^2}{2}~,\\
\vth_{01}(\tau)&\equiv\th_4(\tau)\ceq\sum_{n=-\infty}^{\infty}(-1)^n q^\frac{n^2}{2}~,\\
\vth_{10}(\tau)&\equiv\th_2(\tau)\ceq\sum_{n=-\infty}^{\infty}q^{\frac12(n+\frac12)^2}~.
\end{split}
\end{equation}
These satisfy the Jacobi identity,
\begin{equation}
\vth_{00}(\tau)^4=\vth_{10}(\tau)^4+\vth_{01}(\tau)^4~,
\end{equation}
and under modular transformations they behave as follows,
\begin{equation}
\begin{split}
\vth_{00}(-\tfrac{1}{\tau})&=e^{\frac{\pi i}{4}}(-\tau)^{\frac12}\vth_{00}(\tau)~,\qquad\qquad
\vth_{00}(\tau+1)\,=\,\vth_{01}(\tau)~,\\
\vth_{01}(-\tfrac{1}{\tau})&=e^{\frac{\pi i}{4}}(-\tau)^{\frac12}\vth_{10}(\tau)~,\qquad\qquad
\vth_{01}(\tau+1)\,=\,\vth_{00}(\tau)~,\\
\vth_{10}(-\tfrac{1}{\tau})&=e^{\frac{\pi i}{4}}(-\tau)^{\frac12}\vth_{01}(\tau)~,\qquad\qquad
\vth_{10}(\tau+1)\,=\,-e^{\frac{\pi i}{4}}\vth_{10}(\tau)~.
\end{split}
\end{equation}
Modular forms for many congruence subgroups can be constructed using these theta constants.

\subsection{Eisenstein series}

The ordinary Eisenstein series are modular forms for the full modular group $\G$ of weight $2k$ with $k\geqs2$. There are several conventional normalizations for these series. We define our Eisenstein series, following \cite{Mason:2008zzb}, as
\begin{equation}
\Eb_{2k}(\tau)\ceq -\frac{B_{2k}}{(2k)!}+\frac{2}{(2k-1)!}\sum_{n\geqs1}\frac{n^{2k-1}q^n}{1-q^n}~,
\end{equation}
where $B_{2k}$ is the $2k$'th Bernoulli number. An alternative normalization is natural when defining the function as a Poincar\'e series,
\begin{equation}
G_{2k}(\tau)\ceq\sum_{(m,n)\in\Zb^2\backslash(0,0)}\frac{1}{(m+n\tau)^{2k}}~.
\end{equation}
The two choices of normalization are related according to
\begin{equation}
G_{2k}(\tau)\eqq (2\pi i)^{2k}\Eb_{2k}(\tau)~.
\end{equation}
In addition, one often defines the \emph{normalized} Eisenstein series
\begin{equation}
E_{2k}(\tau)\ceq\frac{1}{2\zeta(2k)}G_{2k}(\tau)=1+\frac{1}{\zeta(1-2k)}\sum_{n=1}^{\infty}\frac{n^{2k-1}q^n}{1-q^n}~.
\end{equation}

The case with $k=1$ is special and does not furnish a modular form of weight two -- there are none. Instead, $\Eb_2(\tau)$ has an anomalous transformation under the modular group given by
\begin{equation}
\Eb_2\left(\frac{a\tau+b}{c\tau+d}\right) = (c\tau+d)^2\Eb_2(\tau) -\frac{c(c\tau+d)}{2\pi i}~.
\end{equation}
The ring of modular forms for the full modular group $\G$ is freely generated by $\Eb_4(\tau)$ and $\Eb_6(\tau)$, so we have
\begin{equation}
\bigoplus_{k=0}^{\infty}M_k(\G,\Cb)=\Cb[\Eb_4(\tau),\Eb_6(\tau)]~.
\end{equation}

\subsection{Twisted Eisenstein series}

We also make use of a class of twisted Eisenstein series that are modular forms for certain congruence subgroups of $\G$. The twisted Eisenstein series of interest are defined as follows (again see \cite{Mason:2008zzb} for more details),
\begin{equation}
\Eb_{k}\left[\genfrac{}{}{0pt}{}{\vph}{\vth}\right](\tau)\eqq-\frac{B_k(\lambda)}{k!}+\frac{1}{(k-1)!}\sum_{r\geqs0}^{\prime}\frac{(r+\lambda)^{k-1}\vth^{-1}q^{r+\lambda}}{1-\vth^{-1}q^{r+\lambda}}+\frac{(-1)^k}{(k-1)!}\sum_{r\geqs1}\frac{(r-\lambda)^{k-1}\vth q^{r-\lambda}}{1-\vth q^{r-\lambda}}~,
\end{equation}
where $\vph=e^{2\pi i \lambda}$ with $\lambda\in[0,1)$ and now $B_k(x)$ is the $k$'th Bernoulli polynomial. The prime in the first summation indicates that the $r=0$ term should be omitted when $\vth=\vph=1$. This class of twisted Eisenstein series transform amongst themselves under general modular transformations,
\begin{equation}
\Eb_k\left[\genfrac{}{}{0pt}{}{\vph^a\vth^b}{\vph^c\vth^d}\right]\left(\frac{a\tau+b}{c\tau+d}\right)=(c\tau+d)^{k}\Eb_{k}\left[\genfrac{}{}{0pt}{}{\vph^{\phantom{a}}\!\!}{\vth^{\phantom{b}}\!\!}\right](\tau)~.
\end{equation}

Relevant to us will be in the cases with $\vth,\vph=\pm1$. For $k\geqs4$, these are modular forms of weight $k$ for $\G^0(2)$, and similarly for $k=2$ when $\vph\neq1$ or $\vth\neq1$.  The weight-two twisted Eisenstein series of this form have the following relations to elliptic theta constants:
\begin{eqnarray}
\Eb_{2}\left[\genfrac{}{}{0pt}{}{-1}{+1}\right]&=&\frac{\th_2(\tau)^4+\th_3(\tau)^4}{24}~,\nn\\
\Eb_{2}\left[\genfrac{}{}{0pt}{}{+1}{-1}\right]&=&\frac{\th_2(\tau)^4-2\th_3(\tau)^4}{24}~,\\
\Eb_{2}\left[\genfrac{}{}{0pt}{}{-1}{-1}\right]&=&\frac{-2\th_2(\tau)^4+\th_3(\tau)^4}{24}~.\nn
\end{eqnarray}
The spaces of modular forms for $\G(2)$, $\G^0(2)=\G^1(2)$, and $\G_0(w)=\G_1(2)$ all admit a simple descriptions in terms of theta functions. In particular, we have
\begin{eqnarray}
\bigoplus_{k=0}^{\infty}M_k(\G(2),\Cb)&=&\Cb[\th_2(\tau)^4,\th_3(\tau)^4]~,\\
\bigoplus_{k=0}^{\infty}M_k(\G^0(2),\Cb)&=&\Cb[\th_2(\tau)^4,\th_3(\tau)^4]^{\SS_2}~,\\
\bigoplus_{k=0}^{\infty}M_k(\G_0(2),\Cb)&=&\Cb[\th_3(\tau)^4,\th_4(\tau)^4]^{\SS_2}~.
\end{eqnarray}
where $\SS_2$ is the symmetric group action that exchanges $\th_2(\tau)\leftrightarrow \th_3(\tau)$ in the second line and $\th_3(\tau)\leftrightarrow \th_4(\tau)$ in the last line. To facilitate the description of elements of $M_k(\G^0(2))$ and $M_k(\G_0(2)$, we define
\begin{equation}
\begin{split}
\label{eq:modular_theta_defs}
\Th_{r,s}(\tau) &\ceq \th_2(\tau)^{4r}\th_3(\tau)^{4s}+\th_2(\tau)^{4s}\th_3(\tau)^{4r}~,\qquad r\leqs s~,\\
\wt{\Th}_{r,s}(\tau) &\ceq (-1)^{r+s}\left(\th_4(\tau)^{4r}\th_3(\tau)^{4s}+\th_4(\tau)^{4s}\th_3(\tau)^{4r}\right)~,\qquad r\leqs s~.
\end{split}
\end{equation}
in terms of which we have
\begin{equation}
\begin{split}
M_{2k}\left(\G^0(2)\right)&={\rm span}\left\{\Th_{r,s}(\tau)~|~ r+s=k\right\}~,\\
M_{2k}\left(\G_0(2)\right)&={\rm span}\left\{\wt{\Th}_{r,s}(\tau)~|~ r+s=k\right\}~.
\end{split}
\end{equation}
Furthermore, we have
\begin{equation}
\label{eq:bigtheta_transform}
\Th_{r,s}\left(\frac{-1}{\tau}\right)=\tau^{2r+2s}\wt{\Th}_{r,s}(\tau)~.
\end{equation}
As a consequence of this transformation rule, $\G^0(2)$-modular differential operators written in terms of $\Th_{r,s}(\tau)$ are related by conjugation by $S\in{\rm PSL}(2,\Zb)$ to $\G_0(2)$-modular differential operators with the same coefficients but with $\Th_{r,s}~\leftrightarrow~\wt{\Th}_{r,s}$.

\subsection{Modular differential operators}

Serre derivatives $\partial_{(k)}$ are improved differential operators that map modular forms of a fixed weight to modular forms of higher weight,
\begin{equation}\label{eq:serre_derivative_map}
\partial_{(k)}:M_{k}(\G,\Cb)\to M_{k+2}(\G,\Cb)~.
\end{equation}
They are defined using the quasi-modular second Eisenstein series,
\begin{equation}\label{eq:serre_derivative_def}
\partial_{(k)}f(q)=\left(q\partial_q+ k\,\Eb_2(\tau)\right)f(q)~.
\end{equation}
When acting on the low-weight Eisenstein series, for example, we have
\begin{equation}
\begin{split}
\partial_{(4)}\Eb_4(\tau)&=14\,\Eb_{6}(\tau)~,\\
\partial_{(6)}\Eb_6(\tau)&=20\,\Eb_{8}(\tau)~,\\
\partial_{(8)}\Eb_8(\tau)&=\tfrac{132}{5}\,\Eb_{10}(\tau)~,\\
\partial_{(10)}\Eb_{10}(\tau)&=\tfrac{600}{77}\,\Eb_{4}(\tau)^3+\tfrac{210}{11}\Eb_6(\tau)^2~,\\
\partial_{(12)}\Eb_{12}(\tau)&=\tfrac{1382}{35}\Eb_{14}(\tau)~.\\
\end{split}
\end{equation}
Using Serre derivatives we can define $k$'th order modular differential operators that naturally act on objects of modular weight zero,
\begin{eqnarray}
\label{eq:modular_op_example}
D^{(k)}_qf(q)&\ceq& \partial_{(2k-2)}\circ\cdots\circ\partial_{(2)}\circ\partial_{(0)}f(q)~,
\end{eqnarray}
where by convention we set $D^{(0)}_qf(q)\ceq f(q)$. These differential operators themselves transform with weight $2k$ under the action of $\G$, and so also under the action of any congruence subgroup $\wt{\G}$, \ie,
\begin{equation}
D^{(k)}_{\g\circ q}=(c\tau+d)^{2k}D^{(k)}_q~,\qquad \g\in\wt{\G}~.
\end{equation}
and from them we can construct a large class of linear modular differential operators of a fixed weight. We will be particularly interested in linear modular differential operators that are holomorphic and monic, so have the form\footnote{By a monic linear modular differential operator we mean an operator of weight $2k$ with unit coefficient for $D^{(k)}$. Any linear modular differential operator can formally be put in this form by simply dividing through by the coefficient of the unit, but this will generally result in an operator for which the coefficient functions $f_r(q)$ are meromorphic, not holomorphic. Such differential operators can be relevant in the study of characters of VOAs --- see, \eg, \cite{Mathur:1988na} --- but we leave deeper investigation of this generalization for future work.}
\begin{equation}
\DD_q^{(k)}\eqq D^{(k)}_q+\sum_{r=1}^k f_r(q) D^{(k-r)}_q~,\qquad f_r(q)\in M_{2r}(\wt{\G},\Cb)~.
\end{equation}
For any such differential operator, if we have a function $g(q)\in{\rm Ker}\,\DD^{(k)}$, then performing a modular transformation we find that
\begin{equation}
(c\tau+d)^{2k}\DD^{(k)}g(\g\circ q)= \DD^{(k)}_{\g\circ q}g(\g\circ q)=0~,
\end{equation}
so the space of solutions to modular differential equations of this type are vector valued modular forms with respect to the appropriate congruence subgroup \cite{Mason:2007vvm}. 

%% file: sections/A2.tex

\section{Trace recursion relations}
\label{app:modular_recursion}

Here we review some technical results regarding the evaluation of torus one-point functions, \ie, traces of vertex operator zero modes in the vacuum module of a VOA. For more complete details, see \cite{ZhuThesis,Zhu:1996,Gaberdiel:2008pr} and especially \cite{Mason:2008zzb}.

\subsection{Vertex operator algebras and torus \tpdf{$n$}{n}-point functions}
\label{subsec:app_torus_n_point}
In this paper we deal only with $\Zb_{\geqs0}$ and $\frac12\Zb_{\geqs0}$ graded vertex operator algebras. Further, we deal exclusively with conformal vertex operator algebras, meaning they will include a subalgebra that is isomorphic to the Virasoro vertex operator algebra. Furthermore, the graded components of the underlying vector space $\VV$ will always be finite dimensional,
\begin{equation}
\VV=\!\!\!\!\!\bigoplus_{n\in \Zb_{\geqs0}~{\rm or}~\frac12\Zb_{\geqs0}}\!\!\!\!\!\VV_n~,\qquad \dim\VV_n<\infty~,
\end{equation}
and the $L_0$ operator from the Virasoro subalgebra acts semi-simply as
\begin{equation}
L_0a=na~,\qquad a\in\VV_n~.
\end{equation}
In the general case we will consider vertex operator superalgebras, which have even and odd parts,
\begin{equation}
\VV=\VV_{\bar{\bf 0}}\oplus\VV_{\bar{\bf 1}}~.
\end{equation}
We say that the parity $p(a)$ of a state $a$ is one if $a\in\VV_{\bar{\bf 1}}$ and zero if $a\in\VV_{\bar{\bf 0}}$. We \emph{do not} assume any correlation between the parity of a state and it conformal dimension modulo one. We define the supertrace of an endomorphism $\OO:\VV\to\VV$ by
\begin{equation}
\STr_{\VV}\OO\eqq\Tr_{\VV_{\bar{\bf 0}}}\OO-\Tr_{\VV_{\bar{\bf 1}}}\OO~.
\end{equation}

The mode expansion for the vertex operator corresponding to a state $a$ of integer weight $h_a$ is given by%
\footnote{As a word of warning, when comparing to the mathematics literature, especially \cite{ZhuThesis,Zhu:1996,Mason:2008zzb}, the convention for the mode numbering is different. In those works, one has
\begin{equation*}
Y(a,z)=\sum_{n}a_{(-1-n)}z^n~.
\end{equation*}
Those conventions have the virtue of making sense even when there is not a good grading by conformal dimension, but this is not relevant in the current work and not standard in the physics literature.}
\begin{equation}
\label{eq:vertex_operator_mode_expansion}
a(z)\eqq Y(a,z)=\sum_{n\in\Zb}a_{(-h_a-n)} z^{n}~,
\end{equation}
where the modes act as endomorphisms that shift grading by a definite amount,
\begin{equation}
\label{eq:graded_endomorphism}
a_{(n)}:\VV_k\rightarrow \VV_{k+n}~.
\end{equation}
When $a$ has half-integer weight $h_a\in\Zb+\frac12$, we define
\begin{equation}
\label{eq:vertex_operator_half_integer_mode_expansion}
a(z)\eqq Y(a,z)=\sum_{n\in\Zb+\frac12}a_{(-h_a-n)} z^{n}~.
\end{equation}
In other words, we always work with the Neveu-Schwarz grading. In the case of integer conformal weight, we further introduce the following notation for the zero mode of $a$,
\begin{equation}
\label{eq:zero_mode_definition}
o(a) \ceq a_0~.
\end{equation}
Torus $n$-point functions are defined as traces of over the space of states as follows,
\begin{equation}
\label{eq:torus_n_point}
\FF((a^{(1)},z_1),\ldots,(a^{(n)},z_n);\tau)=z_1^{h_{1}}\cdots z_n^{h_{n}}\STr_{\VV}\left(Y(a^{(1)},z_1)\cdots Y(a^{(n)},z_n)q^{L_0-\frac{c}{24}}\right)~.
\end{equation}
In the special case of the one point function, by virtue of \eqref{eq:graded_endomorphism}, we then have
\begin{equation}
\label{eq:torus_one_point}
\FF((a,z),\tau)=
\begin{cases}
\STr_{\VV}\left(o(a)q^{L_0-\frac{c}{24}}\right) & \quad{\rm if}~~h_a=0~({\rm mod}~1)~{\rm and}~p(a)=0,\\
0 & \quad{\rm if}~~h_a=\frac12~({\rm mod}~1)~.
\end{cases}
\end{equation}

\subsection{Square brackets}
\label{subsec:app_square_brackets}

An alternative expansion for the same vertex operators is useful in formulating recursion relations for torus $n$-point functions. We define
\begin{equation}
Y[a,z]=e^{z h_a}Y(a,e^{z}-1)~,
\end{equation}
And then introduce the ``square-bracket'' mode expansion
\begin{equation}
Y[a,z]=\sum_{n\in\Zb}a_{[-n-h_a]}z^{n}~.
\end{equation}
This is a reorganization of the original modes under a local change of variables. In particular, the square bracket modes can be expressed in terms of the usual modes according to
\begin{equation}
a_{[n]}=\sum_{j\geqs n}c(j, n; h_a)a_{(j)}~,
\end{equation}
where the coefficients are defined as the coefficients in the following Taylor series,
\begin{equation}
(1+z)^{h-1}\log(1+z)^n=\sum_{j\geqs n}c(j,n;h)z^j~.
\end{equation}
If $a$ is a Virasoro primary state, then the commutation relations for the $a_{[n]}$ modes will be identical to those of the $a_{(n)}$. The stress tensor, however, picks up an anomalous constant since it is not a Virasoro primary. We therefore define instead
\begin{equation}
L_{[n]}=\sum_{j\geqs n}c(j, n; 2)L_{(j)}-\frac{c}{24}\delta_{j,-2}~.
\end{equation}
These square bracket Virasoro generators now have the same commutation relations as the original modes.

From the expression for the coefficients, we see that generally we have
\begin{equation}
a_{[n]} = a_{(n)}+ c(n+1,n; h_a) a_{(n+1)}+c(n+2, n; h_a)a_{(n+2)}+\ldots~,
\end{equation}
so the vacuum state obeys the same highest weight condition with respect to the square bracket modes as it does with respect to the ordinary modes,
\begin{equation}
a_{(n)}\Omega=a_{[n]}\Omega=0~,\qquad n> -h_a~.
\end{equation}
Along with the agreement of commutation relations, this implies that a null vector in the vacuum Verma module of a VOA formulated in terms of the $a_{(n)}$ will still be null after replacing the modes with square bracket modes.

\subsection{Recursion relations for torus one-point functions}
\label{subsec:app_recursion_relations}
The virtue of the square-bracket modes is that torus one point functions (\ie, traces of zero modes) in a VOA obey recursion relations that take an elegant form in terms of the square brackets. The simplest version of such a recursion relation applies to the case when $\VV$ is a $\Zb$ graded VO(S)A. In this case one has \cite{Zhu:1996}
\begin{equation}
\label{eq:not_quite_recursion}
\STr_{\VV}\left(o(a_{[-h_a]}b)q^{L_0-\frac{c}{24}}\right)=\STr_{\VV}\left(o(a)o(b)q^{L_0-\frac{c}{24}}\right)+\sum_{k\geqs1}\Eb_{2k}(\tau)\STr_{\VV}\left(o(a_{[-h_a+2k]}b)q^{L_0-\frac{c}{24}}\right)~,
\end{equation}
where $\Eb_{2k}(\tau)$ is the (unnormalized) Eisenstein series defined in Appendix \ref{app:eisenstein_and_modular}. Specializing to the case where $a$ is a conformal descendant (in which case the zero mode of $a$ vanishes), one finds the following recursion relation,
\begin{equation}
\begin{split}
\label{eq:recursion}
\STr_\VV\left(o(a_{[-h_a-1]}b)q^{L_0-\frac{c}{24}}\right)
&=\sum_{k\geqs1}(1-2k)\Eb_{2k}(\tau)\STr_{\VV}\left(o(a_{[-h_a-1+2k]}b)q^{L_0-\frac{c}{24}}\right)~,\\
&=\sum_{k\geqs2}(1-2k)\Eb_{2k}(\tau)\STr_{\VV}\left(o(a_{[-h_a-1+2k]}b)q^{L_0-\frac{c}{24}}\right)~,
\end{split}
\end{equation}
where the first term in the summation on the first line is zero because $o(a_{-h_a+1}b)$ is a commutator. A generalization of this relation gives a similar relation for zero modes involving higher descendants,
\begin{equation}
\label{eq:recursion_general}
\STr_\VV\left(o(a_{[-h_a-n]}b)q^{L_0-\frac{c}{24}}\right)=
(-1)^{n}\sum_{2k\geqs n+1}^{\prime}\binom{2k-1}{n}\Eb_{2k}(\tau)\STr_{\VV}\left(o(a_{[-h_a-n+2k]}b)q^{L_0-\frac{c}{24}}\right)~,
\end{equation}
with the prime indicating that the $n=k=1$ term is zero. Note that the summation is over traces of zero modes of states whose left most oscillator annihilates the conformal vacuum, so after applying the recursion relation once may further simplify the result by commuting the annihilation operator through to the right. Importantly, the states whose zero mode appears on the right hand sides of \eqref{eq:recursion} and \eqref{eq:recursion_general} have conformal dimension strictly less than those appearing on the left hand sides. This is what allows the recursion algorithm set up using these relations to terminate.

A modified version of these recursion relations holds in the $\frac12\Zb$-graded case; in this case it is the twisted Eisenstein series appear \cite{Mason:2008zzb},
\begin{equation}
\label{eq:twisted_recursion}
\STr_\VV\left(o(a_{[-h_a-1]}b)q^{L_0-\frac{c}{24}}\right)=
\sum_{k\geqs1}^{\prime}(1-2k)\Eb_{2k}\left[\genfrac{}{}{0pt}{}{e^{2\pi i h_a}}{1}\right](\tau)\,\STr_{\VV}\left(o(a_{[-h_a-1+2k]}b)q^{L_0-\frac{c}{24}}\right)~.
\end{equation}
The prime now indicates that $k=1$ term on the right hand side only appears when $h_a$ is half-integral; only the twisted weight-two Eisenstein series appears. Again, this formula admits a generalization for higher descendants
\begin{equation}
\label{eq:twisted_recursion_general}
\STr_\VV\left(o(a_{[-h_a-n]}b)q^{L_0-\frac{c}{24}}\right)=
(-1)^n\sum_{2k\geqs n+1}^{\prime}\binom{2k-1}{n}\Eb_{2k}\left[\genfrac{}{}{0pt}{}{e^{2\pi i h_a}}{1}\right](\tau)\STr_{\VV}\left(o(a_{[-h_a-n+2k]}b)q^{L_0-\frac{c}{24}}\right)~.
\end{equation}
We note that in the case where the supertrace is replaced by an ordinary trace, similar recursion relations hold, but now more general twisted Eisenstein series that relate to whether the fields $a(z)$ and $b(z)$ are parity-even or parity-odd.

\subsection{Stress tensor trace formulae}
\label{subsec:app_trace_formulae}

One point functions of Virasoro descendants of the vacuum can be evaluated in terms of differential operators acting on the vacuum character. Of particular interest are one-point functions of vertex operators corresponding to states of the form $(L_{[-2]})^k\Omega$ for some positive integer $k$. In this case one can directly apply equation \eqref{eq:not_quite_recursion} recursively to express the trace of $o((L_{[-2]})^k\Omega)$ in terms of traces with lower powers of $k$ and additional insertions of $L_{[0]}$. Insertions of $L_{[0]}$ can be turned into derivatives with respect to $q$, so this ultimately leads to pure differential operators acting on the vacuum character,
\begin{equation}
\Tr_\VV\left(o((L_{[-2]})^k\Omega)q^{L_0-\frac{c}{24}}\right)= \PP_k\circ\Tr_{\VV}\left(q^{L_0-\frac{c}{24}}\right)~.
\end{equation}
The first few differential operators $\PP_k$ can be found in \cite{Gaberdiel:2008pr}, and we reproduce them here for convenience (and in our own slightly different conventions),
\begin{equation}
\begin{split}
\PP_2 &= D_q^{(1)}~,\\
\PP_4 &= D_q^{(2)}+\tfrac{c}{2}\Eb_4(\tau)~,\\
\PP_6 &= D_q^{(3)}+\left(8+\tfrac{3c}{2}\right)\Eb_4(\tau)D_q^{(1)}+10 c\,\Eb_6(\tau)~,\\
\PP_8 &= D_q^{(4)}+\left(32+3c\right)\Eb_4(\tau)D_q^{(2)}+\left(160+40c\right)\Eb_6(\tau)D_q^{(1)}+\left(108c + \tfrac34 c^2\right)\Eb_4(\tau)^2~.
\end{split}
\end{equation}
There is no obstruction to going to higher $k$ but it is somewhat tedious. Note also that since only the parity-even stress tensor appears in the calculation, the same differential operators appear in the case where we replace the trace by a supertrace,
\begin{equation}
\STr_\VV\left(o((L_{[-2]})^k\Omega)q^{L_0-\frac{c}{24}}\right)=\PP_k\circ\STr_{\VV}\left(q^{L_0-\frac{c}{24}}\right)~.
\end{equation}

%% file: sections/A3.tex

\section{Characters for the Deligne-Cvitanovi\'c exceptional series}
\label{app:deligne_solutions}

The modular differential equation for the Deligne exceptional series of non-unitary vertex operator algebras described in Section \ref{sec:deligne} is given by \eqref{eq:delinge_series_diffeq}. For convenience we reproduce it here,
\begin{equation}
\label{eq:deligne_modular_equation_app}
\left(D^{(2)}_q-\frac{(h^{\vee}+1)(h^\vee-1)}{144}E_4(q)\right)\goodchi(q)=0~,
\end{equation}
where we switched to the normalized Eisenstein series for later convenience. Second order monic modular equations can be solved explicitly by performing a change of variables to put them in hypergeometric form \cite{VVMF}. This is accomplished by introducing the modular $j$-invariant and its rescaled inverse,
\begin{equation}
j(\tau)=\frac{1728E_4(q)^3}{E_4(q)^3-E_6(q)^2}~,\qquad K(\tau)=\frac{1728}{j(\tau)}=\frac{E_4(q)^3-E_6(q)^2}{E_4(q)^3}~,
\end{equation}
in terms of which there is a simple relation between differential operators,
\begin{equation}
\th_q = \left(\frac{E_6(q)}{E_4(q)}\right)\th_K~,
\end{equation}
where $\th_q\eqq q\partial_q$ and $\th_K\eqq K\partial_K$. Equation \eqref{eq:deligne_modular_equation_app} can then be rewritten as
\begin{equation}
\left(\th_K^2-\left(\frac{1+2K(q)}{6-6K(q)}\right)\th_K-\frac{(h^{\vee}+1)(h^\vee-1)}{144(1-K(q))}\right)\goodchi(q)=0~,
\end{equation}
which is a hypergeometric differential equation in $K$. We can immediately find the most general solution to this differential equation. For generic values of $h^{\vee}$, the two linearly independent solutions are given by
\begin{eqnarray}
\goodchi_1(q)&=&K(q)^{\frac{(1+h^{\vee})}{12}}\,{}_2F_1\left(\tfrac{1+h^\vee}{12},\tfrac{5+h^\vee}{12},1+\tfrac{h^\vee}{6},K(q)\right)~,\\
\goodchi_2(q)&=&K(q)^{\frac{(1-h^{\vee})}{12}}\,{}_2F_1\left(\tfrac{1-h^\vee}{12},\tfrac{5-h^\vee}{12},1-\tfrac{h^\vee}{6},K(q)\right)~.
\end{eqnarray}
Upon normalizing the solution so that the leading coefficient in the $q$ expansion is one, $\goodchi_1(q)$ reproduces the vacuum character for these vertex operator algebras, while $\goodchi_2(q)$ should be a (linear combination of) characters of admissible representations.

When $h^\vee$ is a multiple of six, the corresponding affine current algebra is not at an admissible level, and also the modular differential equation becomes degenerate. The second solution given above then becomes undefined. The new second solution is given in terms of the Meijer G-function as follows,
\begin{equation}
\goodchi_2(q)=
\MeijerG*{2}{0}{2}{2}{\frac23\quad~,\quad1}{\frac{-h+1}{12}~,~~\frac{h+1}{12}}{K(q)}~.
\end{equation}
This is a logarithmic solution. However, we believe these logarithms to be ``fake'' in the sense described briefly in Section \ref{subsec:interpretation} \cite{flavored_characters}. We note here that more beautiful expressions for the vacuum characters of these algebras have been presented in \cite{Arakawa:2016hkg}, using methods described in \cite{KanekoKoike,Kaneko:2013uga}.

%% file: sections/A4.tex

\section{Differential operators for \texorpdfstring{$A_1$}{A1} class \texorpdfstring{$\SS$}{S} indices}
\label{app:class_S_modular}

In this appendix we record the modular and twisted-modular differential operators that annihilate the unflavored Schur indices of $A_1$ type class $\SS$ SCFTs for a variety of low values for the genus and number of (full) punctures.

\medskip

\noindent The unflavored Schur index for all of these examples admits a TQFT expansion given in the text in \eqref{eq:a1_index}, which we reproduce here for convenience,
\begin{equation}
\II^{\af_1}_{g,s}=(q;q)_{\infty}^{2g-2-2s}\sum_{k=0}^{\infty}\left(
\frac{(k + 1)^s q^{\frac{k}{2}(2 g - 2 + s)}}
{(1 - q^{k + 1})^{2 g - 2 + s}}\right)~.
\end{equation}
When the quantity $2g+s$, the expansion is in integer powers of $q$ and we will correspondingly find modular differential operators with respect to the full modular group, while for $2g+s$ odd we will find $\Gamma^0(2)$-modular differential operators.

\subsection*{Genus zero}

\begin{equation*}
\mathmakebox[\textwidth][l]{
\DD_{\CC_{0,3}}^{\mathfrak{a}_1} = D_q^{(1)}-\tfrac16 \Theta_{0,1}(q)~.
}
\end{equation*}

\subsection*{Genus zero, four punctures}

\begin{equation*}
\mathmakebox[\textwidth][l]{
\DD_{\CC_{0,4}}^{\mathfrak{a}_1} = D_q^{(2)}-175\,\Eb_4(q)~.
}
\end{equation*}

\subsection*{Genus zero, five punctures}

\begin{equation*}
\mathmakebox[\textwidth][l]{
\DD_{\CC_{0,5}}^{\mathfrak{a}_1} = 
D_q^{(4)}-
\left(\tfrac{11}{18}\Theta_{0,2}(q)-\tfrac{11}{36}\Theta_{1,1}(q)\right)D_q^{(2)}-
\left(\tfrac{5}{108}\Theta_{0,3}(q)+\tfrac{13}{72}\Theta_{1,2}(q)\right)D_q^{(1)}+
\left(\tfrac{1}{4}\Theta_{1,3}(q)-\tfrac{5}{16}\Theta_{2,2}(q)\right)~.
}
\end{equation*}

\subsection*{Genus zero, six punctures}

\begin{flalign*}
\DD_{\CC_{0,6}}^{\mathfrak{a}_1} = 
D_q^{(6)} &-
545\,\Eb_4(q)\,D_q^{(4)}-
15260\,\Eb_6(q)\,D_q^{(3)}-
164525\,\Eb_4(q)^2\,D_q^{(2)} - 2775500\,\Eb_4(q)\Eb_6(q)\,D_q^{(1)}&&\\
&-
26411000\,\Eb_6(q)^2 + 1483125\,\Eb_4(q)^3~.&&
\end{flalign*}

\subsection*{Genus one, one puncture}

\begin{equation*}
\mathmakebox[\textwidth][l]{
\DD_{\CC_{1,1}}^{\mathfrak{a}_1} = 
D_q^{(2)}-
\tfrac{1}{6}\Theta_{0,1}\,D_q^{(1)}-
\left(\tfrac{5}{144}\Theta_{0,2}-\tfrac{11}{288}\Theta_{1,1}\right)}~.
\end{equation*}

\subsection*{Genus one, two punctures}

\begin{equation*}
\mathmakebox[\textwidth][l]{
\DD_{\CC_{1,2}}^{\mathfrak{a}_1} = 
D_q^{(4)} -
220\,\Eb_4\,D_q^{(2)}-
2380\,\Eb_6\,D_q^{(1)}+
6000\,\Eb_4^2}~.
\end{equation*}

\subsection*{Genus one, three punctures}

\small
\begin{flalign*}
\DD_{\CC_{1,3}}^{\mathfrak{a}_1} = 
D_q^{(6)}&-
\left(\tfrac{-61}{144}\Theta_{0,2}+\tfrac{61}{288}\Theta_{1,1}\right)D_q^{(4)}+
\left(\tfrac{13}{72}\Theta_{0,3}+\tfrac{-19}{48}\Theta_{1,2}\right)D_q^{(3)}&&\\
&+\left(\tfrac{-1205}{20736}\Theta_{0,4}+\tfrac{2501}{10368}\Theta_{1,3}+\tfrac{-3797}{13824}\Theta_{2,2}\right)D_q^{(2)}+
\left(\tfrac{655}{31104}\Theta_{0,5}+\tfrac{-3977}{62208}\Theta_{1,4}+\tfrac{1303}{31104}\Theta_{2,3}\right)D_q^{(1)}&&\\
&+
\left(\tfrac{-15}{4096}\Theta_{0,6}+\tfrac{45}{4096}\Theta_{1,5}+\tfrac{-25}{2048}\Theta_{2,4}+\tfrac{121}{8192}\Theta_{3,3}\right)~.&&
\end{flalign*}
\normalsize

\subsection*{Genus one, four punctures}

\small
\begin{flalign*}
\DD_{\CC_{1,4}}^{\mathfrak{a}_1} = D_q^{(9)} &-
840\,\Eb_4\,D_q^{(7)}-
41160\,\Eb_6\,D_q^{(6)}-
531600\,\Eb_4^2\,D_q^{(5)} - 12516000\,\Eb_4\Eb_6\,D_q^{(4)}&&\\
&-\left(71912400\,\Eb_6^2+3664000\,\Eb_4^3\right)D_q^{(3)}+2466072000\,\Eb_4^2\Eb_6\,D_q^{(2)}&&\\
&+
\left(56026208000\,\Eb_4\Eb_6^2+14324640000\,\Eb_4^4\right)D_q^{(1)}
+
\left(188260352000\,\Eb_6^3+381911040000\,\Eb_4^3\Eb_6\right)~.&&
\end{flalign*}
\normalsize	

\subsection*{Genus two, zero punctures}

\small
\begin{flalign*}
\DD_{\CC_{2,0}}^{\mathfrak{a}_1} = D_q^{(6)} &-
305\,\Eb_4 \,D_q^{(4)}-
4060\,\Eb_6 \,D_q^{(3)}+
20275\,\Eb_4^2 \,D_q^{(2)}+ 
2100\,\Eb_4\Eb_6 \,D_q^{(1)}-
\left(68600\,\Eb_6^2-49125\,\Eb_4^3\right)~.&&
\end{flalign*}
\normalsize	

%% file: sections/A5.tex

\section{Schur indices and differential operators for \texorpdfstring{$\NN=4$}{N=4} super Yang-Mills}
\label{app:N4_modular}

In this appendix we collect the exact expressions for the (unflavored) Schur index of $\NN=4$ super Yang-Mills theory with gauge algebra $\suf(n)$ for $2\leqslant n \leqslant 7$, along with the modular and twisted-modular differential operators that annihilate them. Various pieces of data about these differential operators and their kernels are collected in Table \ref{tab:N4}.

\medskip 

\noindent The unflavored Schur indices are expressed in terms of the complete elliptic integrals 
\begin{equation}
K(k)\colonequals \tfrac{\pi}{2}{}_2F_1\Big(\tfrac12, \tfrac12; 1; k^2 \Big)~,\qquad E(k)\colonequals \tfrac{\pi}{2}{}_2F_1\Big(\tfrac12, -\tfrac12; 1; k^2 \Big)~,
\end{equation}
where the modulus $k$ is given by
\begin{equation}
k^2=\frac{\vartheta_{10}(\tau)}{\vartheta_{01}(\tau)},
\end{equation}
in addition to the Dedekind eta function,
\begin{equation}
\eta(\tau)=q^{\frac{1}{24}}\prod_{n=1}^{\infty}(1-q^n)~.
\end{equation}
A general algorithm for writing the Schur index for any $n$ was given in \cite{Bourdier:2015wda}. We include the results of that prescription here for the reader's convenience.

\subsection*{\texorpdfstring{$\suf(2)$}{su(2)} gauge algebra}

The unflavored Schur index takes the very simple form:
\begin{equation}
\chi_{\suf(2)}(q)=\frac{1}{2\pi^2q^{3/8}}\frac{\eta(q^{1/2})^2}{\eta(q)^4}\Big(K(k)^2-K(k)E(k)\Big)~.
\end{equation}
This is annihilated by a second-order modular differential operator for the modular group $\Gamma^0(2)$,
\begin{equation}
\DD^{\NN=4}_{\suf(2)} \colonequals D_q^{(2)}-\tfrac{1}{12} \Theta_{0,1}(q)D_q^{(1)}-\tfrac{3}{64}\Big(\Theta_{0,1}(q)-\Theta_{1,1}(q)\Big)~.
\end{equation}
The conjugate differential operator whose kernel controls the high temperature behavior of the Schur index is then given by
\begin{equation}
\widetilde{\DD}^{\NN=4}_{\suf(2)} \colonequals D_q^{(2)}-\tfrac{1}{12} \widetilde{\Theta}_{0,1}(q)D_q^{(1)}-\tfrac{3}{64}\Big(\widetilde{\Theta}_{0,1}(q)-\widetilde{\Theta}_{1,1}(q)\Big)~.
\end{equation}

\subsection*{\texorpdfstring{$\suf(3)$}{su(3)} gauge algebra}

The unflavored Schur index takes the form:
\begin{equation}
\chi_{\suf(3)}(q)=\frac{\sqrt{k}}{24\pi^3q}\frac{\eta(q^{1/2})^2}{\eta(q)^4}K\Big(4(2-k^2)K^2-12E\,K+\pi^2)\Big)~.
\end{equation}
This is annihilated by a fourth-order modular differential operator for the full modular group,
\begin{equation}
\DD^{\NN=4}_{\suf(3)} \colonequals D_q^{(4)}-220\,\Eb_4(q)\,D_{q}^{(2)}+700\,\Eb_6(q)\,D_q^{(1)}~.
\end{equation}
We note that the constant function is a solution to the corresponding modular differential equation. It would be somewhat surprising for there to be a representation of the VOSA whose supercharacter is simply a constant. On the other hand, it seems unlikely that the ordinary supercharacters of this VOSA would transform only amongst themselves under the action of the full modular group, since the VOA is secretly $\frac12\Zb-$graded, so this may be a hint that additional modular differential operators are relevant in this example.

\subsection*{\texorpdfstring{$\suf(4)$}{su(4)} gauge algebra}

The unflavored Schur index in this case takes the form:
\begin{equation}
\chi_{\suf(4)}(q)=\frac{1}{24\pi^4q^{15/8}}\frac{\eta(q^{1/2})^2}{\eta(q)^4}K\Big(2k^2K^3+3K\left(K-E\right)^2+\left(K-E\right)\pi^2\Big)~.
\end{equation}
This is annihilated by a sixth-order modular differential operator for the modular group $\Gamma^0(2)$,
\small
\begin{equation}
\begin{split}
\DD^{\NN=4}_{\suf(4)} = D_q^{(6)}
&-\Big(
\tfrac{1}{4}\Theta_{0,1}
\Big)D_q^{(5)}
-\Big(
\tfrac{565}{576}\Theta_{0,2}
-\tfrac{413}{576}\Theta_{1,1}
\Big)D_q^{(4)}
+\Big(
\tfrac{53}{1152}\Theta_{0,3}
-\tfrac{23}{384}\Theta_{1,2}
\Big)D_q^{(3)}\\
&-\Big(
\tfrac{6329}{331776}\Theta_{0,4}
-\tfrac{1261}{82944}\Theta_{1,3}
-\tfrac{4823}{110592}\Theta_{2,2}
\Big)D_q^{(2)}
-\Big(
\tfrac{5515}{3981312}\Theta_{0,5}
-\tfrac{84145}{3981312}\Theta_{1,4}
+\tfrac{42515}{1990656}\Theta_{2,3}
\Big)D_q^{(1)}\\
&+\Big(
\tfrac{405}{262144}\Theta_{0,6}
-\tfrac{1215}{131072}\Theta_{1,5}
+\tfrac{6075}{262144}\Theta_{2,4}
-\tfrac{2025}{131072}\Theta_{3,3}
\Big)~.
\end{split}
\end{equation}
\normalsize
The conjugate differential operator, as usual, is given by
\small
\begin{equation}
\begin{split}
\widetilde{\DD}^{\NN=4}_{\suf(4)} = D_q^{(6)}
&-\Big(
\tfrac{1}{4}\wt\Theta_{0,1}
\Big)D_q^{(5)}
-\Big(
\tfrac{565}{576}\wt\Theta_{0,2}
-\tfrac{413}{576}\wt\Theta_{1,1}
\Big)D_q^{(4)}
+\Big(
\tfrac{53}{1152}\wt\Theta_{0,3}
-\tfrac{23}{384}\wt\Theta_{1,2}
\Big)D_q^{(3)}\\
&-\Big(
\tfrac{6329}{331776}\wt\Theta_{0,4}
-\tfrac{1261}{82944}\wt\Theta_{1,3}
-\tfrac{4823}{110592}\wt\Theta_{2,2}
\Big)D_q^{(2)}
-\Big(
\tfrac{5515}{3981312}\wt\Theta_{0,5}
-\tfrac{84145}{3981312}\wt\Theta_{1,4}
+\tfrac{42515}{1990656}\wt\Theta_{2,3}
\Big)D_q^{(1)}\\
&+\Big(
\tfrac{405}{262144}\wt\Theta_{0,6}
-\tfrac{1215}{131072}\wt\Theta_{1,5}
+\tfrac{6075}{262144}\wt\Theta_{2,4}
-\tfrac{2025}{131072}\wt\Theta_{3,3}
\Big)~.
\end{split}
\end{equation}
\normalsize

\subsection*{\texorpdfstring{$\suf(5)$}{su(5)} gauge algebra}

The unflavored Schur index takes the form:
\begin{eqnarray}
\chi_{\suf(5)}(q)&=&\frac{\sqrt{k}}{1920\pi^5q^{3}}\frac{\eta(q^{1/2})^2}{\eta(q)^4}K\times\\
&&
\Big(
16K^2\left(15 E^2 + 10(k^2-2)E\,K + (k^4-6k^2+6)K^2\right)-40K\left(3E+(k^2-2)K\right)\pi^2+9\pi^4
\Big)~.\nonumber
\end{eqnarray}
This is annihilated by a ninth-order modular differential operator for the full modular group,
\small
\begin{equation}
\begin{split}
\DD^{\NN=4}_{\suf(4)} = 
D_q^{(9)}
&-\Big(
2280\,\Eb_4
\Big)D_q^{(7)}
+\Big(
2520\,\Eb_6
\Big)D_q^{(6)}
+\Big(
447600\,\Eb_4^2
\Big)D_q^{(5)}
+\Big(
10600800\,\Eb_4\Eb_6
\Big)D_q^{(4)}\\
&-\Big(
122245200\,\Eb_6^2
-58544000\,\Eb_4^3
\Big)D_q^{(3)}
-\Big(
798504000\,\Eb_4^2\Eb_6
\Big)D_q^{(2)}\\
&+
\Big(
2626400000\,\Eb_4\Eb_6^2
-732000000\,\Eb_4^4
\Big)D_q^{(1)}~.
\end{split}
\end{equation}
\normalsize
Again we see that the constant function is a solution of the corresponding differential equation.

\subsection*{\texorpdfstring{$\suf(6)$}{su(6)} gauge algebra}

The unflavored Schur index in this case takes the form:
\small
\begin{eqnarray}
\chi_{\suf(6)}(q)=\frac{1}{720 \pi^6q^{35/8}}\frac{\eta(q^{1/2})^2}{\eta(q)^4}&\times&\Bigg(K^3 \Big(
\left(8 k^4-22 k^2+15\right)K^3+45 E^2 K+15\left(2k^2-3\right)E\,K^2 -15 E^3\Big)\nonumber\\
&+&K^2\pi^2\Big(15 E^2-30 E\,K-5\left(2 k^2-3\right) K^2\Big)+4K(K-E)\pi^4\Bigg)~.
\end{eqnarray}
\normalsize
This is annihilated by a twelfth-order modular differential operator for the modular group $\Gamma^0(2)$,
\small
\begin{align}
\DD^{\NN=4}_{\suf(6)} &= 
D_q^{(12)}
-\tfrac12\Theta_{0,1}D_q^{(11)}
-\Big(
\tfrac{2059}{288}\Theta_{0,2}
-\tfrac{1295}{288}\Theta_{1,1}
\Big)D_q^{(10)}
+\Big(
\tfrac{1117}{384}\Theta_{0,3}
-\tfrac{295}{128}\Theta_{1,2}
\Big)D_q^{(9)}\\
&+\Big(
\tfrac{835565}{110592}\Theta_{0,4}
-\tfrac{570589}{27648}\Theta_{1,3}
+\tfrac{633517}{36864}\Theta_{2,2}
\Big)D_q^{(8)}
-\Big(
\tfrac{2334617}{331776}\Theta_{0,5}
-\tfrac{6630635}{331776}\Theta_{1,4}
+\tfrac{2144425}{165888}\Theta_{2,3}
\Big)D_q^{(7)}\nonumber\\
&-\Big(
\tfrac{57188789}{47775744}\Theta_{0,6}
-\tfrac{46000633}{7962624}\Theta_{1,5}
+\tfrac{144715561}{15925248}\Theta_{2,4}
-\tfrac{98045357}{23887872}\Theta_{3,3}
\Big)D_q^{(6)}
+\Big(
\tfrac{153351577}{63700992}\Theta_{0,7}
-\tfrac{70688653}{7077888}\Theta_{1,6}\nonumber\\
&+\tfrac{1094336713}{63700992}\Theta_{2,5}
-\tfrac{640158317}{63700992}\Theta_{3,4}
\Big)D_q^{(5)}
+\Big(
\tfrac{41311042463}{110075314176}\Theta_{0,8}
-\tfrac{24908733703}{13759414272}\Theta_{1,7}
+\tfrac{103788648185}{27518828544}\Theta_{2,6}\nonumber\\
&-\tfrac{123313632241}{13759414272}\Theta_{3,5}
+\tfrac{707664949309}{110075314176}\Theta_{4,4}
\Big)D_q^{(4)}
-\Big(
\tfrac{948081389}{220150628352}\Theta_{0,9}
+\tfrac{1275390935}{73383542784}\Theta_{1,8}
-\tfrac{8914784053}{18345885696}\Theta_{2,7}\nonumber\\
&+\tfrac{23616046103}{18345885696}\Theta_{3,6}
-\tfrac{29820494551}{36691771392}\Theta_{4,5}
\Big)D_q^{(3)}
-\Big(
 \tfrac{5289572539}{31701690482688}\Theta_{0,10}
 -\tfrac{428584583851}{15850845241344}\Theta_{1,9}
 +\tfrac{2038969728437}{10567230160896}\Theta_{2,8}\nonumber\\
&-\tfrac{507660329051}{1320903770112}\Theta_{3,7}
 +\tfrac{740961138619}{1761205026816}\Theta_{4,6}
 -\tfrac{1153481373659}{5283615080448}\Theta_{5,5}
\Big)D_q^{(2)}
+\Big(
\tfrac{487083585625}{380420285792256}\Theta_{0,11}
-\tfrac{7200320745625}{380420285792256}\Theta_{1,10}\nonumber\\
&+\tfrac{19475477545075}{380420285792256}\Theta_{2,9}
-\tfrac{2979240028625}{126806761930752}\Theta_{3,8}
-\tfrac{3611051862425}{63403380965376}\Theta_{4,7}
+\tfrac{2957848559225}{63403380965376}\Theta_{5,6}
\Big)D_q^{(1)}
-\Big(
\tfrac{28704375}{68719476736}\Theta_{0,12}\nonumber\\
&-\tfrac{86113125}{17179869184}\Theta_{,111}
+\tfrac{947244375}{34359738368}\Theta_{10,2}
-\tfrac{1578740625}{17179869184}\Theta_{3,9}
+\tfrac{14208665625}{68719476736}\Theta_{4,8}
-\tfrac{2841733125}{8589934592}\Theta_{5,7}
+\tfrac{6630710625}{34359738368}\Theta_{6,6}
\Big)~.\nonumber
\end{align}
\normalsize
As usual, the conjugate differential operator is obtained by making the replacement $\Theta\leftrightarrow\wt\Theta$.

\subsection*{\texorpdfstring{$\suf(7)$}{su(7)} gauge algebra}
Finally, for this example, which is the largest rank theory we consider, the unflavored Schur index is given by
\small
\begin{eqnarray}
\chi_{\suf(7)}(q)&=&\frac{\sqrt{k}}{322560\pi^7q^{6}}\frac{\eta(q^{1/2})^2}{\eta(q)^4}K\times\\
&&\Bigg(
64 K^3 \Big(-105 E^3-105 E^2 \left(k^2-2\right) K-21 E \left(k^4-6 k^2+6\right) K^2-\left(k^2-2\right) \left(k^4-10 k^2+10\right) K^3\Big)\nonumber\\
&&+560 K^2 \Big(15 E^2+10 E \left(k^2-2\right) K+\left(k^4-6 k^2+6\right) K^2\Big)\pi^2
-1036 K \Big(3 E+\left(k^2-2\right) K\Big)+225\pi^6
\Bigg)~.\nonumber
\end{eqnarray}
\normalsize
This is annihilated by a sixteenth-order modular differential operator for the full modular group $\Gamma$, 
\small
\begin{equation}
\begin{split}
\DD^{\NN=4}_{\suf(7)} &= 
D_q^{(16)}
-\Big(
12080\,\Eb_4
\Big)D_q^{(14)}
-\Big(
80080\,\Eb_6
\Big)D_q^{(13)}
+\Big(
33532000\,\Eb_4^2
\Big)D_q^{(12)}
+\Big(
1026379200\,\Eb_4\Eb_6
\Big)D_q^{(11)}\\
&-\Big(
21787320800\,\Eb_6^2
+21421600000\,\Eb_4^3
\Big)D_q^{(10)}
-\Big(
1222270896000\,\Eb_4^2\Eb_6
\Big)D_q^{(9)}\\
&+\Big(
13105627808000\,\Eb_4\Eb_6^2
+634654880000\,\Eb_4^4
\Big)D_q^{(8)}\\
&+\Big(
507282434848000\,\Eb_6^3
-171337295360000\,\Eb_4^3\Eb_6
\Big)D_q^{(7)}\\
&+\Big(
1103642993600000\,\Eb_4^2\Eb_6^2
+680999091200000\,\Eb_4^5
\Big)D_q^{(6)}\\
&-\Big(
98797003267200000\,\Eb_4\Eb_6^3
-75265677984000000\,\Eb_4^4\Eb_6
\Big)D_q^{(5)}\\
&-\Big(
91303174664800000\,\Eb_6^4
-1480729074496000000\,\Eb_4^3\Eb_6^2
+366906872832000000\,\Eb_4^6
\Big)D_q^{(4)}\\
&-\Big(
17845543203936000000\,\Eb_4^2\Eb_6^3
-7734918175488000000\,\Eb_4^5\Eb_6
\Big)D_q^{(3)}\\
&-\Big(
423397795072000000\,\Eb_4\Eb_6^4
+57206131967040000000\,\Eb_4^4\Eb_6^2
-12425487482880000000\,\Eb_4^7
\Big)D_q^{(2)}\\
&-\Big(
94521954880640000000\,\Eb_6^5
-193140389164800000000\,\Eb_4^3\Eb_6^3
+58270974048000000000\,\Eb_4^6\Eb_6
\Big)D_q^{(1)}~.
\end{split}
\end{equation}
\normalsize
We see that once again, the constant function is a solution of the corresponding differential equation.